\documentclass[OA]{msc}

\usepackage{amsmath}
\usepackage{amsthm}
\usepackage{newtxmath}
\usepackage{mathtools}
\usepackage{url}
\usepackage{xspace}
\usepackage{xifthen}
\usepackage{tikz-cd}
\usepackage{enumitem}
\usepackage[shortcuts]{extdash}

\usepackage{mdframed}
\newenvironment{rules}
{
  \begin{minipage}{\textwidth}
  \tiny
  \begin{mdframed}
  \centering
}
{
  \end{mdframed}
  \end{minipage}
  ~ \\
}

\usepackage{bussproofs}

\newcommand{\proofskip}{0.5em}
\newenvironment{bproof}
  {\leavevmode\hbox\bgroup}
  {\DisplayProof\egroup}
  
\newcommand{\unaryRule}[3]{\begin{bproof}
\AxiomC{\ensuremath{#1}}
\RightLabel{\rulelabel{#3}}
\UnaryInfC{\ensuremath{#2}}
\end{bproof}\vspace{\proofskip}}

\newcommand{\binaryRule}[4]{\begin{bproof}
\AxiomC{\ensuremath{#1}}
\AxiomC{\ensuremath{#2}}
\RightLabel{\rulelabel{#4}}
\BinaryInfC{\ensuremath{#3}}
\end{bproof}\vspace{\proofskip}}

\newcommand{\ternaryRule}[5]{\begin{bproof}
\AxiomC{\ensuremath{#1}}
\AxiomC{\ensuremath{#2}}
\AxiomC{\ensuremath{#3}}
\RightLabel{\rulelabel{#5}}
\TrinaryInfC{\ensuremath{#4}}
\end{bproof}\vspace{\proofskip}}

\newcommand{\quadRule}[6]{\begin{bproof}
\AxiomC{\ensuremath{#1}}
\AxiomC{\ensuremath{#2}}
\AxiomC{\ensuremath{#3}}
\AxiomC{\ensuremath{#4}}
\RightLabel{\rulelabel{#6}}
\QuaternaryInfC{\ensuremath{#5}}
\end{bproof}\vspace{\proofskip}}

\newcommand{\pentRule}[7]{\begin{bproof}
	\AxiomC{\ensuremath{#1}}
	\AxiomC{\ensuremath{#2}}
	\AxiomC{\ensuremath{#3}}
	\AxiomC{\ensuremath{#4}}
	\AxiomC{\ensuremath{#5}}
	\RightLabel{\rulelabel{#7}}
	\QuinaryInfC{\ensuremath{#6}}
	\end{bproof}\vspace{\proofskip}}

\newcommand{\pentRuleTernaryConcl}[9]{\begin{bproof}
		\AxiomC{\ensuremath{#1}}
		\AxiomC{\ensuremath{#2}}
		\AxiomC{\ensuremath{#3}}
		\AxiomC{\ensuremath{#4}}
		\AxiomC{\ensuremath{#5}}
		\RightLabel{\rulelabel{#9}}
		\QuinaryInfC{\ensuremath{#6}}
		\noLine
		\UnaryInfC{\ensuremath{#7}}
		\noLine
		\UnaryInfC{\ensuremath{#8}}
	\end{bproof}\vspace{\proofskip}}

\newcommand{\neworrenewcommand}[1]{\providecommand{#1}{}\renewcommand{#1}}
      
      \newcommand{\pentRuleQuadConcl}[9]{
        \neworrenewcommand{\ffoo}[1]{
        \begin{bproof}
		\AxiomC{\ensuremath{#1}}
		\AxiomC{\ensuremath{#2}}
		\AxiomC{\ensuremath{#3}}
		\AxiomC{\ensuremath{#4}}
		\AxiomC{\ensuremath{#5}}
		\RightLabel{\rulelabel{##1}}
		\QuinaryInfC{\ensuremath{#6}}
		\noLine
		\UnaryInfC{\ensuremath{#7}}
		\noLine
		\UnaryInfC{\ensuremath{#8}}
                \noLine
		\UnaryInfC{\ensuremath{#9}}
	\end{bproof}\vspace{\proofskip}}\ffoo}

\newcommand{\binaryRuleBinaryConcl}[5]{\begin{bproof}
\AxiomC{\ensuremath{#1}}
\AxiomC{\ensuremath{#2}}
\RightLabel{\rulelabel{#5}}
\BinaryInfC{\ensuremath{#3}}
\noLine
\UnaryInfC{\ensuremath{#4}}
\end{bproof}\vspace{\proofskip}}

\newcommand{\ternaryRuleBinaryConcl}[6]{\begin{bproof}
	\AxiomC{\ensuremath{#1}}
	\AxiomC{\ensuremath{#2}}
	\AxiomC{\ensuremath{#3}}
	\RightLabel{\rulelabel{#6}}
	\TrinaryInfC{\ensuremath{#4}}
	\noLine
	\UnaryInfC{\ensuremath{#5}}
	\end{bproof}\vspace{\proofskip}}

\newcommand{\quadRuleBinaryConcl}[7]{\begin{bproof}
		\AxiomC{\ensuremath{#1}}
		\AxiomC{\ensuremath{#2}}
		\AxiomC{\ensuremath{#3}}
		\AxiomC{\ensuremath{#4}}
		\RightLabel{\rulelabel{#7}}
		\QuaternaryInfC{\ensuremath{#5}}
		\noLine
		\UnaryInfC{\ensuremath{#6}}
	\end{bproof}\vspace{\proofskip}}

\newcommand{\quadRuleQuadConcl}[9]{\begin{bproof}
\AxiomC{\ensuremath{#1}}
\AxiomC{\ensuremath{#2}}
\AxiomC{\ensuremath{#3}}
\AxiomC{\ensuremath{#4}}
\RightLabel{\rulelabel{#9}}
\QuaternaryInfC{\ensuremath{#5}}
\noLine
\UnaryInfC{\ensuremath{#6}}
\noLine
\UnaryInfC{\ensuremath{#7}}
\noLine
\UnaryInfC{\ensuremath{#8}}
\end{bproof}\vspace{\proofskip}}

\newcommand{\rulelabel}[1]{\hypertarget{#1}{\textsf{\scriptsize #1}}}
\newcommand{\ruleref}[1]{\hyperlink{#1}{\textsf{#1}}}

\newcommand{\UniMath}{\href{https://github.com/UniMath/UniMath}{\nolinkurl{UniMath}}\xspace}

\newcommand{\longhash}{840ac16608d642606e6ac4d6eab47fb2507a5256}
\newcommand{\shorthash}{840ac16}

\newcommand{\coqdocbasebaseurl}{https://benediktahrens.gitlab.io/sem2dtt/\shorthash}

\newcommand{\coqdocbaseurl}{\coqdocbasebaseurl /UniMath.}
\newcommand{\urlhash}{\#}
\newcommand{\coqdocurl}[2]{\coqdocbaseurl #1.html\urlhash #2}
\newcommand{\nolinkcoqident}[1]{\nolinkurl{#1}} %
\makeatletter
\newcommand{\coqident}{\begingroup\@makeother\#\@coqident}
\newcommand{\@coqident}[3][]{%
  \ifthenelse{\isempty{#2}}%
  {\nolinkcoqident{#3}}%
  {\ifthenelse{\isempty{#1}}%
  {\href{\coqdocurl{#2}{#3}}{\nolinkcoqident{#3}}}%
  {\href{\coqdocurl{#2}{#3}}{\nolinkcoqident{#1}}}}%
\endgroup}
\newcommand{\coqfile}[2]{%
  \ifthenelse{\isempty{#1}}%
  {\href{\coqdocbaseurl #2.html}{\nolinkcoqident{#2.v}}}%
  {\href{\coqdocbaseurl #1.#2.html}{\nolinkcoqident{#2.v}}}}
\makeatother

\newcommand{\jeq}{\equiv}
\newcommand{\peq}{=}
\newcommand{\defeq}{:=}

\newcommand{\constfont}[1]{\ensuremath{\mathsf{#1}}}
\newcommand{\cat}[1]{\ensuremath{\constfont{#1}}\xspace}
\newcommand{\CC}{\cat{C}}
\newcommand{\D}{\cat{D}}
\newcommand{\E}{\cat{E}}
\newcommand{\B}{\cat{B}}
\newcommand{\CAT}{\cat{CAT}}

\newcommand{\dob}[2]{\ensuremath{{#1}_{#2}}} %
\newcommand{\dmor}[3]{#1 \xrightarrow{#3} #2} %
\newcommand{\dtwo}[3]{#1 \xRightarrow{#3} #2} %
\newcommand{\total}[2][]{\ensuremath{\textstyle \int_{#1}{#2}}} %
\newcommand{\dproj}{\pi}

\newcommand{\disp}[1]{\overline{#1}}
\newcommand{\ff}{\disp{f}}
\renewcommand{\gg}{\disp{g}}
\newcommand{\hh}{\disp{h}}
\renewcommand{\aa}{\disp{a}}
\newcommand{\bb}{\disp{b}}
\newcommand{\cc}{\disp{c}}

\newcommand{\mytwocell}{\Rightarrow}
\newcommand{\invtwocell}[2]{#1 \Rightarrow #2}

\newcommand{\whiskerl}{\vartriangleleft}
\newcommand{\whiskerr}{\vartriangleright}
\newcommand{\vcomp}{\bullet}
\newcommand{\id}{\operatorname{id}}

\newcommand{\arrows}[1]{#1^{\downarrow}}
\newcommand{\arrowbicat}[1]{#1^{\rightarrow}}
\newcommand{\trivial}[2]{\ensuremath{{#1}^{+{#2}}}}
\newcommand{\Cat}{\cat{Cat}}
\newcommand{\Bicat}{\cat{Bicat}}
\newcommand{\StrictCat}{\cat{StrictCat}}
\newcommand{\CatOfStrictCat}{\underline{\cat{StrictCat}}}
\newcommand{\Terminal}{\Cat_{\cat{Terminal}}}
\newcommand{\BinProd}{\Cat_{\cat{BinProd}}}
\newcommand{\Pullback}{\Cat_{\cat{Pullback}}}
\newcommand{\FinLim}{\Cat_{\cat{FinLim}}}
\newcommand{\Initial}{\Cat_{\cat{Initial}}}
\newcommand{\Sums}{\Cat_{\cat{BinSum}}}
\newcommand{\DTerminal}{\D_{\cat{Terminal}}}
\newcommand{\DBinProd}{\D_{\cat{BinProd}}}
\newcommand{\DPullback}{\D_{\cat{Pullback}}}
\newcommand{\DFinLim}{\D_{\cat{FinLim}}}
\newcommand{\DInitial}{\D_{\cat{Initial}}}
\newcommand{\DSums}{\D_{\cat{BinSum}}}

\newcommand{\Fib}{\cat{Cleav}}
\newcommand{\Grpd}{\cat{Grpd}}
\newcommand{\OpFib}{\cat{OpCleav}}
\newcommand{\SplitOpFib}{\cat{IndexedCat}}
\newcommand{\IsoFib}{\cat{IsoFib}}
\newcommand{\SFib}[1]{\cat{SFib}(#1)}
\newcommand{\SOpFib}[1]{\cat{SOpFib}(#1)}

\newcommand{\subbicat}{\cat{SubBicat}}
\newcommand{\cod}{\constfont{cod}}
\newcommand{\dom}{\constfont{dom}}

\newcommand{\homC}[3]{\underline{#1(#2,#3)}}

\newcommand{\op}[1]{#1^{\cat{op}}}
\newcommand{\co}[1]{#1^{\cat{co}}}

\newcommand{\tc}{\theta}

\newcommand{\tcC}{\tau}
\newcommand{\assoc}{\alpha}

\newcommand{\cartesianlift}[2]{L_{#1}(#2)}
\newcommand{\opcartesianliftmor}[2]{\mu_{#1}(#2)}

\newcommand{\Factone}[2]{\ensuremath{\mathsf{F_1}(#1,#2)}}
\newcommand{\Facttwo}[2]{\ensuremath{\mathsf{F_2}(#1,#2)}}
\newcommand{\Eqtwo}[2]{\ensuremath{\mathsf{E}(#1,#2)}}
\newcommand{\PB}[2]{\ensuremath{{#1}^*{#2}}}
\newcommand{\PF}[2]{\ensuremath{{#1}_!{#2}}}

\newcommand{\leftunitor}[1]{\ell_{#1}}
\newcommand{\rightunitor}[1]{r_{#1}}

\newcommand{\reindex}[2]{#1^*(#2)}
\newcommand{\compbicat}[3]{
  \begin{tikzcd}[ampersand replacement=\&]
    {#1} \ar[rr, "{#2}"]  \&\&  \arrows{#3}
    \\
    \&
    {#3}
  \end{tikzcd}
}
\newcommand{\textcompbicat}[3]{\begin{tikzcd}[ampersand replacement=\&] {#1} \ar[r, "{#2}"]  \&  \arrows{#3}\end{tikzcd}\text{over }{#3}}

\newcommand{\defemph}[1]{\textbf{#1}}
\newcommand{\ie}{\emph{i.\,e.,} }

\newcommand{\eg}{\emph{e.\,g.,} }

\newcommand{\Coq}{\textsf{Coq}\xspace}

\newcommand{\Id}{\constfont{Id}}

\newcommand{\ctx}{\ \mathsf{ctx}}

\newcommand{\type}{\ \mathsf{type}}
\newcommand{\ctxbar}{\hspace{0.01em} \mid \hspace{0.01em} }
\newcommand{\Epsilon}{\Theta}
\newcommand{\Zeta}{\mathrm{Z}}
\newcommand{\Eta}{\mathrm{H}}
\newcommand{\rewrite}{\mathrel{\rightsquigarrow}}
\newcommand{\ctxzerocell}[1]{#1 \ctx}
\newcommand{\ctxonecell}[3]{#2 \vdash #1 : #3}
\newcommand{\typezerocell}[2]{#1 \vdash #2 \type}
\newcommand{\typeonecell}[4]{#1 \ctxbar #3 \vdash #2 : #4}
\newcommand{\typeonecellalt}[3]{#1 \ctxbar #3 \vdash #2}
\newcommand{\gtypezerocell}[1]{\typezerocell{\Gamma}{#1}}
\newcommand{\gtypeonecell}[3]{\typeonecell{\Gamma}{#1}{#2}{#3}}
\newcommand{\gtypeonecellalt}[3]{\Gamma \ctxbar #2 \vdash #1} %
\newcommand{\horcomp}[2]{#1 #2}
\newcommand{\horcompt}[3]{#1 #2  #3}

\newcommand{\vertcompwrongway}[2]{\vertcomp{#2}{#1}}
\newcommand{\vertcomp}[2]{#1 \vcomp #2}
\newcommand{\vertcompthree}[3]{#1 \vcomp #2 \vcomp #3}
\newcommand{\vertcompfour}[4]{#1 \vcomp #2 \vcomp #3 \vcomp #4}
\newcommand{\vertcompfive}[5]{#1 \vcomp #2 \vcomp #3 \vcomp #4 \vcomp #5}

\newcommand{\vdashtilde}{%
  \>
   \mathbin{\vbox{\offinterlineskip\ialign{%
    \hfil##\hfil\cr
    $\scriptstyle\sim$\cr
    $\vdash$\cr
  }}}
\>
}
\newcommand{\isoctxonecell}[3]{#2 \vdashtilde #1 : #3}
\newcommand{\isotypeonecell}[4]{#1 \ctxbar #3 \vdashtilde #2 : #4}

\newcommand{\isorewrite}{\mathrel{\tilde{\rewrite}}}

\newcommand{\proj}[2]{\pi_{#1.#2}}
\newcommand{\comm}[2]{c_{#1.#2}}

\newcommand{\comprid}[1]{\chi^{\cat{id}}_{#1}}
\newcommand{\comprcomp}[2]{\chi^{\cat{comp}}_{#1, #2}}

\newcommand{\substTy}[2]{#1[#2]}
\newcommand{\substTyTwo}[3]{#1[#2][#3]}
\newcommand{\substTyThree}[4]{#1[#2][#3][#4]}
\newcommand{\substTm}[2]{#1[#2]}
\newcommand{\substTmtwo}[3]{#1[#2][#3]}
\newcommand{\substRed}[2]{#1[#2]}
\newcommand{\substRedTwo}[3]{#1[#2][#3]}
\newcommand{\rewTm}[2]{#1[#2]}
\newcommand{\mapoperator}[0]{\cat{map}}
\newcommand{\map}[2]{\mapoperator \ #1 \ #2}

\newcommand{\suboperator}[0]{\cat{sub}}

\newcommand{\subidoperator}[0]{\suboperator_{\cat{id}}}
\newcommand{\subid}{\subidoperator}
\newcommand{\invsubid}[1]{\subidoperator^{-1} \ #1}

\newcommand{\subcompoperator}[0]{\suboperator_{\cat{comp}}}
\newcommand{\subcomp}[2]{\subcompoperator(#1, #2)}
\newcommand{\invsubcomp}[2]{\subcompoperator^{-1}(#1, #2)}

\newcommand{\subcompmapsubidl}[2]{\suboperator_{\cat{compidl}}(#1,#2)}
\newcommand{\subcompmapsubidr}[2]{\suboperator_{\cat{compidr}}(#1,#2)}
\newcommand{\subcompcomp}[4]{\suboperator_{\cat{2comp}}(#1,#2,#3,#4)}

\newcommand{\mapidoperator}[0]{\mapoperator_{\cat{id}}}
\newcommand{\mapid}[2]{\mapidoperator^{#1} \ #2}
\newcommand{\mapcompoperator}[0]{\mapoperator_{\cat{comp}}}
\newcommand{\mapcomp}[3]{\mapcompoperator^{#1}(#2, #3)}
\newcommand{\maplwhiskeroperator}[0]{\mapoperator_{\cat{l}}}
\newcommand{\maplwhisker}[3]{\maplwhiskeroperator^{#1}(#2, #3)}
\newcommand{\maprwhiskeroperator}[0]{\mapoperator_{\cat{r}}}
\newcommand{\maprwhisker}[3]{\maprwhiskeroperator^{#1}(#2, #3)}

\newcommand{\sem}[1]{\llbracket #1 \rrbracket}

\newcommand{\ruleExtendTy}{extend-con-Ty}
\newcommand{\ruleExtendTm}{extend-con-Tm}
\newcommand{\ruleExtendRed}{extend-con-Red}
\newcommand{\ruleExtendId}{extend-con-id}
\newcommand{\ruleExtendComp}{extend-con-comp}

\newcommand{\tmSubId}[1]{\cat{STm_I}(#1)}
\newcommand{\tmSubComp}[3]{\cat{STm_C}(#1, #2, #3)}
\newcommand{\subOnId}[1]{\cat{Sub_I}(#1)}
\newcommand{\subOnComp}[3]{\cat{Sub_C}(#1, #2, #3)}

\newcommand{\ruleSubTy}{sub-ty}
\newcommand{\ruleSubTm}{sub-tm}
\newcommand{\ruleSubRed}{sub-red}
\newcommand{\ruleMap}{map}
\newcommand{\ruleRewTm}{rew-tm}
\newcommand{\ruleRewRed}{rew-red}
\newcommand{\ruleRewOnId}{rew-pres-id}
\newcommand{\ruleRewOnComp}{rew-pres-comp}
\newcommand{\ruleSubId}{sub-id}
\newcommand{\ruleSubComp}{sub-comp}
\newcommand{\ruleTmSubId}{tm-sub-id}
\newcommand{\ruleTmSubComp}{tm-sub-comp}
\newcommand{\ruleSubOnId}{sub-pres-id}
\newcommand{\ruleSubOnComp}{sub-pres-comp}
\newcommand{\ruleMapId}{map-id}
\newcommand{\ruleMapComp}{map-comp}
\newcommand{\ruleMapWhiskerL}{map-lwhisker}
\newcommand{\ruleMapWhiskerR}{map-rwhisker}
\newcommand{\ruleSubRedId}{sub-red-id}
\newcommand{\ruleSubRedComp}{sub-red-comp}
\newcommand{\ruleSubRedOnId}{sub-red-pres-id}
\newcommand{\ruleSubRedOnComp}{sub-red-pres-comp}
\newcommand{\ruleSubLunitor}{sub-pres-lunitor}
\newcommand{\ruleSubRunitor}{sub-pres-runitor}
\newcommand{\ruleSubAssoc}{sub-pres-assoc}
\newcommand{\ruleSubLWhisker}{sub-pres-lwhisker}
\newcommand{\ruleSubRWhisker}{sub-pres-rwhisker}

\newcommand{\ruleSubIdPrevId}{sub-id-pres-id}
\newcommand{\ruleSubIdPrevComp}{sub-id-pres-comp}

\newcommand{\ruleSubCompPrevId}{sub-comp-id}
\newcommand{\ruleSubCompPrevComp}{sub-comp-comp}

\newcommand{\ruleSubCompMapSubIdL}{sub-comp-map-id-l}
\newcommand{\ruleSubCompMapSubIdR}{sub-comp-map-id-r}
\newcommand{\ruleSubCompMapSubAssoc}{sub-comp-map-assoc}

\newcommand{\BTT}{\cat{BTT}}
\newcommand{\MLTT}{\cat{MLTT}}

\newcommand{\twodimensional}{two-di\-men\-sion\-al\xspace}
\newcommand{\higherdimensional}{higher-di\-men\-sion\-al\xspace}
\newcommand{\inftygroupoids}{$\infty$-group\-oids\xspace}
\newcommand{\simplytyped}{sim\-ply-typed\xspace}
\newcommand{\metatheoretic}{me\-ta-the\-o\-ret\-ic\xspace}
\newcommand{\proofirrelevant}{proof-ir\-rel\-e\-vant\xspace}
\usepackage[shortcuts]{extdash}

\newcommand{\isocomma}[2]{#1 \mathbin{/_\simeq} #2}
\newcommand{\isocommafst}{\pi_1^\simeq}
\newcommand{\isocommasnd}{\pi_2^\simeq}
\newcommand{\isocommacomm}{n^\simeq}

\newcounter{enumcounter}

\newtheorem{example}[therm]{Example}

\newtheorem{problem}[therm]{Problem}%
\newtheorem{constrInternal}[therm]{Construction}%

\newenvironment{construction}[2][]
 {\pushQED{\qed}\begin{constrInternal}[{for Problem~\ref{#2}\ifthenelse{\isempty{#1}}{}{; #1}}]}
 	{\popQED\end{constrInternal}}

\begin{document}

\lefttitle{Bicategorical type theory: semantics and syntax}
\righttitle{Ahrens, North, and Van der Weide}

\papertitle{Article}

\jnlPage{1}{00}
\jnlDoiYr{2019}
\doival{10.1017/xxxxx}

\title{Bicategorical type theory: semantics and syntax}

\begin{authgrp}
  
\author{Benedikt Ahrens}
\affiliation{
  Delft University of Technology}
\affiliation{
  University of Birmingham\\
  \email{B.P.Ahrens@tudelft.nl}
}
\author{Paige Randall North}
\affiliation{
  Utrecht University\\
  \email{p.r.north@uu.nl}
 }
\author{ Niels van der Weide}
\affiliation{
  Radboud University\\
  \email{nweide@cs.ru.nl}
}

\end{authgrp}

\history{(Received xx xxx xxx; revised xx xxx xxx; accepted xx xxx xxx)}

\begin{abstract}

  We develop semantics and syntax for bicategorical type theory.
  Bicategorical type theory features contexts, types, terms, and directed reductions between terms.
  This type theory is naturally interpreted in a class of structured bicategories.
   We start by developing the semantics, in the form of 
  \emph{comprehension bicategories}.
  Examples of comprehension bicategories are plentiful;
  we study both specific examples as well as classes of examples constructed from other data.
  From the notion of comprehension bicategory, we extract the syntax of bicategorical type theory, that is, judgment forms and structural inference rules.
  We prove soundness of the rules by giving an interpretation in any comprehension bicategory.
  The semantic aspects of our work are fully checked in the Coq proof assistant, based on the UniMath library.

\end{abstract}

\begin{keywords}
  directed type theory; dependent types; comprehension bicategory; computer-checked proof
\end{keywords}

\maketitle

\section{Introduction}

In recent years, efforts have been made to develop \emph{directed} type theory.
Roughly, directed type theory should correspond to Martin-Löf type theory ($\MLTT$) as $\infty$-categories correspond to \inftygroupoids.
Besides theoretical interest in directed type theory, it is hoped that such a type theory can serve as a framework for synthetic directed homotopy theory and synthetic $\infty$-category theory.
Applications of those, in turn, include reasoning about concurrent processes \cite[]{DBLP:books/sp/FajstrupGHMR16}.

Several proposals for \emph{syntax} for directed type theory have been given (reviewed in Section~\ref{sec:directed-type-theory}),
but are ad-hoc and are not always semantically justified.
The \emph{semantic} aspects of directed type theory are particularly underdeveloped;
a general notion of model of a directed type theory is still lacking. Indeed these proposals often provide an interpretation of their syntax in categories. They either employ pre-existing $1$-categorical semantical tools, thus forcing them to interpret their syntax into a $1$-category of categories, as in the work by \cite{DBLP:journals/entcs/North19}, or, in the absence of $2$-categorical tools, give an ad-hoc interpretation of their syntax into the $2$-category of categories, as done by \cite{LicataH11}. We rectify this situation by providing a $2$-categorical semantical structure.

Specifically, in this work we introduce \emph{comprehension bicategories} as a suitable mathematical structure for higher-dimensional (directed) type theory. Approaching the development of directed type theory from the semantic side, we extract from this the core syntax---judgment forms and structural inference rules --- of a \twodimensional dependent type theory that can accommodate directed type theory.
We also give a soundness proof of our structural rules.
In future work, we will equip our syntax and semantics with variances and type and term formers for directed type theory.

To motivate our approach, we analyze in Section~\ref{sec:judg-and-typal} how higher-groupoidal structure arises in \MLTT through an interplay of judgmental equality and typal identity.
Our analysis thus leads to the desiderata listed in Section~\ref{sec:intro-deriv-syn-from-sem}.
In Section~\ref{sec:foundations} we discuss the foundations we work in, and aspects of the computer formalization of some of our results.

\subsection{Judgmental and Typal Higher Dimensions}
\label{sec:judg-and-typal}

The judgment forms of traditional \MLTT specify types, contexts, terms, and judgmental equality (conversion) between types and terms.
There is, prima facie, nothing higher-dimensional about these judgments, and an interpretation of types as sets and terms as elements of sets seems perfectly adequate. In this sense, Martin-Löf type theory is 1-dimensional.
However, \MLTT is often said to be $\infty$-dimensional.
The higher dimensions are generated by the identity type, which internalizes the judgmental equality;
specifically, the well-known \textbf{reflexivity} rule generates a typal identity from a judgmental equality.
Since the identity type can be iterated, judgmental equality then also becomes available for terms of the identity type itself.
This mutual interaction between judgmental equality and typal identity provides the infrastructure to ``lift'' judgmental equality to higher dimensions without extending the judgmental structure of \MLTT.
The tower of types $(A, \Id_A,\Id_{\Id_A},\ldots)$ then can be given the structure of an $\infty$-groupoid, as shown by \cite{van2011types} and \cite{lumsdaine2010weak}.

When developing a \emph{directed} type theory, with models in $\infty$-\emph{categories}, the analogous ingredients are the following:

\begin{enumerate}[label=\bfseries I\arabic*:, align=left,
  ref={I\arabic*},topsep=0pt]
\item \label{ingredient:judg} A \emph{judgment} of \emph{directed reductions} between types and terms, analogous to judgmental equality;
\item \label{ingredient:type} A \emph{type former} for \emph{homomorphisms} between terms, analogous to identity types;
\item \label{ingredient:sem} A notion of \emph{model} in which to interpret the judgments and type formers.
\end{enumerate}

In the present work, we propose a judgmental framework (\ref{ingredient:judg}), and a suitable general notion of semantics (\ref{ingredient:sem}), for \higherdimensional and directed type theory. 
In a separate work we will expand this core by a system of \emph{variances} suitable for accommodating a type former akin to North's hom-types (\ref{ingredient:type}), to build a fully functional higher-dimensional type theory.

\subsection{Deriving Syntax from Semantics}
\label{sec:intro-deriv-syn-from-sem}

Previous work on higher-dimensional and directed type theory --- reviewed in detail in Section~\ref{sec:relwork} --- has focused on syntax (\ref{ingredient:judg}/\ref{ingredient:type}).
\cite{DBLP:conf/popl/LicataH12,LicataH11} and \cite{nuyts_dtt} devise judgmental structure for higher-dimensional and directed type theory.
\cite{DBLP:journals/entcs/North19} devises a type former for directed homomorphisms between terms, on top of the judgmental structure of \MLTT.
None of these works propose a general definition of \emph{model} of directed type theory.
\cite{DBLP:journals/mscs/Garner09} defines a notion of higher-di\-men\-sional model, but considers only \emph{un}directed type theory.

Our approach is different from that of previous work on directed and higher-dimensional type theory.
Specifically, we choose to approach the challenge from the other direction:
we start by devising a suitable categorical structure for directed type theory, and extract a syntax from it.

We refine the ingredients above to the following list of desiderata for our work:
\begin{enumerate}[label=\bfseries D\arabic*:, align=left,
  ref={D\arabic*},topsep=0pt]
\item \label{goal:syn} A system of inference rules for dependent types with directed reductions;
\item \label{goal:sem} A definition of \emph{mathematical structures} suitable for the mathematical modelling of the syntactic rules;
\item \label{goal:interp} An \emph{interpretation} of the inference rules in such a mathematical structure;
\item \label{goal:syn-typeformers} A syntax for type and term formers on top of \ref{goal:syn};
\item \label{goal:sem-typeformers} A semantic structure for the interpretation of type and term formers.
\end{enumerate}

\noindent
In the present work, we achieve desiderata \ref{goal:syn}, \ref{goal:sem}, and \ref{goal:interp}.
The study of variances and type and term constructors will be reported on elsewhere.
In Section~\ref{sec:relwork}, we refer back to these desiderata for describing related work.

The semantics we propose are described in Section~\ref{sec:comp-bicats},
and the extracted syntax is described in Section~\ref{sec:syntax}.
Both our syntax and semantics are quite general;
for instance, our reductions are \emph{proof-relevant} --- like those considered by \cite{DBLP:conf/popl/LicataH12,LicataH11}, and unlike judgmental equality in MLTT, which is proof-irrelevant.
Syntax and semantics could reasonably be simplified or specialized.
Crucially, our work provides a framework to modify syntax and semantics \emph{in lockstep}, with a clear mechanism to analyze changes to the syntax on the semantic side and vice versa.
We suggest some possible variants in Section~\ref{sec:variations-syntax}.

\subsection{Foundations and Formalization in UniMath}
\label{sec:foundations}

The main results presented here are agnostic to foundations:
they can be formalized in both set theory and type theory.

However, some of the notions we employ can economically be formulated using dependent types.
In particular, we work with cloven (Grothendieck) fibrations of (bi)categories, whose formulation in set theory usually relies on postulating equality of objects.
Using dependent types, a formulation of such concepts can be given that avoids any reasoning about equality of objects; instead, these concepts are formulated in terms of fibers.
For this reason, we use type-theoretic language throughout the paper;
see also, \eg Remark~\ref{rem:cartesian-via-dep-types}.

We carefully distinguish data and property; in particular, we postulate entities in categories (\eg limits) to be explicitly given as data rather than to merely exist. 
We do not rely on any choice axioms or on excluded middle.

The results of Sections~\ref{sec:prelims} to \ref{sec:disp_map_bicat} of this work are checked in \Coq~\cite[]{the_coq_development_team_2022_5846982}, based on the \UniMath library \cite[]{UniMath} of univalent mathematics.
In univalent mathematics, we distinguish \emph{strict} (bi)categories and \emph{univalent} (bi)categories \cite[Section~4]{ahrens_frumin_maggesi_veltri_van_der_weide_2022}.
Our definitions are agnostic to this difference; hence, our definition of bicategory (Definition~\ref{def:bicat}) does not make a commitment to either strictness or univalence.
Specific examples can then be strict (such as Example~\ref{ex:split-opfib}) or univalent (such as Example~\ref{exa:sopfib-dispmapbicat}); see also Remark~\ref{rem:not-assume-univalence}.

Our code has been integrated into \UniMath in commit
\href{https://github.com/UniMath/UniMath/commit/\longhash}{\shorthash}.
The computer-checked definitions and results are accompanied by a link (\eg \coqident{Bicategories.Core.Bicat}{bicat})
to the corresponding definition in an HTML version of that commit.
The code written specifically for this work comprises approximately 33,600 lines of code;
specifically, the \verb!coqwc! tool counts as follows:

\begin{verbatim}
 spec    proof comments
13143    20553      449 total
\end{verbatim}

We build upon an existing library of (bi)category theory by \cite{AhrensKS15,ahrens_frumin_maggesi_veltri_van_der_weide_2022},
and use heavily the \emph{displayed} machinery, developed for 1-categories by \cite{AhrensL19} and extended to bicategories by \cite{ahrens_frumin_maggesi_veltri_van_der_weide_2022}.
In particular, the formalized notions of cloven Grothendieck fibration we are using (in the 1-categorical case) and developing (in the bicategorical case) are based on displayed (bi)categories;
we can thus discuss these notions without postulating equality of objects.

The syntax presented in Section~\ref{sec:syntax} and its interpretation given in Section~\ref{sec:soundn-interpr-compr} are not computer-checked;
we therefore give these constructions in more detail in the paper.

\subsection{Synopsis}
\label{sec:synopsis}

In Section~\ref{sec:relwork} we review related work.
In Section~\ref{sec:prelims} we review (displayed) bicategories and functors.
In Section~\ref{sec:fibrations} we define cloven global and local (op/iso)fibrations of bicategories, and we use these notions to define comprehension bicategories.
In Section~\ref{sec:comp-bicats} we discuss some examples of comprehension bicategories.
In Section~\ref{sec:internal-sfibs} we discuss Street (op)fibrations internal to bicategories, which form our main examples of comprehension bicategories.
In Section~\ref{sec:disp_map_bicat} we define display map bicategories, and we show how any display map bicategory gives rise to a comprehension bicategory.
In Section~\ref{sec:syntax} we present structural type-theoretic rules for the syntax of a two-dimensional type theory, which we call $\BTT$.
In Section~\ref{sec:soundn-interpr-compr} we give an interpretation of $\BTT$ in any comprehension bicategory.
In Section~\ref{sec:variations-syntax} we discuss variations of $\BTT$ and the semantic structures these variations correspond to.
In Section~\ref{sec:terms} we explain the difference between terms in \BTT and terms in other approaches to directed type theory.

\subsection{Version History}

A shorter version of this paper was published in the proceedings of Logic in Computer Science under a different name \cite[]{DBLP:conf/lics/AhrensNW22}.
Compared to that shorter version, the following material has been added in the present version:
\begin{itemize}
\item We provide more instantiations of Example~\ref{ex:streetopfib-compcat}. In particular, we show that several bicategories of structured categories have pullbacks (Example~\ref{ex:pb_in_structured_categories}).
\item We give more detail on the specific comprehension bicategory given by functors into the 1-category of strict categories, in Example~\ref{ex:splitopfibincat-compcat}. We also present a formalization of that example.
\item We introduce display map bicategories and show that any display map bicategory gives rise to a comprehension bicategory.
\item We present the type theory $\BTT$ in more detail, and specify more of its type-theoretic rules.
\item We give more detail in the interpretation of the rules of $\BTT$ in any comprehension bicategory.
\end{itemize}
Furthermore, compared to the short version, we have added the definition of \emph{weak} comprehension bicategory (see Definition~\ref{def:comp-bicat} and Remark~\ref{rem:no-preservation-of-cart}), where the comprehension $\chi$ is not required to preserve (op)cartesian cells.
Our interpretation of \BTT works in any such weak comprehension bicategory.

\section{Related Work}
\label{sec:relwork}

In this section, we review work with a similar goal to ours, as well as work we rely on.
We pay particular attention to the desiderata outlined in Section~\ref{sec:intro-deriv-syn-from-sem} and to the difference between judgmental and typal dimensions.

\subsection{Non-dependent type theories}

The following works satisfy a non-dependent variant of \ref{goal:syn}, together with suitable adaptations of \ref{goal:sem} and \ref{goal:interp}.
However, due to the absence of type dependency, they do not immediately compare to our work.

\cite{Seely87} presents a syntax for a \twodimensional \simplytyped lambda calculus, consisting of types, terms, and reductions between terms.
Seely then constructs a 2-category out of that syntax.
\cite{DBLP:conf/foal/Tabareau11} frames aspect-oriented programming in a 2-categorical way,
developing a lambda calculus that provides an internal language for 2-categories.
\cite{DBLP:journals/corr/Hirschowitz13} constructs a 2-ad\-junction between 2-signatures for lambda calculi (where such signatures specify types, terms, and reductions) and the category of Cartesian closed 2-categories.
\cite{DBLP:conf/lics/FioreS19} construct an internal language for cartesian closed bicategories; the result is a class (parametrized by a notion of signature for constants) of \emph{simple} 2-dimensional type theories or lambda calculi.
This last work shares one aspect with ours that the others do not: it uses (weak) \emph{bi}categorical structure, rather than (strict) \emph{2-}categorical structure.

\subsection{Theories for Higher Categories}

There is a body of work on designing type theories for $\omega$-groupoids and $\omega$-categories.
In these type theories, one works, semantically speaking, \emph{within one fixed} $\infty$-groupoid (or $\omega$-category).
Compare this to, \eg Martin-Löf type theory, where one manipulates $\infty$-groupoids (types and identity types) and $\infty$-functors (functions) between them.
Analogously, in our type theory, each type can be thought of as a category.
Despite these different goals, we mention some of the work in this area.

\cite{brunerie_phd} constructs a type theory whose models are weak $\infty$-groupoids. \cite{DBLP:journals/corr/abs-2106-04475} (see also the work by \cite{DBLP:conf/lics/FinsterM17}) design a type theory whose models are precisely $\omega$-categories à la Grothendieck--Maltsiniotis.
\cite{tt-for-st-unital} study \metatheoretic properties of a language for strictly unital $\infty$-categories.
There are also computer tools implementing such type theories, see, \eg
the work by \cite{DBLP:journals/lmcs/BarKV18} and \cite{DBLP:conf/lics/ReutterV19}.

\subsection{Theories with Dependent Types}
\label{sec:theor-with-depend}

In this section we review work on higher-dimensional and directed type theory with dependent types.
We start with a review of work on \emph{un}directed type theory.

\subsubsection{Undirected Type Theory}
\label{sec:undir-type-theory}

The idea of considering \higherdimensional interpretations of type theory stems from Hofmann and Streicher's groupoid interpretation of Martin-Löf type theory by \cite{DBLP:conf/lics/HofmannS94}.
This interpretation is generalized to stacks (poset-indexed groupoids satisfying a sheaf condition) in order to prove the independence of several logical principles by \cite{DBLP:conf/lics/CoquandMR17}.
It is furthermore generalized, from different angles, to higher dimensions, see, \eg work by \cite{van2011types}, \cite{lumsdaine2010weak}, and \cite{kapulkin2021simplicial}.
All of this work considers the Martin-Löf identity type, which is undirected.

\cite{DBLP:conf/popl/LicataH12} develop a two-dimensional dependent type theory with a judgment for \emph{equivalences} $\Gamma \vdash \alpha : M \simeq_A N$ between terms $M,N:A$. These equivalences are postulated to have (strict) inverses.
The authors give an interpretation of types as groupoids: terms are (interpreted as) objects in the interpreting groupoid, and equivalences are morphisms, necessarily invertible.
No general notion of semantic structure is discussed;
this work hence satisfies an \emph{un}directed version of \ref{goal:syn}, but not \ref{goal:sem}.

\cite{https://doi.org/10.48550/arxiv.2111.09438} introduce judgmental theories and calculi for them as a general framework to present and study deductive systems.
They instantiate their framework to obtain a type theory that describes an internal language of 2-toposes \cite[Section~5.3]{https://doi.org/10.48550/arxiv.2111.09438}; however, they only consider a one-dimensional type theory à la Martin-Löf, and thus their work does not satisfy \ref{goal:syn}.

\cite{DBLP:journals/mscs/Garner09} 
studies a typal two-dimensional type theory à la Martin-Löf: the forms of judgment are the same as in Martin-Löf type theory. Garner calls a type $X$ ``discrete'' if it satisfies identity reflection (that is, if any identity $p : x \peq y$ between elements $x,y:X$ induces a judgmental equality $x \jeq y$. Rules are then added that turn any identity type into a discrete type, effectively making every type into a 1-type, in the sense of the h-levels of homotopy type theory. 
Garner defines a notion of two-dimensional model based on (strict) \emph{comprehension 2-categories}.
Exploiting the restriction to 1-truncated types, a sound and complete interpretation of that two-dimensional type theory in any model is then given.
The Martin-Löf identity type is undirected;
correspondingly, Garner defines their comprehension 2-categories to consist of \emph{locally groupoidal} 2-categories.
Thus, Garner's work satisfies \ref{goal:syn} for \emph{undirected} reductions, using the identity type for this purpose. Garner also considers type constructors such as dependent pair types and dependent product types, thus satisfying \ref{goal:syn-typeformers} and \ref{goal:sem-typeformers} in this case.%
\footnote{Garner also relies on Hermida's \cite[]{HERMIDA199983} slightly incomplete definition of fibration of 2-categories; see Buckley's work \cite[Remark~2.1.9]{buckley2014fibred} for details.
  We have not checked if Garner's work extends to Buckley's corrected definition of 2-fibration.}

\subsubsection{Directed Type Theory}
\label{sec:directed-type-theory}

\cite{LicataH11} (see also \cite[Chapter~7]{10.5555/2338432}) also design a \emph{directed} two-dimensional type theory and give an interpretation for it in the strict 2-category of categories.
Their syntax has a judgment for \emph{substitutions} between contexts, written $\Gamma \vdash \theta : \Delta$, and \emph{transformations} between parallel substitutions.
An important aspect of their work is \emph{variance} of contexts/types, built into the judgments.
The type formers there have a certain variance --- \emph{co}variance or \emph{contra}variance --- in each of the arguments.
They do not define a general notion of model for their theory;
this work hence satisfies \ref{goal:syn}, but not \ref{goal:sem}.

Nuyts~\cite[Section~1.3.1]{nuyts_dtt} observes that the type theory developed by \cite{LicataH11} does not allow for a non-trivial Martin-Löf identity type --- any such type would coincide with the directed transformations.
Nuyts thus attempts to generalize the treatment of variance by Licata and Harper,
and designs a directed type theory with additional variances, such as isovariance and invariance.
Nuyts does not provide any interpretation of their syntax, and thus no proof of (relative) consistency;
the work hence does not satisfy \ref{goal:sem}.

\cite{DBLP:journals/entcs/North19} develops a type former for \emph{directed} types of morphisms, resulting in a typal higher-dimensional directed type theory based on the judgments of \MLTT.
North's work thus does not satisfy \ref{goal:syn}.
The model given by North is in the 2-category of categories, similar to the model by \cite{LicataH11}.

\cite{shulman_2-topos}, in unfinished work, aims to develop 2-categorical logic, including a two-dimensional notion of topos and a suitable internal language for such toposes.
Specifically, Shulman sketches two internal languages for 2-toposes.
The first language \cite[]{shulman_internal-logic} is undirected, consisting only of types and terms.
The second language \cite[]{shulman_functorially-dep-types} is only described in a short sketch; it is a kind of directed type theory featuring, in particular, variances.
Our work is similar to Shulman's in the sense that both start from a (bi)categorical notion and extract a language from it,
with the goal of developing a precise correspondence between extensions of the syntax and additional structure on the semantics.
Unfortunately, Shulman's work is unfinished, which makes a more complete evaluation difficult.
However, it contains several ideas that have influenced the present work.
For instance, \cite{shulman_fibrational-slice} emphasizes the usefulness of restricting to (op)fibrations instead of considering all 1-cells when constructing bicategories of arrows --- we do this in our main examples of comprehension bicategory, Examples~\ref{ex:opfibincat-compcat} and \ref{ex:streetopfib-compcat}.

\cite{RS17} design a \emph{simplicial} type theory (STT) featuring a directed interval type, as a synthetic theory of $(\infty, 1)$-categories.
As a notion of model, they introduce ``comprehension categories with shapes'' \cite[Def.~A.5]{RS17}.
These are (1-categorical) towers of fibrations accounting for several layers of contexts.
Further work on STT was done, among others, by \cite{DBLP:conf/lics/WeaverL20} and \cite{DBLP:journals/corr/abs-2105-01724}.
STT is not higher-dimensional in the sense of \cite{LicataH11} or the present work;
in particular, reductions, both in the tope layer and in the type layer, are undirected.
This work thus does not satisfy \ref{goal:syn}.

\paragraph*{Summary}

In the present work, we define bicategorical semantics for the interpretation of types, terms, and reductions, and derive from it a system of inference rules; our work thus satisfies \ref{goal:syn}, \ref{goal:sem}, and \ref{goal:interp}.
We do not handle \ref{goal:syn-typeformers} and \ref{goal:sem-typeformers} in this work.

Among the described related work, our work is closest to work by \cite{LicataH11} and \cite{DBLP:journals/mscs/Garner09}.
Compared to \cite{LicataH11}, we add a general definition of ``model'' of a \emph{directed} two-dimensional type theory, and provide many examples of models.
Compared to \cite{DBLP:journals/mscs/Garner09}, we cover \emph{directed} reductions, and provide many instances of our general definition of model.
Compared to both works, we do \emph{not} handle type and term formers.

\section{Preliminaries}
\label{sec:prelims}

Here, we sketch some definitions used later on.
Many would be very long if given in full;
instead, we try to convey some intuition while pointing to the formalized definitions.
As a reference for bicategory theory, see \cite{10.1007/BFb0074299}.
We use here the vocabulary and notation introduced by \cite{ahrens_frumin_maggesi_veltri_van_der_weide_2022}.

\begin{definition}[\coqident{Bicategories.Core.Bicat}{bicat}]
  \label{def:bicat}
  A \defemph{bicategory} consists of a type $\B_0$ of 0-cells (or objects), a type
$a \to b$ of \emph{1-cells from $a$ to $b$} for every $a,b: \B_0$,
and a \emph{set} $f \mytwocell g$ of \emph{2-cells from $f$ to $g$} for every $a,b: \B_0$ and $f,g: a \to b$.
We have identity $\id_1(a) : a \to a$ and composition of 1-cells $f \cdot g : a \to c$ (also written $fg$), which we write in diagrammatical order.
These operations do \emph{not} satisfy the axioms for a 1-category.
Instead, we have, for instance, the \emph{left unitors}, that is, invertible 2-cells $\leftunitor{f} : \id_1(a) \cdot f \mytwocell f$ for any 1-cell $f$,
and similarly right unitors $\rightunitor{f} : f \cdot \id_1(b) \mytwocell f$.
Analogously, we have the \emph{associators}, a family of invertible 2-cells  $\alpha(f, g, h) : f \cdot (g \cdot h) \to (f \cdot g) \cdot h$.
For 2-cells $\tc : f \mytwocell g$ and $\tcC : g \mytwocell h$ (where $f,g,h : a \to b$ for some $a,b : \B_0$), we have a \emph{vertical composition} $\tc \vcomp \tcC : f \mytwocell h$.
For any 1-cell $f : a \to b$, we have an \emph{identity 2-cell} $\id_2(f) : f \mytwocell f$,
which is neutral with respect to vertical composition: $\id_2(f) \vcomp \tc = \tc$.
For any two objects $a$ and $b$, the 1-cells from $a$ to $b$, and 2-cells between them, form the objects and morphisms of the hom-category $\homC{\B}{a}{b}$, with composition given by vertical composition of $\B$.
We also have \emph{left} and \emph{right whiskering};
given a 2-cell $\tc : f \mytwocell g : b \to c$ and a 1-cell $e : a \to b$, we have the \emph{left whiskering} $e \whiskerl \tc : e \cdot f \mytwocell e \cdot g$, and, similarly, the \emph{right whiskering}
$\tc \whiskerr h : f \cdot h \mytwocell g \cdot h$ for $h : c \to d$.
We do not list the axioms that these operations satisfy; the interested reader can consult, \eg Def.~2.1 of \cite{ahrens_frumin_maggesi_veltri_van_der_weide_2022}.

We occasionally write $1_a$ for $\id_1(a)$ and $1_f$ for $\id_2(f)$.

\end{definition}

\begin{remark}
\label{rem:not-assume-univalence}
We do \emph{not} generally require that our bicategories (and displayed bicategories, see Definition~\ref{def:disp-bicat}) are univalent in the sense of \cite{ahrens_frumin_maggesi_veltri_van_der_weide_2022}.

Sometimes it is still interesting to assume that a given (displayed) bicategory is univalent;
in such cases, fibrations are better behaved, since lifts can be shown to be actually unique rather than just essentially unique (see, for instance, Proposition~\ref{prop:univ-cart-1cell-unique}).
Whenever we state such a result, the assumption on the (displayed) bicategory to be univalent is stated explicitly.
\end{remark}

We denote by $\Cat$ the bicategory of categories, and by $\Grpd$ the bicategory of groupoids.
The bicategory $\op{\B}$ has the same objects as $\B$, but 1-cells from $x$ to $y$ in $\op{\B}$ are 1-cells $f : y \rightarrow x$ in $\B$.
The 2-cells in $\op{\B}$ are 2-cells in $\B$.
The bicategory $\co{\B}$ has the same objects and 1-cells as $\B$, but 2-cells from $f$ to $g$ in $\co{\B}$ are the same as 2-cells from $g$ to $f$ in $\B$.

\begin{definition}[\coqident{Bicategories.PseudoFunctors.PseudoFunctor}{psfunctor}]
  \label{def:pseudofunctor}
  Given two bicategories $\B$ and $\B'$, a \defemph{pseudofunctor} $F : \B \to \B'$ is given by
maps $F_0 : \B_0 \to \B'_0$, $F_1 : (a\to b) \to (F_0a \to F_0b)$,%
\footnote{Note that $\to$ is used for both 1-cells and function types.}
and $p_2 : (f \mytwocell g) \to (F_1f \mytwocell F_1g)$, preserving structure on 1-cells up to invertible 2-cells in $\B'$ (specified as part of the functor $F$) and preserving structure on 2-cells up to equality.

\end{definition}

We build complicated bicategories from simpler ones by adding structure at all dimensions.
The additional structure should come with its own composition and identity, which should lie suitably over composition and identity of the original bicategory.
This idea is formalized in the notion of \emph{displayed bicategory} --- a layer of data over a base bicategory --- and the resulting \emph{total bicategory} --- the bicategory of pairs $(b,\disp{b})$ of a cell $b$ in the base and a cell $\disp{b}$ ``over'' $b$.
We also obtain a pseudofunctor from the total bicategory into the base, given at all dimensions by the first projection.

\begin{definition}[{\cite[Def.~6.1]{ahrens_frumin_maggesi_veltri_van_der_weide_2022}}, \coqident{Bicategories.DisplayedBicats.DispBicat}{disp_bicat}]
  \label{def:disp-bicat}
  Let $\B$ be a bicategory. A \defemph{displayed bicategory $\D$ over $\B$}
  consists of
  \begin{enumerate}
  \item for any $b : \B_0$, a type $\dob{\D}{b}$ of \defemph{objects over $b$};
  \item for any $f : a\to b$ and $x : \dob{\D}{a}$ and $y : \dob{\D}{b}$, a type $\dmor{x}{y}{f}$ of \defemph{1-cells over $f$};
  \item for any $\tc : f \mytwocell g$ and $\disp{f} : \dmor{x}{y}{f}$ and $\disp{g} : \dmor{x}{y}{g}$, a \emph{set} $\dtwo{\disp{f}}{\disp{g}}{\tc}$ of \defemph{2-cells over $\tc$};
  \end{enumerate}
  together with suitably typed composition (over composition in $\B$) and identity (over identity in $\B$) for both 1- and 2-cells.
  These operations are subject to ``axioms over axioms in $\B$''.
\end{definition}

\begin{definition}[{\cite[Def.~6.2]{ahrens_frumin_maggesi_veltri_van_der_weide_2022}}, \coqident{Bicategories.DisplayedBicats.DispBicat}{total_bicat}]
  Given a displayed bicategory $\D$ over $\B$, we define the \defemph{total bicategory} $\total{\D}$
  to have, as cells at dimension $i$, pairs $(b,\disp{b})$ where $b$ is a cell of $\B$ at dimension $i$ and $\disp{b}$ is a cell of $\D$ over $b$, with the obvious source and target.

  We define the \defemph{projection} pseudofunctor $\dproj : \total{\D} \to \B$ to be given, on any cell, by $(b,\disp{b})\mapsto b$.
\end{definition}

\begin{definition}[\coqident{Bicategories.DisplayedBicats.Examples.Sub1Cell}{disp_subbicat}]
\label{def:subbicat}
Suppose that we have a bicategory $\B$, a predicate $P_0$ on the 0-cells (by which we mean a proposition $P_0(a)$ for every 0-cell $a$), and a predicate $P_1$ on the 1-cells.
Furthermore, we assume that $P_1$ is closed under identity and composition; that is, $P_1(\id(x))$ holds for every $x$ satisfying $P_0$ and that for all $f : x \rightarrow y$ and $g : y \rightarrow z$ between 0-cells satisfying $P_0$ we have $P_1(f \cdot g)$ if we have both $P_1(f)$ and $P_1(g)$.
Then we define a displayed bicategory $\subbicat(P_0, P_1)$ on $\B$ such that
\begin{itemize}
	\item the type of displayed objects over $x$ is $P_0(x)$;
	\item the type of displayed 1-cells over $f : x \rightarrow y$ is $P_1(f)$; and
	\item the type of displayed 2-cells over $\tc : f \Rightarrow g$ is the unit type.
\end{itemize}
\end{definition}

\noindent
The total bicategory of this bicategory selects 0-cells and 1-cells from the original bicategory $\B$.
Its objects are objects $x : \B$ such that $P_0(x)$, its 1-cells are 1-cells $f : x \rightarrow y$ in $\B$ such that $P_1(f)$, and its 2-cells are the same as 2-cells $\tau : f \Rightarrow g$ in $\B$.
The projection pseudofunctor $\dproj : \subbicat(P_0, P_1) \to \B$ is the inclusion.

\begin{remark}
  Note that Definition \ref{def:subbicat} is defined slightly differently in the formalization. In \coqident{Bicategories.DisplayedBicats.Examples.Sub1Cell}{disp_subbicat}, we do not require that $P_0$ and $P_1$ are propositions. Instead, we only require this in the theorem \coqident{Bicategories.DisplayedBicats.Examples.Sub1Cell}{is_univalent_2_subbicat} that shows the univalence of this displayed bicategory.
\end{remark}
  
We instantiate Definition \ref{def:subbicat} to define bicategories of categories with a certain structure.

\begin{example}[\coqfile{Bicategories.Core.Examples}{StructuredCategories}]
\label{example:structured_cats}
We define the following displayed bicategories over $\Cat$:
\begin{itemize}
	\item $\DTerminal$ where $P_0(\CC)$ says that $\CC$ has a terminal object and $P_1(F)$ says that $F$ preserves terminal objects.
	We denote its total bicategory by $\Terminal$.
	\item $\DBinProd$ where $P_0(\CC)$ says that $\CC$ has binary products and $P_1(F)$ says that $F$ preserves binary products.
	We denote its total bicategory by $\BinProd$.
	\item $\DPullback$ where $P_0(\CC)$ says that $\CC$ has pullbacks and $P_1(F)$ says that $F$ preserves pullbacks.
	We denote its total bicategory by $\Pullback$.
	\item $\DFinLim$ where $P_0(\CC)$ says that $\CC$ has finite limits and $P_1(F)$ says that $F$ preserves finite limits.
	We denote its total bicategory by $\FinLim$.
	\item $\DInitial$ where $P_0(\CC)$ says that $\CC$ has an initial object and $P_1(F)$ says that $F$ preserves initial objects.
	We denote its total bicategory by $\Initial$.
	\item $\DSums$ where $P_0(\CC)$ says that $\CC$ has binary coproducts and $P_1(F)$ says that $F$ preserves binary coproducts.
	We denote its total bicategory by $\Sums$.
\end{itemize}
\end{example}

\noindent
The above examples can be defined using Definition \ref{def:subbicat} assuming that the categories involved are univalent, since then the type families $P_0$ and $P_1$ stating a choice of limits are indeed predicates.
For non-univalent categories, one could consider instead the truncated predicates stating mere existence of limits.

The total bicategories of the displayed bicategories in Example~\ref{example:structured_cats} are bicategories of categories with certain (co)limits.
In addition, we can combine the structure of these by taking the product of the suitable displayed bicategories.
Note that we could use similar methods to construct bicategories of extensive categories, regular categories, exact categories, and (pre)toposes.

The following (displayed) bicategories are used:
\begin{example}[\coqident{Bicategories.DisplayedBicats.Examples.Trivial}{trivial_displayed_bicat}]
Given bicategories $\B_1$ and $\B_2$, we define a displayed bicategory $\trivial{\B_1}{\B_2}$ over $\B_1$ as follows:
\begin{itemize}
\item The displayed 0-cells over $x : \B_1$ are 0-cells $y : \B_2$.
\item The displayed 1-cells over $f : x_1 \to x_2$ from $y_1 : \B_2$ to $y_2 : \B_2$ are 1-cells $g : y_1 \to y_2$ in $\B_2$.
\item The displayed 2-cells over $\tc : f \Rightarrow g$ from $g_1 : y_1 \rightarrow y_2$ to $g_2 : y_1 \rightarrow y_2$ are 2-cells $\tcC : g_1 \Rightarrow g_2$ in $\B_2$.
\end{itemize}
The total bicategory is $\total{\trivial{\B_1}{\B_2}} \simeq \B_1 \times \B_2$ with projection $\dproj : \B_1 \times \B_2 \to \B_1$.
\end{example}

\begin{example}[\coqident{Bicategories.DisplayedBicats.Examples.Codomain}{cod_disp_bicat}]
\label{example:arrows}
Let $\B$ be a bicategory.
Define a displayed bicategory $\arrows{\B}$ over $\B$ as follows:
\begin{itemize}
\item The displayed objects over $y : \B$ are 1-cells $x \rightarrow y$.
\item The displayed 1-cells over $g : y_1 \rightarrow y_2$ from $h_1 : x_1 \rightarrow y_1$ to $h_2 : x_2 \rightarrow y_2$ are pairs consisting of a 1-cell $f : x_1 \rightarrow x_2$ and an invertible 2-cell $\gamma : \invtwocell{g \cdot h_2}{h_1 \cdot f}$.
\item Given displayed 1-cells $f_1 : x_1 \rightarrow x_2$ with $\gamma_1 : \invtwocell{g_1 \cdot h_2}{h_1 \cdot f_1}$, and $f_2 : x_1 \rightarrow x_2$ with $\gamma_2 : \invtwocell{g_2 \cdot h_2}{h_1 \cdot f_2}$, we define the displayed 2-cells over $\tc : g_1 \Rightarrow g_2$ from $(f_1, \gamma_1)$ to $(f_2, \gamma_2)$ as 2-cells $\tcC : f_1 \Rightarrow f_2$ such that $\gamma_1 \vcomp (h_1 \whiskerl \tcC) = (\tc \whiskerr h_2) \vcomp \gamma_2$.
\end{itemize}
The generated total bicategory is the \emph{arrow bicategory}, $\total{\arrows{\B}} = \arrowbicat{\B}$ with projection given by the codomain, $\cod : \arrowbicat{\B} \to \B$.
\end{example}

In the next example, we write $\StrictCat$ for the bicategory of strict categories and $\CatOfStrictCat$ for the category of strict categories.

\begin{example}[\coqident{Bicategories.DisplayedBicats.Examples.FunctorsIntoCat}{disp_bicat_of_functors_into_cat}]
\label{ex:split-opfib}
We define a displayed bicategory $\SplitOpFib$ over $\StrictCat$ as follows:
\begin{itemize}
	\item The displayed objects over $\CC : \StrictCat$ are functors $G : \CC \rightarrow \CatOfStrictCat$.
	\item The displayed 1-cells over $F : \CC_1 \rightarrow \CC_2$ from $G_1 : \CC_1 \rightarrow \CatOfStrictCat$ to $G_2 : \CC_2 \rightarrow \CatOfStrictCat$ are natural transformations $\gamma : G_1 \Rightarrow F \cdot G_2$.
	\item The displayed 2-cells over $n : F_1 \Rightarrow F_2$ from $\gamma_1 : G_1 \Rightarrow F_1 \cdot G_2$ to $\gamma_2 : G_1 \Rightarrow F_2 \cdot G_2$ are proofs that for all $x : \CC$ we have $\gamma_1(x) \cdot G_2(n(x)) = \gamma_2(x)$.
\end{itemize}
The associated projection pseudofunctor $\dproj : \total{\SplitOpFib} \to \StrictCat$ maps functors $\CC \rightarrow \CatOfStrictCat$ to their domain.
\end{example}

\begin{example}[\coqident{Bicategories.DisplayedBicats.Examples.DispBicatOfDispCats}{disp_bicat_of_opcleaving}]
\label{ex:opfib}
We define a displayed bicategory $\OpFib$ over $\Cat$ as follows:
\begin{itemize}
\item The displayed objects over $\CC : \Cat$ are displayed categories $\D$ over $C$ together with an opcleaving.
\item The displayed 1-cells over $F : \CC_1 \rightarrow \CC_2$ from $\D_1$ to $\D_2$ are displayed functors $\disp{F}$ from $\D_1$ to $\D_2$ that preserve opcartesian morphisms.
\item The displayed 2-cells over $\tc : F \Rightarrow G$ from $\disp{F} : \dmor{\D_1}{\D_2}{F}$ to $\disp{G} : \dmor{\D_1}{\D_2}{G}$ are displayed natural transformations from $\disp{F}$ to $\disp{G}$ over $\tc$.
\end{itemize}
The associated projection pseudofunctor $\dproj : \total{\OpFib} \to \Cat$ maps any opcleaving to its codomain category.
\end{example}

Similarly, we define displayed bicategories $\Fib$ and $\IsoFib$ of cleavings and isocleavings, respectively.

The idea of displayed (bi)categories transfers to functors:
\begin{definition}[{\cite[Def.~8.2]{ahrens_frumin_maggesi_veltri_van_der_weide_2022}}, \coqident{Bicategories.DisplayedBicats.DispPseudofunctor}{disp_psfunctor}]
Given $F : \B \to \B'$ and $\D$ and $\D'$ displayed bicategories over $\B$ and $\B'$, respectively, a \defemph{displayed pseudofunctor} $\disp{F}$ over $F$ consists of
\begin{itemize}
	\item for all objects $x : \B$ and $\disp{x} : \dob{\D}{x}$ an object $\disp{F}(\disp{x}) : \dob{\D'}{F(x)}$;
	\item for all displayed morphisms $\ff : \dmor{\disp{x}}{\disp{y}}{f}$, a displayed 1-cell $\disp{F}(\ff) : \dmor{\disp{F}(\disp{x})}{\disp{F}(\disp{y})}{F(f)}$;
	\item for all displayed 2-cells $\disp{\tc} : \dtwo{\ff}{\gg}{\tc}$, a displayed 2-cell $\disp{F}(\disp{\tc}) : \dtwo{\disp{F}(\ff)}{\disp{F}(\gg)}{F(\tc)}$;
        \item coherence isomorphisms for identity and composition of displayed 1-cells, over the analogous isomorphisms in the base;
        \item satisfying suitable equations over the corresponding equations in the base.
\end{itemize}
We denote by $\total{\disp{F}} : \total{\D} \to \total{\D'}$ the induced \defemph{total pseudofunctor}.
\end{definition}
\begin{remark}
  \label{rem:total-square}
The square of pseudofunctors
\begin{equation*}%
 \begin{tikzcd}
    \total{\D} \ar[r, "\total{\disp{F}}"] \ar[d, "\dproj_{\D}"']
    &
    \total{\D'} \ar[d, "\dproj_{\D'}"]
    \\
    \B \ar[r, "F"]
    &
    \B'
  \end{tikzcd}
\end{equation*}
induced by $\disp{F}$ over $F$
commutes \emph{up to judgmental equality}. %
\end{remark}

Furthermore, we need pullbacks and products in bicategories.
\begin{definition}[\coqident{Bicategories.Limits.Pullbacks}{has_pb}]
\label{def:pullback}
Let $\B$ be a bicategory, and suppose we have two 1-cells $f : a \rightarrow c$ and $g : b \rightarrow c$.
A \defemph{pullback structure for $f$ and $g$} on an object $x : \B$ together with two 1-cells $\pi_1 : x \rightarrow a$ and $\pi_2 : x \rightarrow b$ and an invertible 2-cell $\gamma : \invtwocell{p \cdot f}{q \cdot g}$ is given by the following data:
\begin{itemize}
	\item for all 1-cells $p' : z \rightarrow a$ and $q' : z \rightarrow b$ and invertible 2-cells $\gamma' : \invtwocell{p' \cdot f}{q' \cdot g}$, we have a 1-cell $u : z \rightarrow x$ together with invertible 2-cells $\tc : \invtwocell{u \cdot p}{p'}$ and $\tcC : \invtwocell{u \cdot q}{q'}$ such that 
	\[
	\assoc \vcomp \tc \whiskerr f \vcomp \gamma' = u \whiskerl \gamma \vcomp \assoc^{-1} \vcomp \tcC \whiskerr g.
	\]
	\item for all 1-cells $u_1, u_2 : z \rightarrow x$ and 2-cells $\tc : u_1 \vcomp p \mytwocell u_2 \vcomp p$ and $\tcC : u_1 \vcomp q \mytwocell u_2 \vcomp q$ such that 
	\[
	u_1 \whiskerl \gamma \vcomp \assoc \vcomp \tcC \whiskerr q \vcomp \assoc^{-1}
	=
	\assoc \vcomp \tc \whiskerr f \vcomp \assoc^{-1} \vcomp u_2 \whiskerl \gamma,
	\]
        we have a unique 2-cell $\nu : u_1 \mytwocell u_2$ such that $\nu \whiskerr p = \tc$ and $\nu \whiskerr q = \tcC$.
\end{itemize}
\end{definition}

In Definition~\ref{def:pullback} it is irrelevant if we postulate data to be given explicitly or to merely exist, provided the bicategory $\B$ is univalent:

\begin{proposition}[\coqident{Bicategories.Limits.Pullbacks}{isaprop_has_pb_ump}]
If $\B$ is univalent, then the type of pullback structures on $(x,\pi_1,\pi_2, \gamma)$ is a proposition.
\end{proposition}

\begin{remark}
There are different notions of pullback in bicategories depending on whether $p \cdot f$ and $q \cdot g$ are postulated to be related up to an equality, invertible 2-cell or even just a 2-cell.
In Definition~\ref{def:pullback}, the square commutes up to invertible 2-cell.
One could also define \emph{strict pullbacks}: this is done similarly to Definition~\ref{def:pullback}, but all involved squares must commute up to equality rather than just up to invertible 2-cell.
\end{remark}

\begin{example}[\coqident{Bicategories.Limits.Examples.OneTypesLimits}{one_types_has_pb}, \ \coqident{Bicategories.Limits.Examples.BicatOfUnivCatsLimits}{has_pb_bicat_of_univ_cats}]
\label{ex:pullbacks}
The bicategory of groupoids has pullbacks.

The bicategory $\Cat$ also has pullbacks, and they are given by iso-comma categories.
These are defined as follows:
given categories $\CC_1$, $\CC_2$, and $\CC_3$, and functors $F : \CC_1 \rightarrow \CC_3$ and $G : \CC_2 \rightarrow \CC_3$, we define the \textbf{iso-comma category} $\isocomma{F}{G}$
\begin{itemize}
	\item Its objects consist of objects $x : \CC_1$ and $y : \CC_2$ together with an isomorphism $f : F(x) \rightarrow G(y)$
	\item The morphisms from $(x_1, y_1, f_1)$ to $(x_2, y_2, f_2)$ consists of morphisms $g : x_1 \rightarrow x_2$ and $h : y_1 \rightarrow y_2$ such that the following diagram commutes:
	\[
	\begin{tikzcd}
		{F(x_1)} && {G(y_1)} \\
		\\
		{F(x_2)} && {G(y_2)}
		\arrow["{F(g)}"', from=1-1, to=3-1]
		\arrow["{f_1}", from=1-1, to=1-3]
		\arrow["{G(h)}", from=1-3, to=3-3]
		\arrow["{f_2}"', from=3-1, to=3-3]
	\end{tikzcd}
	\]
\end{itemize}

We define functors $\isocommafst : \isocomma{F}{G} \rightarrow \CC_1$ and $\isocommasnd : \isocomma{F}{G} \rightarrow \CC_2$ and a natural isomorphism $\isocommacomm : \isocommafst \cdot F \Rightarrow \isocommasnd \cdot G$.
The category $\isocomma{F}{G}$ together with the functors $\isocommafst,\isocommasnd$ and natural isomorphism $\isocommacomm$ is universal among such data, making it the pullback of $F$ and $G$.
\end{example}

\begin{example}[\coqfile{Bicategories.Limits.Examples}{LimitsStructuredCategories}]
\label{ex:pb_in_structured_categories}
The bicategories defined in Example~\ref{example:structured_cats} all have pullbacks as well.
This is because taking the iso-comma category preserves the desired (co)limits and the projections and pairing functors preserve them.
\end{example}

As a special case of pullbacks in the presence of terminal objects, we can define products in bicategories (\coqident{Bicategories.Limits.Products}{has_binprod_ump}).
If $\B$ has chosen products, we write $x \times y$ for the product of $x$ and $y$, and we denote the projections by $\pi_1 : x \times y \rightarrow x$ and $\pi_2 : x \times y \rightarrow y$.

\begin{notation}
  In the following, given a cloven fibration, we notate the pullback (or reindexing) of $A$ along $f$ by $\PB{f}{A}$.
  Given a cloven opfibration, we notate the pushforward of $A$ along $f$ by $\PF{f}{A}$.
\end{notation}

\section{Comprehension Bicategories}
\label{sec:fibrations}
In this section, we define the notion of global cleaving and local (op)cleavings of bicategories.
Afterwards, we use these notions to define comprehension bicategories.
We are guided by \cite{buckley2014fibred}, where local and global fibrations are defined, and we add definitions for local cloven iso- and opfibrations.
However, there is an important difference: while Buckley works in a set-theoretic setting,
we reformulate the definitions in a type-theoretic setting using the displayed technology developed by \cite{ahrens_frumin_maggesi_veltri_van_der_weide_2022} and reviewed in Section~\ref{sec:prelims} --- see also Remark~\ref{rem:cartesian-via-dep-types}.

Throughout this section, we assume that $\B$ is a bicategory and $\D$ is a displayed bicategory over $\B$.

\begin{definition}[{\cite[Def.~3.1.1]{buckley2014fibred}}, \coqident{Bicategories.DisplayedBicats.CleavingOfBicat}{cartesian_1cell}]
\label{def:cart-1-cell}
Let $f : a \to b$ be a 1-cell in $\B$, and let $\ff : \dmor{\aa}{\bb}{f}$ be a displayed 1-cell over $f$ in $\D$.
A \defemph{cartesian structure} on $\ff$  consists of the following data.
Note that we draw diagrams in the displayed category on the left side and diagrams in the base category on the right side.
\begin{enumerate}
\item \label{item:def-cartesian-1cell-1cell} For any $\gg : \dmor{\cc}{\bb}{h \cdot f}$, a choice of a displayed morphism $\hh : \dmor{\cc}{\aa}{h}$ and a displayed isomorphism $\tc$ over the identity isomorphism on $h \cdot f$ in $\B$.
\[
\begin{tikzcd}[column sep=large, row sep = large]
 \cc \ar[dr, "\gg", ""'{name=U}] \ar[d, dashed, "\hh"']
 \\
 \aa \ar[r,"\ff"'] \ar[Rightarrow, dashed, "\tc", to=U,]
 &
 \bb
\end{tikzcd}
\quad \quad
\begin{tikzcd}[column sep=large, row sep = large]
 c \ar[d, "h"'] \ar[dr, "h \cdot f", ""'{name=V}]
 \\
 a \ar[r, "f"']  \ar[Rightarrow, "\id_2", to=V]
 &
 b
\end{tikzcd}
\]
We call $(\hh,\tc)$ the \textbf{lift} of $(h,\gg)$.
\item \label{item:def-cartesian-1cell-2cell}
Given lifts $(\hh_1,\tc_1)$ and $(\hh_2, \tc_2)$ of $(h_1,\gg_1)$ and $(h_2,\gg_2)$, respectively, and $\delta : h_1 \mytwocell h_2$, and a 2-cell $\disp{\sigma} : \gg_1 \mytwocell \gg_2$ over $\delta \whiskerr f$,
we have a unique 2-cell $\disp{\delta} : \hh_1 \mytwocell \hh_2$ over $\delta$ such that
$\disp{\delta}\whiskerr \ff \vcomp \tc_2 = \tc_1 \vcomp \disp{\sigma}$.
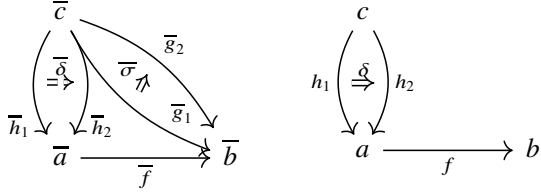
\[
\begin{tikzcd}[column sep = huge, row sep = huge]
 \cc \ar[dr, "\gg_1" near end, ""{name=U1}, bend right=20] \ar[dr, "\gg_2", ""'{name=U2}, bend left=20]\ar[d, "\hh_1"'near end, ""{name=Uhh1}, bend right] \ar[d, "\hh_2"near end, ""'{name=Uhh2}, bend left]
 \\
 \aa \ar[r,"\ff"'] %
 &
 \bb
 \ar[from=U1,to=U2,Rightarrow, "\disp{\sigma}"]
 \ar[from=Uhh1,to=Uhh2,Rightarrow, dashed, "\disp{\delta}"]
\end{tikzcd}
\quad \quad
\begin{tikzcd}[column sep = huge, row sep = huge]
 c \ar[d, "h_1"', ""{name=UU}, bend right=30]\ar[d, "h_2", ""'{name=VV}, bend left=30] 
 \\
 a \ar[r, "f"']  %
 &
 b
  \ar[from=UU, to=VV, Rightarrow, "\delta"]
\end{tikzcd}
\]

\end{enumerate}
Given a cartesian structure on $\ff$, we call $\Factone{\ff}{\gg} \defeq \hh$ and $\Facttwo{\ff}{\gg} \defeq \theta$.
We also call $\Eqtwo{\delta}{\disp{\sigma}} \defeq \disp{\delta}$.
\end{definition}

\begin{proposition}
\label{prop:example_carts}
We have the following cartesian 1-cells:
\begin{itemize}
	\item (\coqident{Bicategories.DisplayedBicats.Cartesians}{cartesian_1cell_id}) The identity $\disp{\id_1} : \aa \rightarrow \aa$ is cartesian.
	\item (\coqident{Bicategories.DisplayedBicats.Cartesians}{comp_cartesian_1cell}) If $\ff : \aa \rightarrow \bb$ and $\gg : \bb \rightarrow \cc$ are cartesian, then so is $\ff \cdot \gg$.
\end{itemize}
\end{proposition}

\begin{problem}
\label{prob:equiv_carts}
Given two cartesian 1-cells $\ff : \aa \rightarrow \bb$ and $\ff' : \aa' \rightarrow \bb$, to construct an adjoint equivalence $e : \aa \rightarrow \aa'$ over the identity and an invertible 2-cell from $\ff$ to $e \cdot \ff'$.
\end{problem}

\begin{construction}[\coqfile{Bicategories.DisplayedBicats}{EquivalenceBetweenCartesian}]{prob:equiv_carts}
\label{const:equiv_carts}
The 1-cells of the adjoint equivalence are constructed as cartesian factorizations (first item of Definition~\ref{def:cart-1-cell}). 
In that way, we also obtain the desired invertible 2-cell.
\end{construction}

\begin{remark}\label{rem:cartesian-via-dep-types}
  Recall that a displayed 1-cell $\ff : \dmor{\aa}{\bb}{f}$ in $\D$ gives rise to the 1-cell $(f,\ff)$ in the total bicategory $\total{\D}$.
  The definition of cartesian structure on $\ff$ in $\D$ of Definition~\ref{def:cart-1-cell} gives rise to a notion of cartesian structure for $(f,\ff)$ in $\total{\D}$.
  By expressing the definition of cartesian 1-cell in the displayed bicategory (instead of in the resulting projection $\dproj : \total{\D} \to \B$), we can postulate that a lift in $\total{\D}$ lies \emph{definitionally} over a given cell in $\B$, not just modulo an invertible 2-cell.

  \cite{buckley2014fibred} shows, in Remark 3.1.6, that these two definitions are equivalent.
\end{remark}  

\begin{proposition}[\coqident{Bicategories.DisplayedBicats.Cartesian}{isaprop_cartesian_1cell}]
  \label{prop:univ-cart-1cell-unique}
Suppose that $\D$ is a univalent displayed bicategory over $\B$, and let $\ff$ be a displayed 1-cell in $\D$ over $f$ in $\B$.
Then the type of cartesian 1-cell structures on $\ff$ is a proposition.
\end{proposition}

\begin{definition}[\coqident{Bicategories.DisplayedBicats.CleavingOfBicat}{global_cleaving}]
\label{def:global-cleaving}
A \defemph{global cleaving on $\D$} is a choice,
for any $f : a \to b$ in $\B$ and $\disp{b} : \dob{\D}{b}$, of
\begin{enumerate}
\item a displayed object $\disp{a}$ over $a$;
\item a displayed 1-cell $\disp{f} : \dmor{\disp{a}}{\disp{b}}{f}$;
\item a cartesian structure on $\disp{f}$.
\end{enumerate}
Given a global cleaving on $\D$, we use the notation $\substTy{\disp{b}}{f} \defeq \disp{a}$ and $\cartesianlift{f}{\disp{b}} \defeq \disp{f}$ to denote the choice given by the cleaving.
\end{definition}
\begin{remark}
\label{rem:global-cleaving-classic}
The notion of global cleaving as in Definition~\ref{def:global-cleaving} gives rise to a notion of cloven fibration on the total bicategory $\total{\D}$. 
\end{remark}

Next we look at local cleavings and opcleavings.
A 2-cell is opcartesian if and only if it is opcartesian in the 1-categorical sense for the hom-functor. 
However, in the formalization we give the following direct definition not relying on hom-categories, and prove the characterization via hom-categories afterwards (\coqident{Bicategories.DisplayedBicats.CleavingOfBicat}{opcartesian_2cell_weq_opcartesian}).
Similarly, we give an unfolded definition of local opcleaving.

\begin{definition}[\coqident{Bicategories.DisplayedBicats.CleavingOfBicat}{is_opcartesian_2cell}]
Suppose we have 1-cells $f, g : x \rightarrow y$, a 2-cell $\tc : f \mytwocell g$, and displayed objects $\disp{x}$ and $\disp{y}$ over $x$ and $y$, respectively.
Given displayed 1-cells $\ff : \dmor{\disp{x}}{\disp{y}}{f}$ and $\gg : \dmor{\disp{x}}{\disp{y}}{g}$ and a displayed 2-cell $\disp{\tc} : \dtwo{\ff}{\gg}{\tc}$, 
we say that $\disp{\tc}$ is \defemph{2-opcartesian} (or just \defemph{opcartesian}) if for all 1-cells $h : x \rightarrow y$, displayed 1-cells $\disp{h} : \dmor{\disp{x}}{\disp{y}}{h}$, 2-cells $\tcC : g \mytwocell h$, and displayed 2-cells $\disp{\gamma} : \dtwo{\ff}{\hh}{\tcC \vcomp \tc}$, there is a unique displayed 2-cell $\disp{\tcC} : \dtwo{\gg}{\hh}{\tcC}$ such that $\disp{\tc} \vcomp \disp{\tcC} = \disp{\gamma}$.
\end{definition}

Being an opcartesian 2-cell is always a property.
The notion of \emph{cartesian 2-cells} is analogous.

\begin{definition}[\coqident{Bicategories.DisplayedBicats.CleavingOfBicat}{local_opcleaving}]
\label{def:local-opcleaving}
A \defemph{local opcleaving on $\D$} is given by, for every $\tc : f \mytwocell g$ and $\disp{f} : \dmor{\disp{x}}{\disp{y}}{f}$, a displayed 1-cell $\disp{g} : \dmor{\disp{x}}{\disp{y}}{g}$ and an opcartesian 2-cell $\disp{\tc} : \dtwo{\disp{f}}{\disp{g}}{\tc}$.

The notions of \defemph{local cleaving} and \defemph{local isocleaving} are defined analogously.
\end{definition}

\begin{remark}
\label{rem:local-iso-cleaving}
Every displayed bicategory on a univalent bicategory has a local isocleaving.
The construction is the same as for categories \cite[Construction 5.12]{AhrensL19}.
\end{remark}

\begin{remark}
  \label{rem:local-opfib-classic}
The notions of opcartesian 2-cell and of local opcleaving as in Definition~\ref{def:local-opcleaving} give rise to notions of opcartesian 2-cell and of cloven local opfibration on the total bicategory $\total{\D}$. %
\end{remark}

Note that one can construct a local isocleaving from either a local cleaving or a local opcleaving.

\begin{proposition}[\coqfile{Bicategories.DisplayedBicats}{CleavingOfBicatIsAProp}]\label{prop:isaprop_cleav}
  Suppose that $\B$ is a univalent bicategory and $\D$ is a univalent displayed bicategory over $\D$.
  Then the types of local (resp.\ global) (op)cleavings on $\D$ are propositions.
\end{proposition}

Now let us look at some examples of these notions.

\begin{example}[\coqfile{Bicategories.DisplayedBicats.ExamplesOfCleavings}{TrivialCleaving}]
The trivial displayed bicategory $\trivial{\B_1}{\B_2}$ over $\B_1$ comes equipped with a cloven global fibration.
Cartesian 1-cells in $\trivial{\B_1}{\B_2}$ correspond to adjoint equivalences in $\B_2$.
As such, we can take the identity 1-cell as the global lift.
In addition, $\trivial{\B_1}{\B_2}$ also has both a cloven local fibration and a cloven local opfibration.
\end{example}

\begin{example}[\coqfile{Bicategories.DisplayedBicats.ExamplesOfCleavings}{CodomainCleaving}]
\label{ex:arrows_cleaving}
Suppose that $\B$ is a locally groupoidal bicategory with pullbacks.
Since cartesian 1-cells in $\arrows{\B}$ correspond to pullback squares, we can construct a global cleaving for $\arrows{\B}$ by taking pullbacks.
Note that all 2-cells in $\arrows{\B}$ are cartesian, because $\B$ is locally groupoidal, and thus $\arrows{\B}$ also has a local cleaving and a local opcleaving.
\end{example}

\begin{example}[\coqfile{Bicategories.DisplayedBicats.ExamplesOfCleavings}{FunctorsIntoCatCleaving}]
\label{example:split-opfib}
The displayed bicategory $\SplitOpFib$ has a global cleaving and a local opcleaving.
To construct these, we first observe that a displayed 1-cell $\gamma : G_1 \Rightarrow F_1 \cdot G_2$ from $G_1 : \CC_1 \rightarrow \CatOfStrictCat$ to $G_2 : \CC_2 \rightarrow \CatOfStrictCat$ is cartesian if and only if it is a natural isomorphism.
Now we construct a global cleaving for $\SplitOpFib$ as follows: whenever we have functors $F : \CC_1 \rightarrow \CC_2$ and $G : \CC_2 \rightarrow \CatOfStrictCat$, then we also get a functor $F \cdot G : \CC_1 \rightarrow \CatOfStrictCat$.

Every displayed 2-cell in $\SplitOpFib$ is opcartesian.
To find local opcartesian lifts, suppose that we have functors $F_1, F_2 : \CC_1 \rightarrow \CC_2$, $G_1 : \CC_1 \rightarrow \CatOfStrictCat$ and $G_2 : \CC_2 \rightarrow \CatOfStrictCat$, and natural transformations $n : F_1 \Rightarrow F_2$ and $\gamma : G_1 \Rightarrow F_1 \cdot G_2$.
Our goal is to construct a natural transformation $G_1 \Rightarrow F_2 \cdot G_2$, and for the desired transformation we take $\gamma \vcomp (n \whiskerr G_2)$.
Hence, $\SplitOpFib$ has a local opcleaving.
\end{example}

\begin{example}[\coqfile{Bicategories.DisplayedBicats.ExamplesOfCleavings}{OpFibrationCleaving}]
\label{example:opfib}
The displayed bicategory $\OpFib$ has a global cleaving and a local opcleaving.
Given a functor $F : \CC_1 \rightarrow \CC_2$ and a displayed category $\D_2$ over $\CC_2$, we construct a displayed category $\reindex{F}{\D_2}$ over $\CC_1$:
\begin{itemize}
	\item The displayed objects over $x : \CC_1$ are displayed objects in $\D_2$ over $F(x)$.
	\item The displayed morphisms over $f : x \rightarrow y$ from $\disp{x}$ to $\disp{y}$ are displayed morphisms over $\dmor{\disp{x}}{\disp{y}}{F(f)}$.
\end{itemize}
Note that $\reindex{F}{\D_2}$ inherits any opcleaving from $\D_2$.
In addition, we have a displayed functor over $F$ from $\reindex{F}{\D_2}$ to $\D_2$ that preserves cartesian morphisms.

Opcartesian 2-cells in $\OpFib$ correspond to displayed natural transformations of which all components are opcartesian.
We form local lifts pointwise.
Since displayed 1-cells in $\OpFib$ preserve cartesian morphisms, cartesian 2-cells are preserved under both left and right whiskering.
\end{example}

Similarly, we can show that the displayed bicategory $\Fib$ has a global and a local cleaving (\coqfile{Bicategories.DisplayedBicats.ExamplesOfCleavings}{FibrationCleaving}). 
We finish this section by defining comprehension bicategories.
Examples of those are given in Section~\ref{sec:comp-bicats}.

\begin{definition}[\coqident{Bicategories.Logic.ComprehensionBicat}{comprehension_bicat}]
\label{def:comp-bicat}
A \defemph{weak comprehension bicategory} is given by a bicategory $\B$, a displayed bicategory $\D$ over $\B$, and a displayed pseudofunctor $\chi$ over the identity on $\B$ as pictured below, 
\[
\compbicat{\D}{\chi}{\B}
\]
satisfying the following properties (see also Proposition~\ref{prop:isaprop_cleav}):
\begin{enumerate}
\item $\D$ is equipped with a cloven global fibration;
\item \label{item:local-opcleaving} $\D$ has a cloven local opfibration;
\item \label{item:preserve-opcart-hcomp} opcartesian 2-cells of $\D$ are preserved under both left and right whiskering.   \setcounter{enumcounter}{\value{enumi}}
\end{enumerate}
A \defemph{comprehension bicategory} is a weak comprehension bicategory such that
\begin{enumerate}
  \setcounter{enumi}{\value{enumcounter}}
\item $\chi$ preserves cartesian 1-cells and opcartesian 2-cells.\label{item:chi-preserves}
\end{enumerate}

\end{definition}

\begin{remark}
Recall from Remarks~\ref{rem:total-square}, \ref{rem:global-cleaving-classic}, and \ref{rem:local-opfib-classic} that the displayed pseudofunctor $\chi : \D \to \arrows{\B}$ of Definition~\ref{def:comp-bicat} gives rise to a \emph{strictly} commuting diagram of pseudofunctors
\[
\begin{tikzcd}
	\total{\D} \ar[rr, "\total{\chi}"] \ar[dr]
	&
	&
	\total{\arrows{\B}} = \arrowbicat{\B} \ar[dl, "\cod", start anchor={south west}]
	\\
	&
	\B
\end{tikzcd}
\]
with the structure of a global fibration and local opfibration in the sense of \cite{buckley2014fibred}.
\end{remark}

\begin{remark}
  \label{rem:no-preservation-of-cart}
In the definition of comprehension categories, one usually requires that the functor $\chi$ preserves cartesian morphisms.
The analogous requirement for comprehension bicategories is that the pseudofunctor $\chi$ preserves both cartesian 1-cells and opcartesian 2-cells --- this is condition \ref{item:chi-preserves} in Definition~\ref{def:comp-bicat}.

We have factored out this requirement, since for our interpretation of the language \BTT (introduced in Section~\ref{sec:syntax}), we do not need this condition;
that is, we can interpret \BTT in any \emph{weak} comprehension bicategory.
The reason has to do with how we interpret terms of the language \BTT  in a (weak) comprehension bicategory.
Traditionally, in a comprehension category, terms of \MLTT are interpreted as sections of projections.
Specifically, every type $A$ in context $\Gamma$ gives rise, in the interpretation, to a morphism $\pi_A : \Gamma . A \rightarrow \Gamma$ in the arrow category.
A term of type $A$ in context $\Gamma$ in \MLTT is then interpreted as a section of $\pi_A$.

In our setting, terms of \BTT will be interpreted differently.
Instead of interpreting \BTT terms as sections of projections, we interpret them as morphisms over the identity.
More concretely, in a bicategory $\D$ displayed over $\B$,
a term in context $\Gamma$ of type $B$ with source $A$ of \BTT is interpreted as a morphism from $A$ to $B$ over the identity of $\Gamma$.

In the 1-categorical case, these two ways of interpreting terms are equivalent in many examples.
The reason is that, in practice, most comprehension categories are full, which means that $\chi$ is fully faithful, and that every fiber has a terminal object.
From these two assumptions, one can conclude that these two ways of interpreting terms are actually equivalent.
However, in the bicategorical setting, these two ways are often \emph{not} equivalent: for example, in the comprehension bicategory of displayed categories with opcleavings, i.e. opfibrations (Examples~\ref{ex:opfib}, \ref{example:opfib}, and \ref{ex:opfibincat-compcat}).
There, morphisms over the identity are required to preserve opcartesian morphisms, but such a requirement is not present for sections of the projection.
If one were to include \MLTT style terms (which are interpreted as sections of projections) in \BTT, then to give an interpretation one would also need that $\chi$ preserves cartesian 1-cells and opcartesian 2-cells in order to interpret substitution.
This is why we present here both definitions.

In short, all the examples presented in Section~\ref{sec:comp-bicats} are comprehension bicategories; the interpretation of \BTT given in Section~\ref{sec:syntax} only requires a \emph{weak} comprehension bicategory.
\end{remark}

\begin{remark}
\label{rem:contravariant-compbicats}
\defemph{Contravariant} and \defemph{isovariant} comprehension bicategories are defined analogously with the notions of (iso)cleaving and (iso)cartesian taking the place of opcleaving and opcartesian
in Items~\ref{item:local-opcleaving} and \ref{item:preserve-opcart-hcomp} of Definition~\ref{def:comp-bicat}.
Some of our examples of comprehension bicategories can similarly be equipped with a contravariant and isovariant comprehension structure.
\end{remark}

\section{Examples of Comprehension Bicategories}
\label{sec:comp-bicats}

In this section, we give several (classes of) instances of comprehension bicategories.
In some of these instances, we recognize structures studied previously in the context of higher-dimensional and directed type theory.

\begin{example}[\coqident{Bicategories.Logic.Examples.TrivialComprehensionBicat}{trivial_comprehension_bicat}]
\label{ex:trivial-compcat}
Suppose we have a bicategory $\B$ with products.
Then we have the following comprehension bicategory
\[
\compbicat{\trivial{\B}{\B}}{\chi}{\B}
\]
The displayed pseudofunctor $\chi : \trivial{\B}{\B} \rightarrow \arrows{\B}$ sends the object $y_2$ over $y_1$ to the product projection $y_1 \times y_2 \to y_1$, that is, the corresponding total pseudofunctor $\B \times \B \to \arrows{\B}$ is defined by $(y_1,y_2)\mapsto \pi_1 : y_1 \times y_2 \to y_1$.

\end{example}
Example~\ref{ex:trivial-compcat} corresponds roughly to the semantics studied by \cite{DBLP:conf/lics/FioreS19}, but without the type formers.

Another example of a comprehension bicategory comes from locally groupoidal bicategories, such as $\Grpd$.

\begin{example}[\coqident{Bicategories.Logic.Examples.PullbackComprehensionBicat}{locally_grpd_comprehension_bicat}]
\label{ex:pb-compcat}
Let $\B$ be a locally groupoidal bicategory with pullbacks.
Then we get the following comprehension bicategory
\[
\compbicat{\arrows{\B}}{\id}{\B}
\]
Since every 2-cell is opcartesian in $\arrows{\B}$ if $\B$ is locally groupoidal, the displayed bicategory $\arrows{\B}$ has a local opcleaving.
\end{example}
Example~\ref{ex:pb-compcat} models \emph{undirected} reductions between terms.
It is thus related to the groupoid model of type theory by \cite{DBLP:conf/lics/HofmannS94}
and to the definition of comprehension 2-category by \cite{DBLP:journals/mscs/Garner09}.
A more detailed comparison to Garner's comprehension 2-categories is given in Remark~\ref{rem:comparison-garner}.

We can also consider \emph{directed} versions of this example by using categories instead of groupoids.

\begin{example}[\coqident{Bicategories.Logic.Examples.FunctorsIntoCatComprehensionBicat}{functors_into_cat_comprehension_bicat}]
\label{ex:splitopfibincat-compcat}
We construct the following comprehension bicategory using cloven split opfibrations.
\[
\compbicat{\SplitOpFib}{\chi}{\StrictCat}
\]
To define $\chi$, we use the Grothendieck construction.
Specifically, from a functor $G : \CC \rightarrow \CatOfStrictCat$ we construct a category $\int G$ as follows
\begin{enumerate}
	\item The objects of $\int G$ are pairs $(x, y)$ where $x : \CC$ and $y : G(x)$.
	\item The morphisms from $(x_1, y_1)$ to $(x_2, y_2)$ are pairs $(f, g)$ where $f : x_1 \rightarrow x_2$ and $g : G(f)(y_1) \rightarrow y_2$.
\end{enumerate}
Note that we also have a projection $\pi_1 : \int G \rightarrow \CC$.
In addition, from a displayed 1-cell $\gamma : G_1 \Rightarrow F \cdot G_2$, we get a functor $\int \gamma : \int G_1 \rightarrow \int G_2$ and a natural isomorphism $\gamma_c : \pi_1 \cdot F \Rightarrow \int \gamma \cdot \pi_1$.
If we have $p : \gamma_1(x) \cdot G_2(n(x)) = \gamma_2(x)$ for every $x : \CC_1$, then we get a natural transformation $\int p$ from $\int \gamma_1$ to $\int \gamma_2$.

\end{example}

Example~\ref{ex:splitopfibincat-compcat} is related to the interpretations given by \cite{LicataH11} and \cite{DBLP:journals/entcs/North19} in their works on directed type theories, albeit without considering type formers.
However, their notion of term, in both the syntax and the interpretation, is different from ours;
see Section~\ref{sec:variations-syntax} for details.

In Example~\ref{ex:splitopfibincat-compcat}, contexts are categories and types over a context $\CC$ are cloven split opfibrations on $\CC$.
We also consider a version of this interpretation where there are more types: instead of looking at only those opfibrations that are split, \emph{all} opfibrations are considered.

\begin{example}[\coqident{Bicategories.Logic.Examples.OpfibrationsComprehensionBicat}{opcleaving_comprehension_bicat}]
\label{ex:opfibincat-compcat}
From opcleavings we build the following comprehension bicategory:
\[
\compbicat{\OpFib}{\chi}{\Cat}
\]
The pseudofunctor $\chi$ sends a displayed category $\D$ over $\CC$ to the functor $\pi_1 : \total{\D} \rightarrow \CC$. 
\end{example}

Similarly, we can define a \emph{contravariant} comprehension bicategory (\coqident{Bicategories.ComprehensionBicat}{cleaving_contravariant_comprehension_bicat}) using cleavings instead of opcleavings.

\section{Internal Street (Op)Fibrations}
\label{sec:internal-sfibs}

In this section, we discuss Street (op)fibrations internal to a fixed bicategory $\B$.
They will yield, in Section~\ref{sec:disp_map_bicat}, many examples of comprehension bicategories, see Example~\ref{ex:streetopfib-compcat} and Remark~\ref{rem:interest-ex-internal-sfib}.
The examples of Street opfibrations internal to bicategories of stacks are particularly interesting (see Remark~\ref{rem:interest-ex-internal-sfib}).

Note that $\arrows{\B}$ comes equipped with a cloven \emph{global} fibration if $\B$ has pullbacks.
However, to obtain a \emph{local} (op)cleaving, we used that $\B$ is locally groupoidal in Example~\ref{ex:arrows_cleaving}.
This assumption is avoided in Example~\ref{example:opfib} where $\B = \Cat$: instead of looking at arbitrary functors, one only considers the opfibrations.
We can generalize this idea to arbitrary bicategories by using \emph{internal Street (op)fibrations} \cite[]{buckley2014fibred}.

The displayed bicategory $\OpFib$ has a local opcleaving where the desired lifts are constructed pointwise.
To generalize this to arbitrary bicategories, we need to adjust this definition so that we can lift arbitrary 2-cells.
Furthermore, the notion of cloven Grothendieck opfibration of categories is stricter than appropriate for bicategories.
If we have $x \rightarrow y$ and an object over $y$, then a cloven Grothendieck fibration gives an object \emph{strictly} living over $x$, while a Street fibration only gives an object living \emph{weakly} over $x$ (\ie up to an isomorphism). 
More information can be found in work by Loregian and Riehl \cite[Example 4.1.2]{LOREGIAN2020496}.

\begin{definition}[\coqident{Bicategories.Morphisms.InternalStreetFibration}{internal_sfib}]
\label{def:internal_sfib}
Let $p : e \rightarrow b$ be a 1-cell in a bicategory $\B$.
Then $p$ is an \defemph{internal Street fibration} if
\begin{itemize}
	\item For every $x : \B$, the functor $p_* : \homC{\B}{x}{e} \rightarrow \homC{\B}{x}{b}$ of hom-categories is a Street fibration.
	\item For every $f : x \rightarrow y$, the following square is a morphism of Street fibrations:
	\[
	\begin{tikzcd}
          \homC{\B}{y}{e} \ar[d, "p_*"'] \ar[r, "f^*"]
          &
          \homC{\B}{x}{e} \ar[d, "p_*"]
          \\
          \homC{\B}{y}{b} \ar[r, "f^*"]
          &
          \homC{\B}{x}{b}
	\end{tikzcd}
	\]
\end{itemize}

A 1-cell is called an \defemph{internal Street opfibration} if it is an internal Street fibration in $\co{\B}$ (\coqident{Bicategories.Core.InternalStreetOpFibration}{internal_sopfib}). 
\end{definition}

In $\Cat$, internal Street fibrations are the same as Street fibrations of categories.
However, the notion of internal Street fibrations can be applied in a wider variety of settings: for example, one could also look at internal Street fibrations in the bicategories from Example~\ref{example:structured_cats} or in presheaves or stacks valued in $\Cat$.
A classical result on internal Street (op)fibrations is that they are closed under taking pullbacks, see \cite{gray_fib_and_cofib_cats} and \cite{street1980fibrations}.

\begin{proposition}[\coqident{Bicategories.Morphisms.Properties.ClosedUnderPullback}{pb_of_sfib_cleaving}]
\label{prop:sfib_pb}
Street (op-)fibrations are closed under pullback. Concretely, given a pullback square
\[
\begin{tikzcd}
  e_1 \ar[r] \ar[d, "p_1"']
  &
  e_2 \ar[d, "p_2"]
  \\
  b_1 \ar[r]
  &
  b_2
\end{tikzcd}
\]
where $p_2$ is a Street (op)fibration, then $p_1$ is so, too.
\end{proposition}

\section{Display Map Bicategories}
\label{sec:disp_map_bicat}

In this section, we introduce ``display map bicategories'' as a convenient way to build comprehension bicategories.

Let $\B$ be a bicategory with pullbacks.
We aim to construct a displayed bicategory $\SOpFib{\B}$ of Street opfibrations over $\B$ with both a global cleaving and a local opcleaving.
The construction should be done in a modular and general way,
since using the same techniques we aim to construct a displayed bicategory $\SFib{\B}$ of Street fibrations over $\B$ with a global and local cleaving.
One can imagine more examples, such as the intersection of Street opfibrations with discrete or fully faithful 1-cells.

The common pattern among all these examples is that the displayed bicategory actually represents a subbicategory of the arrow bicategory of $\B$.
In the 1-categorical case, such examples are captured by \textbf{display map categories} \cite[]{DBLP:books/daglib/0031002}.
We adapt that notion to the bicategorical setting.

However, there is a difference between the 1-categorical and the bicategorical case.
A display map category on a 1-category $\CC$ is a \textbf{full} subcategory of the arrow category on $\CC$ satisfying some requirements.
Such a notion would not be useful in the bicategorical case, since neither $\SOpFib{\B}$ nor $\SFib{\B}$ are full subbicategories of $\arrows{\B}$. Indeed, the 1-cells in $\SOpFib{\B}$ and $\SFib{\B}$ are required to preserve (op)cartesian cells, which is not required in $\arrows{\B}$.
As such, to obtain a notion of display map bicategory that captures these two examples, some care is needed in the definition of display map bicategories.

For that reason, our notion of display map bicategory comes in three different flavors.\footnote{The definitions in the formalization differ slightly from the definitions in the paper. This is because the focus in the formalization is more general as it considers arbitrary display map bicategories that are not necessarily covariant.}

\begin{definition}[\coqident{Bicategories.Logic.DisplayMapBicat}{disp_map_bicat}]
\label{def:dispmapbicat}
Let $\B$ be a bicategory.
A \textbf{display map bicategory} on $\B$ consists of
\begin{enumerate}
\item a predicate $P$ on the 1-cells of $\B$, and
\item a choice of pullbacks along 1-cells satisfying $P$ \label{enum:pb-choice}
  \setcounter{enumcounter}{\value{enumi}}
\end{enumerate}
such that
\begin{enumerate}
  \setcounter{enumi}{\value{enumcounter}}
\item if a morphism satisfies $P$, then it is an internal Street opfibration, and\label{enum:opfib}
\item $P$ is closed under pullback.\label{enum:closed-pb}
\end{enumerate}
Given a display map bicategory $\B$, a \textbf{display map} of $\B$ is a 1-cell of $\B$ together with a proof of the predicate $P$.
\end{definition}

We also define notions of \textbf{contravariant display map bicategory} and \textbf{isovariant display map bicategory}: for the former, condition (\ref{enum:opfib}) above is replaced by the condition that the display maps are required to be internal Street fibrations, while for the latter, condition (\ref{enum:opfib}) is omitted.
The reason for this terminology is that every display map bicategory of a given variance gives rise to a comprehension bicategory with the same variance.

\begin{example}[\coqident{Bicategories.Logic.DisplayMapBicat}{sopfib_disp_map_bicat_is_covariant}]
\label{exa:sopfib-dispmapbicat}
Let $\B$ be a locally univalent bicategory with pullbacks.
Since $\B$ has pullbacks, pullbacks exist along all 1-cells, and in particular, along Street opfibrations.
In addition, Street opfibrations are closed under pullbacks by Proposition~\ref{prop:sfib_pb}.
Hence we have a display map bicategory in which the predicate $P$ says that a 1-cell is an internal Street opfibration.
\end{example}

\begin{remark}
Note that in Example~\ref{exa:sopfib-dispmapbicat}, we assume $\B$ to be locally univalent.
This is to guarantee that the notion of being a Street opfibration is actually a proposition as required in Definition~\ref{def:dispmapbicat}.
However, we expect that this assumption can be dropped by suitably truncating the notion of Street opfibration.
\end{remark}

Analogously, one can define a contravariant display map bicategory of Street fibrations.
There are other examples of display map bicategories as well, for instance \textbf{discrete Street opfibrations}.
To define those, we first need the notion of \textbf{discrete morphisms}.\footnote{In the formalization, we actually use unfolded versions of these definitions, and we prove these two versions are equivalent.}

\begin{definition}
A 1-cell $f : a \rightarrow b$ in a bicategory $\B$ is called
\begin{itemize}
	\item (\coqident{Bicategories.Morphisms.FullyFaithful}{faithful_1cell}) \textbf{faithful} if for every object $x$, the functor $f_* : \homC{\B}{x}{a} \rightarrow \homC{\B}{x}{b}$ given by postcomposition with $f$ is faithful;
	\item (\coqident{Bicategories.Morphisms.DiscreteMorphisms}{conservative_1cell}) \textbf{conservative} if for every object $x$, the functor $f_* : \homC{\B}{x}{a} \rightarrow \homC{\B}{x}{b}$ is conservative;
	\item (\coqident{Bicategories.Morphisms.DiscreteMorphisms}{discrete_1cell}) \textbf{discrete} if  for every object $x$, the functor $f_* : \homC{\B}{x}{a} \rightarrow \homC{\B}{x}{b}$ is faithful and conservative.
\end{itemize}
\end{definition}

\begin{example}[\coqident{Bicategories.Logic.DisplayMapBicat}{discrete_sopfib_disp_map_bicat_is_covariant}]
Let $\B$ be a bicategory with pullbacks.
Since both faithful and conservative 1-cells are closed under pullbacks, discrete 1-cells are as well.
Hence, we get a display map bicategory of discrete Street opfibrations.
\end{example}

Every display map bicategory gives rise to a displayed bicategory as follows.

\begin{definition}[\coqident{Bicategories.DisplayedBicats.Examples.DisplayMapBicatToDispBicat}{disp_map_bicat_to_disp_bicat}]
Let $\B$ be a bicategory and let $\D$ be a display map bicategory on $\B$.
We define a displayed bicategory $\overline{\D}$ over $\B$ as follows:
\begin{itemize}
	\item The objects over $b : B$ are pairs $e : \B$ and $p : e \rightarrow b$ such that $p$ is a display map;
	\item The 1-cells over $f : b_1 \rightarrow b_2$ from $p_1 : e_1 \rightarrow b_1$ to $p_2 : e_2 \rightarrow b_2$ are pairs of a 1-cell $g : e_1 \rightarrow e_2$ and an invertible 2-cell $\invtwocell{g \cdot p_2}{p_1 \cdot f}$ such that $g$ preserves opcartesian 2-cells;
	\item The 2-cells are as in Example~\ref{example:arrows}.
\end{itemize}
\end{definition}

Similarly, every contravariant display map bicategory gives rise to a displayed bicategory. However, the 1-cells are required to preserve cartesian 2-cells.
For isovariant display map bicategories, we do not make any restriction on the 1-cells.
This way, we obtain displayed bicategories $\SOpFib{\B}$ and $\SFib{\B}$ over a bicategory $\B$ with pullbacks.
In $\SOpFib{\B}$, 2-cells are the same as 2-cells in $\arrowbicat{\B}$. However, 1-cells are a bit different: while 1-cells in $\arrowbicat{\B}$ are squares
\[
\begin{tikzcd}
	e_1 \ar[r, "f_e"] \ar[d, "p_1"']
	&
	e_2 \ar[d, "p_2"]
	\\
	b_1 \ar[r, "f_b"']
	&
	b_2
\end{tikzcd}
\]
that commute up to invertible 2-cell, 1-cells in $\SOpFib{\B}$ have the additional requirement that whiskering with $f_e$ preserves opcartesian 2-cells.
Similarly, we can define the bicategory $\SFib{\B}$ where the objects are internal Street fibrations and where the 1-cells preserve cartesian 2-cells.

Now we can construct the desired cleavings.

\begin{example}[\coqfile{Bicategories.DisplayedBicats.ExamplesOfCleavings}{DisplayMapBicatCleaving}]
\label{ex:disp_map_bicat}
Let $\B$ be a bicategory and let $\D$ be a display map bicategory.
As in Example~\ref{ex:arrows_cleaving}, cartesian 1-cells are the same as pullback squares.
Hence, we can construct a global cleaving for $\D$ using pullbacks and Proposition~\ref{prop:sfib_pb}.
To construct a local opcleaving for $\D$, we use that it is contained in $\SOpFib{\B}$, and then we can use the same construction as for $\SFib{\B}$ \cite[Example 3.4.6]{buckley2014fibred}.
\end{example}

Similarly, given a contravariant display map bicategory, one obtains both a global and a local cleaving.
However, from an isovariant display map bicategory, one only gets a global cleaving and a local isocleaving.
One can instantiate Example~\ref{ex:disp_map_bicat} to internal Street opfibrations to obtain a global cleaving and a local opcleaving for $\SOpFib{\B}$ if $\B$ has pullbacks.

Using Example~\ref{ex:disp_map_bicat}, we can generalize the example of (op)fibra\-tions to arbitrary display map bicategories.
We discuss the interest in this generalization in Remark~\ref{rem:interest-ex-internal-sfib}. 

\begin{problem}
\label{prob:dispmapbicat-compcat}
Given a display map bicategory $\D$ on a bicategory $\B$, to define a comprehension bicategory.
\end{problem}

\begin{construction}[\coqident{Bicategories.Logic.Examples.DisplayMapComprehensionBicat}{disp_map_bicat_comprehension_bicat}]{prob:dispmapbicat-compcat}
\label{ex:dispmapbicat-compcat}
From bicategory $\B$ and a display map bicategory $\D$ on $\B$, we  construct the comprehension bicategory
\[
\compbicat{\D}{\chi}{\B}
\]
The (displayed) pseudofunctor $\chi$ is the inclusion of $\D$ into $\arrows{\B}$.
\end{construction}

Similarly, we get a contravariant comprehension bicategory from a contravariant display map bicategory, and an isovariant comprehension bicategory from an isovariant display map bicategory.
We can instantiate Example~\ref{ex:dispmapbicat-compcat} to internal Street opfibrations to get the following comprehension bicategory.

\begin{example}[\coqident{Bicategories.Logic.Examples.DisplayMapComprehensionBicat}{internal_sopfib_comprehension_bicat}]
  \label{ex:streetopfib-compcat}
For a bicategory $\B$ with pullbacks (see, \eg Example~\ref{ex:pb_in_structured_categories}),
we  construct the comprehension bicategory
\[
\compbicat{\SOpFib{\B}}{\chi}{\B}
\]
The (displayed) pseudofunctor $\chi$ forgets that the morphisms in $\SOpFib{\B}$ are internal Street opfibrations.
\end{example}

\begin{example}
  \label{ex:interest-ex-internal-sfib}
  By Example~\ref{ex:streetopfib-compcat} any bicategory with pullbacks, thus in particular any
  bicategory of stacks \cite[]{Street-characterization-stacks}, gives rise to a comprehension bicategory.
\end{example}
\begin{remark}
	\label{rem:interest-ex-internal-sfib}
  Recall from Section~\ref{sec:undir-type-theory} that \cite{DBLP:conf/lics/CoquandMR17} constructed models of \MLTT in stacks valued in groupoids. Using those models, they proved the independence of countable choice in univalent foundations.
  
 With our comprehension bicategories of stacks (not just ones valued in groupoids), one could follow \cite{DBLP:conf/lics/CoquandMR17} and study the validity and independence of logical principles in \emph{directed} type theory.
\end{remark}

Note that we can specialize Example~\ref{ex:streetopfib-compcat} to each of the examples given in Example~\ref{example:structured_cats}.
This is because we showed in Example~\ref{ex:pb_in_structured_categories} that those bicategories have pullbacks.
In the case of $\Terminal$, we get a comprehension bicategory in which
\begin{itemize}
	\item the ``contexts'' are categories with a terminal object; and
	\item the ``types'' are cloven opfibrations of categories which preserve terminal objects.
\end{itemize}
Similarly, we can instantiate this to the other categories mentioned in Example~\ref{example:structured_cats}.

\begin{remark}
  Note that, similarly, we can construct a contravariant comprehension bicategory from $\SFib{\B}$.
\end{remark}

\section{The Type Theory \BTT}
\label{sec:syntax}

In this section, we extract a core syntax for two-dimensional type theory from our semantic model.
We call the resulting type theory \emph{Bicategorical Type Theory} (\BTT).
In Section~\ref{sec:soundn-interpr-compr}, we prove soundness of our syntax by giving an interpretation of the syntax in any weak comprehension bicategory.

The syntax extracted here is \emph{maximally general}, in the sense that it reflects the structure of a general comprehension bicategory.
In Section~\ref{sec:variations-syntax}, we propose several orthogonal simplifications to the syntax, along with the corresponding semantic structure and properties.
  As such, our syntax and semantics are to be viewed as a framework to study different semantic structures and their corresponding internal languages, rather than as one particular pair of syntax and semantics.

In Section~\ref{sec:judgments} we present the judgment forms of \BTT, as well as the rules extracted from the bicategories of contexts and of types, respectively.
In Sections~\ref{sec:compr-struct} and \ref{sec:subst-struct} we present the comprehension and substitution rules, respectively.

For reasons of space we omit here several rules, namely the naturality rules in Figures~\ref{fig:rules-local-substitution2} and \ref{fig:rules-coherence-substitution}, and the coherencies of trifunctors.
Such rules, written out in the ``linear'' syntax used here, would take up several lines and would be difficult to understand.
Consequently, the syntax presented here is not \emph{complete}.
The presentation of the complete syntax and a suitable completeness theorem is left for future work.

\subsection{Judgments and Basic Rules}
\label{sec:judgments}

\BTT features contexts, substitutions, types, \emph{generalized} terms, and reductions between terms.
As befits the bicategorical semantics, judgmental equality is only postulated between parallel reductions.

There are eight kinds of \emph{judgments} in \BTT:
\begin{enumerate}
	\item $\Gamma \ctx$, which is read as ``$\Gamma$ is a context'';
	\item $\ctxonecell{s}{\Delta}{\Gamma}$ (given $\Delta, \Gamma \ctx$), which is read as ``$s$ is a substitution from $\Delta$ to $\Gamma$'';
	\item $\ctxonecell{r : s \rewrite t}{\Delta}{\Gamma}$ (where $\ctxonecell{s,t}{\Delta}{\Gamma}$), which is read as ``$r$ is a reduction from $s$ to $t$'';
	\item $\ctxonecell{r \equiv r' : s \rewrite t}{\Delta}{\Gamma}$ (where $\ctxonecell{r,r' : s \rewrite t}{\Delta}{\Gamma}$), which is read as ``$r$ is equal to $r'$'';
	\item $\Gamma \vdash T \type$ (where $\Gamma \ctx$), which is read as ``$T$ is a type in context $\Gamma$'';
	\item $\typeonecell{\Gamma}{t}{S}{T}$ (where $\Gamma \vdash S, T \type$), which is read as ``$t$ is a term in $T$ depending on $S$ in context $\Gamma$'';
	\item $\typeonecell{\Gamma}{\rho : t \rewrite t'}{S}{T}$ (where $\typeonecell{\Gamma }{t,t'}{S}{T}$), which is read as ``$\rho$ is a reduction from $t$ to $t'$'';
	\item $\typeonecell{\Gamma}{\rho \equiv \rho' : t \rewrite t'}{S}{T}$ (where $\typeonecell{\Gamma }{\rho , \rho' : t \rewrite t'}{S}{T}$), which is read as ``$\rho$ is equal to $\rho'$''.
\end{enumerate}

\noindent
We often abbreviate the above judgments and write, \eg just $\rho : t \rewrite t'$ instead of $\typeonecell{\Gamma }{\rho : t \rewrite t'}{S}{T}$. %
For these judgments, we have rules that express the bicategorical structure of contexts and types.
Rules are given in Figure~\ref{fig:rules-base} for the bicategory of contexts, and in Figure~\ref{fig:rules-total} for the bicategory of types.

\begin{figure*}[tb] \footnotesize
\begin{rules}
\unaryRule
	{\ctxzerocell{\Gamma}}
	{\ctxonecell{1_\Gamma}{\Gamma}{\Gamma}}
	{}
\binaryRule
	{\ctxzerocell{\Gamma,\Delta}}
	{\ctxonecell{s}{\Delta}{\Gamma}}
	{\ctxonecell{1_s: s \rewrite s}{\Delta}{\Gamma}}
	{}
\ternaryRule
	{\ctxzerocell{\Gamma,\Delta}}
	{\ctxonecell{s,s'}{\Delta}{\Gamma}}
	{\ctxonecell{\rho : s \rewrite s'}{\Delta}{\Gamma}}
	{\ctxonecell{\rho \equiv \rho : s \rewrite s'}{\Delta}{\Gamma}}
	{}
\ternaryRule
	{\ctxzerocell{\Gamma,\Delta,\Epsilon}}
	{\ctxonecell{s}{\Epsilon}{\Delta}}
	{\ctxonecell{t}{\Delta}{\Gamma}}
	{\ctxonecell{\horcomp{s}{t}}{\Epsilon}{\Gamma}}
	{}
\quadRule
	{\ctxzerocell{\Gamma,\Delta,\Epsilon}}
	{\ctxonecell{s}{\Epsilon}{\Delta}}
	{\ctxonecell{t,t'}{\Delta}{\Gamma}}
	{\ctxonecell{\rho : t \rewrite t'}{\Delta}{\Gamma}}
	{\ctxonecell{s \whiskerl \rho :  \horcomp{s}{t} \rewrite \horcomp{s}{t'}}{\Epsilon}{\Gamma}}
	{}
\quadRule
	{\ctxzerocell{\Gamma,\Delta,\Epsilon}}
	{\ctxonecell{s,s'}{\Epsilon}{\Delta}}
	{\ctxonecell{t}{\Delta}{\Gamma}}
	{\ctxonecell{\sigma : s \rewrite s'}{\Epsilon}{\Delta}}
	{\ctxonecell{\sigma \whiskerr t :  \horcomp{s}{t} \rewrite \horcomp{s'}{t}}{\Epsilon}{\Gamma}}
	{}
\quadRule
	{\ctxzerocell{\Gamma,\Delta}}
	{\ctxonecell{s,s',s''}{\Delta}{\Gamma}}
	{\ctxonecell{\rho : s \rewrite s'}{\Delta}{\Gamma}}
	{\ctxonecell{\sigma : s' \rewrite s''}{\Delta}{\Gamma}}
	{\ctxonecell{\vertcomp{\rho}{\sigma} : s \rewrite s''}{\Delta}{\Gamma}}
	{}
\ternaryRuleBinaryConcl
	{\ctxzerocell{\Gamma,\Delta,\Epsilon}}
	{\ctxonecell{s}{\Epsilon}{\Delta}}
	{\ctxonecell{t}{\Delta}{\Gamma}}
	{\ctxonecell{1_s \whiskerr t \equiv 1_{\horcomp{s}{t}} : \horcomp{s}{t} \rewrite \horcomp{s}{t}}{\Epsilon}{\Gamma}}
	{\ctxonecell{s \whiskerl 1_t \equiv 1_{\horcomp{s}{t}} : \horcomp{s}{t} \rewrite \horcomp{s}{t}}{\Epsilon}{\Gamma}}
	{}
\pentRule
	{\ctxzerocell{\Gamma,\Delta,\Epsilon}}
	{\ctxonecell{s}{\Epsilon}{\Delta}}
	{\ctxonecell{t,t',t''}{\Delta}{\Gamma}}
	{\ctxonecell{\rho : t \rewrite t'}{\Delta}{\Gamma}}
	{\ctxonecell{\rho' : t' \rewrite t''}{\Delta}{\Gamma}}
	{\ctxonecell{\vertcomp{(s \whiskerl \rho)}{(s \whiskerl \rho')} \equiv s \whiskerl (\vertcomp{\rho}{\rho'}): \horcomp{s}{t} \rewrite \horcomp{s}{t''}}{\Epsilon}{\Gamma}}
	{}
\pentRule
	{\ctxzerocell{\Gamma,\Delta,\Epsilon}}
	{\ctxonecell{s,s',s''}{\Epsilon}{\Delta}}
	{\ctxonecell{t}{\Delta}{\Gamma}}
	{\ctxonecell{\sigma : s \rewrite s'}{\Epsilon}{\Delta}}
	{\ctxonecell{\sigma' : s' \rewrite s''}{\Epsilon}{\Delta}}
	{\ctxonecell{\vertcomp{(\sigma \whiskerr t)}{(\sigma' \whiskerr t)} \equiv (\vertcomp{\sigma}{\sigma'}) \whiskerr t : \horcomp{s}{t} \rewrite \horcomp{s''}{t}}{\Epsilon}{\Gamma}}
	{}
\pentRule
	{\ctxzerocell{\Gamma,\Delta,\Epsilon}}
	{\ctxonecell{s,s'}{\Epsilon}{\Delta}}
	{\ctxonecell{t,t'}{\Delta}{\Gamma}}
	{\ctxonecell{\sigma : s \rewrite s'}{\Epsilon}{\Delta}}
	{\ctxonecell{\rho : t \rewrite t'}{\Delta}{\Gamma}}
	{\ctxonecell{\vertcomp{(\sigma \whiskerr t)}{(s' \whiskerl \rho)} \equiv \vertcomp{(s \whiskerl \rho)}{(\sigma \whiskerr t')} : \horcomp{s}{t} \rewrite \horcomp{s'}{t'}}{\Epsilon}{\Gamma}}
	{}
\binaryRuleBinaryConcl
	{\ctxzerocell{\Gamma,\Delta}}
	{\ctxonecell{s}{\Delta}{\Gamma}}
	{\ctxonecell{\ell_s : \horcomp{1_\Delta}{s} \isorewrite s}{\Delta}{\Gamma}}
        {\ctxonecell{r_s : \horcomp{s}{1_\Gamma} \isorewrite s}{\Delta}{\Gamma}}
	{}
\quadRule
	{\ctxzerocell{\Gamma,\Delta, \Epsilon, \Zeta}}
	{\ctxonecell{r}{\Zeta}{\Epsilon}}
	{\ctxonecell{s}{\Epsilon}{\Delta}}
	{\ctxonecell{t}{\Delta}{\Gamma}}
	{\ctxonecell{\alpha_{r,s,t}: \horcomp{r}{(\horcomp{s}{t})} \isorewrite \horcomp{(\horcomp{r}{s})}{t}}{\Zeta}{\Gamma}}
	{}
	\ternaryRuleBinaryConcl
	{\ctxzerocell{\Gamma,\Delta}}
	{\ctxonecell{s,s'}{\Delta}{\Gamma}}
	{\ctxonecell{\rho : s \rewrite s'}{\Delta}{\Gamma}}
	{\ctxonecell{\vertcompwrongway{r_{s'}}{\vertcompwrongway{(\rho \whiskerr 1_\Gamma)}{r_s^{-1}}} \equiv \rho : s \rewrite s'}{\Delta}{\Gamma}}
	{\ctxonecell{\vertcompwrongway{\ell_{s'}}{\vertcompwrongway{(1_\Delta \whiskerl \rho)}{\ell_s^{-1}}} \equiv \rho : s \rewrite s'}{\Delta}{\Gamma}}
	{}
	\ternaryRule
	{\ctxzerocell{ \Gamma, \Delta, \Epsilon, \Zeta}}
	{\ctxonecell{s}{\Zeta }{\Epsilon}
		\hspace{1em}
		\ctxonecell{t}{\Epsilon}{\Delta}
		\hspace{1em}
		\ctxonecell{u,u'}{\Delta}{\Gamma}}
	{\ctxonecell{\rho : u \rewrite u'}{\Delta}{\Gamma}}
	{\ctxonecell{\vertcompwrongway{\alpha_{s,t,u'}}{\vertcompwrongway{(s \whiskerl (t \whiskerl \rho))}{\alpha^{-1}_{s,t,u}}}\equiv \horcomp{s}{t} \whiskerl \rho : \horcomp{(\horcomp{s}{t})}{u} \rewrite \horcomp{(\horcomp{s}{t})}{u'}}{\Zeta}{\Gamma}}
	{}
	\ternaryRule
	{\ctxzerocell{ \Gamma, \Delta, \Epsilon, \Zeta}}
	{\ctxonecell{s}{\Zeta }{\Epsilon}
		\hspace{1em}
		\ctxonecell{t,t'}{\Epsilon}{\Delta}
		\hspace{1em}
		\ctxonecell{u}{\Delta}{\Gamma}}
	{\ctxonecell{\rho : t \rewrite t'}{\Epsilon}{\Delta}}
	{\ctxonecell{ \vertcompwrongway{\alpha_{s,t',u}}{\vertcompwrongway{(s \whiskerl (\rho \whiskerr u))}{\alpha^{-1}_{s,t,u}}}
			\equiv 
			(s \whiskerl \rho) \whiskerr u : \horcomp{(\horcomp{s}{t})}{u} \rewrite \horcomp{(\horcomp{s}{t'})}{u}}{\Zeta }{\Gamma}}
	{}
	\ternaryRule
	{\ctxzerocell{ \Gamma, \Delta, \Epsilon, \Zeta}}
	{\ctxonecell{s,s'}{\Zeta }{\Epsilon}
		\hspace{1em}
		\ctxonecell{t}{\Epsilon}{\Delta}
		\hspace{1em}
		\ctxonecell{u}{\Delta}{\Gamma}}
	{\ctxonecell{\rho : s \rewrite s'}{\Zeta}{\Epsilon}}
	{\ctxonecell{\vertcompwrongway{\alpha_{s',t,u}}{\vertcompwrongway{(\rho \whiskerr \horcomp{t}{u})}{\alpha^{-1}_{s,t,u}}} \equiv (\rho \whiskerr t) \whiskerr u : \horcomp{(\horcomp{s}{t})}{u} \rewrite \horcomp{(\horcomp{s'}{t})}{u}}{\Zeta}{\Gamma}}
	{}
	\ternaryRuleBinaryConcl
	{\ctxzerocell{\Gamma, \Delta}}
	{\ctxonecell{s,s'}{\Delta }{\Gamma}}
	{\ctxonecell{\rho : s \rewrite s'}{\Delta }{\Gamma}}
	{\ctxonecell{\vertcompwrongway{\rho}{1_s} \equiv \rho : s \rewrite s'}{\Delta }{\Gamma}}
	{\ctxonecell{\vertcompwrongway{1_{s'}}{\rho} \equiv \rho : s \rewrite s'}{\Delta }{\Gamma}}
	{}
	\pentRule
	{\ctxzerocell{\Gamma, \Delta}}
	{\ctxonecell{s,s',s'',s'''}{\Delta }{\Gamma}}
	{\ctxonecell{\rho : s \rewrite s'}{\Delta }{\Gamma}}
	{\ctxonecell{\sigma : s' \rewrite s''}{\Delta }{\Gamma}}
	{\ctxonecell{\tau : s'' \rewrite s'''}{\Delta }{\Gamma}}
	{\ctxonecell{\vertcompwrongway{\tau}{(\vertcompwrongway{\sigma}{\rho})} \equiv \vertcompwrongway{(\vertcompwrongway{\tau}{\sigma})}{\rho} : s \rewrite s'''}{\Delta }{\Gamma}}
	{}
	\ternaryRule
	{\ctxzerocell{\Gamma, \Delta, \Epsilon}}
	{\ctxonecell{s}{\Epsilon}{\Delta}}
	{\ctxonecell{t}{\Delta}{\Gamma}}
	{\ctxonecell{\vertcompwrongway{(r_s \whiskerr t)}{\alpha_{s,1_\Delta,t}} \equiv s \whiskerl \ell_t : \horcomp{s}{(\horcomp{{1_\Delta}}{t})} \rewrite \horcomp{s}{t} }{\Epsilon}{\Gamma}}
	{}
	\pentRule
	{\ctxzerocell{\Gamma,\Delta, \Epsilon, \Zeta, \Eta}}
	{\ctxonecell{s}{\Eta}{\Zeta}}
	{\ctxonecell{t}{\Zeta}{\Epsilon}}
	{\ctxonecell{u}{\Epsilon}{\Delta}}
	{\ctxonecell{v}{\Delta}{\Gamma}}
	{\ctxonecell{\vertcomp{\alpha_{s,t,\horcomp{u}{v}}}{\alpha_{\horcomp{s}{t}, u, v}} \equiv \vertcomp{(s \whiskerl \alpha_{t,u,v})}{\vertcomp{\alpha_{s, \horcomp{t}{u}, v}}{(\alpha_{s,t,u} \whiskerr v)}} : \horcomp{s}{(\horcomp{t}{(\horcomp{u}{v})})} \rewrite \horcomp{(\horcomp{(\horcomp{s}{t})}{u})}{v}}{\Eta}{\Gamma}}
	{}
\end{rules}
\caption{Rules for the bicategory of contexts}
\label{fig:rules-base}
\end{figure*}

\begin{figure*}[t] \footnotesize
\begin{rules}
\unaryRule
	{\gtypezerocell{T}}
	{\gtypeonecell{1_T}{T}{T}}
	{}
\binaryRule
	{\gtypezerocell{S,T}}
	{\gtypeonecell{t}{S}{T}}
	{\gtypeonecell{1_t: t \rewrite t}{S}{T}}
	{}
\ternaryRule
	{\gtypezerocell{S,T}}
	{\gtypeonecell{t,t'}{S}{T}}
	{\gtypeonecell{\rho : t \rewrite t'}{S}{T}}
	{\gtypeonecell{\rho \equiv \rho : t \rewrite t'}{S}{T}}
	{}
\ternaryRule
	{\gtypezerocell{R,S,T}}
	{\gtypeonecell{s}{R}{S}}
	{\gtypeonecell{t}{S}{T}}
	{\gtypeonecell{\horcomp{s}{t}}{R}{T}}
	{}
\quadRule
	{\gtypezerocell{R,S,T}}
	{\gtypeonecell{s}{R}{S}}
	{\gtypeonecell{t,t'}{S}{T}}
	{\gtypeonecell{\rho : t \rewrite t'}{S}{T}}
	{\gtypeonecell{s \whiskerl \rho:  \horcomp{s}{t} \rewrite \horcomp{s}{t'}}{R}{T}}
	{}
\quadRule
	{\gtypezerocell{R,S,T}}
	{\gtypeonecell{s,s'}{R}{S}}
	{\gtypeonecell{t}{S}{T}}
	{\gtypeonecell{\sigma : s \rewrite s'}{R}{S}}
	{\gtypeonecell{\sigma \whiskerr t :  \horcomp{s}{t} \rewrite \horcomp{s'}{t}}{R}{T}}
	{}
\quadRule
	{\gtypezerocell{S,T}}
	{\gtypeonecell{t,t',t''}{S}{T}}
	{\gtypeonecell{\rho : t \rewrite t'}{S}{T}}
	{\gtypeonecell{\sigma : t' \rewrite t''}{S}{T}}
	{\gtypeonecell{\vertcomp{\rho}{\sigma} : t \rewrite t''}{S}{T}}
	{}
\ternaryRuleBinaryConcl
	{\gtypezerocell{R,S,T}}
	{\gtypeonecell{s}{R}{S}}
	{\gtypeonecell{t}{S}{T}}
	{\gtypeonecell{1_s \whiskerr t \equiv 1_{\horcomp{s}{t}} : \horcomp{s}{t} \rewrite \horcomp{s}{t}}{R}{T}}
	{\gtypeonecell{s \whiskerl 1_t \equiv 1_{\horcomp{s}{t}} : \horcomp{s}{t} \rewrite \horcomp{s}{t}}{R}{T}}
	{}
\pentRule
	{\gtypezerocell{R,S,T}}
	{\gtypeonecell{s}{R}{S}}
	{\gtypeonecell{t,t',t''}{S}{T}}
	{\gtypeonecell{\rho : t \rewrite t'}{S}{T}}
	{\gtypeonecell{\rho' : t' \rewrite t''}{S}{T}}
	{\gtypeonecell{\vertcomp{(s \whiskerl \rho)}{(s \whiskerl \rho')} \equiv  s \whiskerl (\vertcomp{\rho}{\rho'}) : st \rewrite st''}{R}{T}}
	{}
\pentRule
	{\gtypezerocell{R,S,T}}
	{\gtypeonecell{s,s',s''}{R}{S}}
	{\gtypeonecell{t}{S}{T}}
	{\gtypeonecell{\sigma : s \rewrite s'}{R}{S}}
	{\gtypeonecell{\sigma' : s' \rewrite s''}{R}{S}}
	{\gtypeonecell{\vertcomp{(\sigma \whiskerr t)}{(\sigma' \whiskerr t)} \equiv  (\vertcomp{\sigma}{\sigma'}) \whiskerr t : \horcomp{s}{t} \rewrite \horcomp{s''}{t}}{R}{T}}
	{}
\pentRule
	{\gtypezerocell{R,S,T}}
	{\gtypeonecell{s,s'}{R}{S}}
	{\gtypeonecell{t,t'}{S}{T}}
	{\gtypeonecell{\sigma : s \rewrite s'}{R}{S}}
	{\gtypeonecell{\rho : t \rewrite t'}{S}{T}}
	{\gtypeonecell{\vertcomp{(\sigma \whiskerr t)}{(s' \whiskerl \rho)} \equiv \vertcomp{(s \whiskerl \rho)}{(\sigma \whiskerr t')} : \horcomp{s}{t} \rewrite \horcomp{s'}{t'}}{R}{T}}
	{}
\binaryRuleBinaryConcl
	{\gtypezerocell{S,T}}
	{\gtypeonecell{s}{S}{T}}
	{\gtypeonecell{\ell_s : \horcomp{1_S}{s} \isorewrite s}{S}{T}}
	{\gtypeonecell{r_s : \horcomp{s}{1_T} \isorewrite s}{S}{T}}
	{}
\quadRule
	{\gtypezerocell{Q,R,S,T}}
	{\gtypeonecell{r}{Q}{R}}
	{\gtypeonecell{s}{R}{S}}
	{\gtypeonecell{t}{S}{T}}
	{\gtypeonecell{\alpha_{r,s,t}: \horcomp{r}{(\horcomp{s}{t})} \isorewrite \horcomp{(\horcomp{r}{s})}{t}}{Q}{T}}
	{}
	\ternaryRuleBinaryConcl
	{\gtypezerocell{S,T}}
	{\gtypeonecell{s,s'}{S}{T}}
	{\gtypeonecell{\rho : s \rewrite s'}{S}{T}}
	{\gtypeonecell{\vertcompwrongway{r_{s'}}{\vertcompwrongway{(\rho \whiskerr 1_T)}{r_s^{-1}}} \equiv \rho : s \rewrite s'}{S}{T}}
	{\gtypeonecell{\vertcompwrongway{\ell_{s'}}{\vertcompwrongway{(1_S \whiskerl \rho)}{\ell_s^{-1}}} \equiv \rho : s \rewrite s'}{S}{T}}
	{}
	\ternaryRule
	{\gtypezerocell{Q,R,S,T}}
	{\gtypeonecell{s}{Q }{R}
		\hspace{1em}
		\gtypeonecell{t}{R}{S}
		\hspace{1em}
		\gtypeonecell{u,u'}{S}{T}}
	{\gtypeonecell{\rho : u \rewrite u'}{S}{T}}
	{\gtypeonecell{\vertcompwrongway{\alpha_{s,t,u'}}{\vertcompwrongway{(s \whiskerl (t \whiskerl \rho))}{\alpha^{-1}_{s,t,u}}} \equiv \horcomp{s}{t} \whiskerl \rho : \horcomp{(\horcomp{s}{t})}{u} \rewrite \horcomp{(\horcomp{s}{t})}{u'}}{Q}{T}}
	{}
	\ternaryRule
	{\gtypezerocell{Q,R,S,T}}
	{\gtypeonecell{s}{Q }{R}
		\hspace{1em}
		\gtypeonecell{t,t'}{R}{S}
		\hspace{1em}
		\gtypeonecell{u}{S}{T}}
	{\gtypeonecell{\rho : t \rewrite t'}{R}{S}}
	{\gtypeonecell{ \vertcompwrongway{\alpha_{s,t',u}}{\vertcompwrongway{(s \whiskerl (\rho \whiskerr u))}{\alpha^{-1}_{s,t,u}}}
			\equiv 
			(s \whiskerl \rho) \whiskerr u : \horcomp{(\horcomp{s}{t})}{u} \rewrite \horcomp{(\horcomp{s}{t'})}{u}}{Q }{T}}
	{}
	\ternaryRule
	{\gtypezerocell{Q,R,S,T}}
	{\gtypeonecell{s,s'}{Q }{R}
		\hspace{1em}
		\gtypeonecell{t}{R}{S}
		\hspace{1em}
		\gtypeonecell{u}{S}{T}}
	{\gtypeonecell{\rho : s \rewrite s'}{Q}{R}}
	{\gtypeonecell{\vertcompwrongway{\alpha_{s',t,u}}{\vertcompwrongway{(\rho \whiskerr \horcomp{t}{u})}{\alpha^{-1}_{s,t,u}}} \equiv (\rho \whiskerr t) \whiskerr u : \horcomp{(\horcomp{s}{t})}{u} \rewrite \horcomp{(\horcomp{s'}{t})}{u}}{Q}{T}}
	{}

\ternaryRuleBinaryConcl
	{\gtypezerocell{S,T}}
	{\gtypeonecell{s,s'}{S }{T}}
	{\gtypeonecell{\rho : s \rewrite s'}{S }{T}}
	{\gtypeonecell{\vertcompwrongway{\rho}{1_s} \equiv \rho : s \rewrite s'}{S }{T}}
	{\gtypeonecell{\vertcompwrongway{1_{s'}}{\rho} \equiv \rho : s \rewrite s'}{S }{T}}
	{}
\pentRule
	{\gtypezerocell{S,T}}
	{\gtypeonecell{s,s',s'',s'''}{S }{T}}
	{\gtypeonecell{\rho : s \rewrite s'}{S }{T}}
	{\gtypeonecell{\sigma : s' \rewrite s''}{S }{T}}
	{\gtypeonecell{\tau : s'' \rewrite s'''}{S }{T}}
	{\gtypeonecell{\vertcomp{\rho}{(\vertcomp{\sigma}{\tau})} \equiv \vertcomp{(\vertcomp{\rho}{\sigma})}{\tau} : s \rewrite s'''}{S }{T}}
	{}
\ternaryRule
	{\gtypezerocell{R,S,T}}
	{\gtypeonecell{s}{R}{S}}
	{\gtypeonecell{t}{S}{T}}
	{\gtypeonecell{\vertcomp{\alpha_{s,1_S,t}}{(r_s \whiskerr t)} \equiv s \whiskerl \ell_t : \horcomp{s}{(\horcomp{1_S}{t})} \rewrite \horcomp{s}{t}}{R}{T}}
	{}
\pentRule
	{\gtypezerocell{P,Q,R,S,T}}
	{\gtypeonecell{q}{P}{Q}}
	{\gtypeonecell{r}{Q}{R}}
	{\gtypeonecell{s}{R}{S}}
	{\gtypeonecell{t}{S}{T}}
	{\gtypeonecell{\vertcomp{\alpha_{q,r,\horcomp{s}{t}}}{\alpha_{\horcomp{q}{r}, s, t}} \equiv \vertcomp{(q \whiskerl \alpha_{r, s, t})}{\vertcomp{\alpha_{q, \horcomp{r}{s}, t}}{(\alpha_{q, r, s} \whiskerr t)}} : \horcomp{q}{(\horcomp{r}{(\horcomp{s}{t})})} \rewrite \horcomp{(\horcomp{(\horcomp{q}{r})}{s})}{t}}{P}{T}}
	{}
\end{rules}
\caption{Rules for the the bicategory of types}
\label{fig:rules-total}
\end{figure*}

We also introduce symbols which read like judgments but stand for several judgments, using the composition and identities
introduced in Figures~\ref{fig:rules-base} and \ref{fig:rules-total}.

\begin{enumerate}
	\item $\ctxonecell{\rho : s \isorewrite t}{\Delta}{\Gamma}$ stands for the following four judgments.
	\begin{itemize}
               \item $\ctxonecell{\rho : s \rewrite t}{\Delta}{\Gamma}$
               \qquad $\ctxonecell{\rho^{-1} : t \rewrite s}{\Delta}{\Gamma}$
               \item $\vertcomp{\rho}{\rho^{-1}} \equiv 1_s$
               \qquad $\vertcomp{\rho^{-1}}{\rho} \equiv 1_t$
	\end{itemize}
	\item $\typeonecell{\Gamma}{\rho : t \isorewrite t'}{S}{T}$ stands for the following four judgments.
	\begin{itemize}
		\item $\typeonecell{\Gamma}{\rho : t \rewrite t'}{S}{T}$
		\qquad $\typeonecell{\Gamma}{\rho ^{-1}: t' \rewrite t}{S}{T}$
		\item $\vertcomp{\rho}{\rho^{-1}} \equiv 1_t$
		\qquad $\vertcomp{\rho^{-1}}{\rho} \equiv 1_{t'}$
	\end{itemize}
	\item $\isoctxonecell{s}{\Delta}{\Gamma}$ stands for the following four judgments.
	\begin{itemize}
		\item $\ctxonecell{s}{\Delta}{\Gamma}$
		\qquad $\ctxonecell{s^{-1}}{\Gamma}{\Delta}$
		\item $s^\ell: s \horcomp s^{-1} \isorewrite 1_\Delta$
		\qquad $s^\rho: s^{-1} \horcomp s \isorewrite 1_\Gamma$
	\end{itemize}
	We also require that $s^\ell$ and $s^\rho$ form an adjoint equivalence.
	This can be specified by formulating the usual triangle equalities as equality judgments. 
	\item $\isotypeonecell{\Gamma}{t}{S}{T}$ stands for the following four judgments.
	\begin{itemize}
		\item $\typeonecell{\Gamma}{t}{S}{T}$
		\qquad $\typeonecell{\Gamma}{t^{-1}}{T}{S}$
		\item $t^\ell: t \horcomp t^{-1} \isorewrite 1_S$
		\qquad $t^\rho : t^{-1} \horcomp t \isorewrite 1_T$
	\end{itemize}
\end{enumerate} 

\begin{remark}
  By abuse of notation, we write several topically related rules that share all the same hypotheses as one rule with several conclusions.
  These rules then also share the same name, \eg \ruleref{\ruleExtendTy}.
  When referring to a rule by name, it will be clear from the context which of the possible rules we refer to.
  The names of inference rules in the text are hyperlinks to the corresponding rules (\eg \ruleref{map}).
  The equality $\equiv$ is assumed to be a congruence for every other constructor and judgment. For brevity, we have not recorded here the resulting rules.
  When parentheses are omitted, everything is associated to the left: that is, $rst$ stands for $((rs)t)$.
  Note also that in several rules in which it is necessary to re-associate several four or more terms or substitutions, we have written $\alpha$ instead of a long composition of whiskered associators $\alpha_{\bullet, \bullet, \bullet}$ in the interest of readability.

\end{remark}

\subsection{Comprehension Structure}
\label{sec:compr-struct}
Comprehension, that is, context extension, is extracted from the pseudofunctor $\chi$.
The rules for comprehension are given in Figure~\ref{fig:rules-comprehension}.
There are some notable differences to comprehension in \MLTT.
First, the rule \ruleref{\ruleExtendTm}, which forms a substitution, comes together with a reduction that expresses the commutativity of a triangle.
Second, we also have a rule \ruleref{\ruleExtendRed} that extends a substitution with a reduction.
Since reductions are proof-relevant, this rule comes with a coherency on the commutativity.

\begin{figure*}[bt] \footnotesize
\begin{rules}
\binaryRuleBinaryConcl
	{\ctxzerocell{\Gamma}}
	{\gtypezerocell{T}}
	{\Gamma. T \ctx}
	{\ctxonecell{\proj{\Gamma}{T}}{\Gamma.T}{\Gamma}}
	{\ruleExtendTy}
\ternaryRuleBinaryConcl
	{\ctxzerocell{\Gamma}}
	{\gtypezerocell{S,T}}
	{\gtypeonecell{t}{S}{T}}
	{\ctxonecell{\Gamma.t}{\Gamma.S}{\Gamma.T}}
	{\ctxonecell{\comm{\Gamma}{t} : \proj{\Gamma}{S} \isorewrite (\Gamma.t) \proj{\Gamma}{T} }{\Gamma.S}{\Gamma}}
	{\ruleExtendTm}
\quadRuleBinaryConcl
	{\ctxzerocell{\Gamma}}
	{\gtypezerocell{S,T}}
	{\gtypeonecell{t,t'}{S}{T}}
	{\gtypeonecell{r: t \rewrite t'}{S}{T}}
	{\ctxonecell{\Gamma.r: \Gamma.t \rewrite \Gamma.t'}{\Gamma.S}{\Gamma.T}}
	{\ctxonecell{\comm{\Gamma}{t'} \equiv \vertcomp{\comm{\Gamma}{t}}{(\Gamma.r \whiskerr \proj{\Gamma}{T}  )} : \proj{\Gamma}{S} \rewrite  (\Gamma.t') \proj{\Gamma}{T}}{\Gamma.S}{\Gamma}}
	{\ruleExtendRed}
\binaryRuleBinaryConcl
	{\ctxzerocell{\Gamma}}
	{\gtypezerocell{T}}
	{\ctxonecell{\comprid{T} : \Gamma . 1_T \isorewrite 1_{\Gamma . T}}{\Gamma . T}{\Gamma . T}}
	{\ctxonecell{\vertcompthree{\comm{\Gamma}{1_T}}{(\comprid{T} \whiskerr \proj{\Gamma}{T})}{\ell_{\proj{\Gamma}{T}}}  \equiv 1_{\proj{\Gamma}{T}}: \proj{\Gamma}{T} \rewrite \proj{\Gamma}{T}}{\Gamma . T}{\Gamma }}
	{\ruleExtendId}
\quadRuleBinaryConcl
	{\ctxzerocell{\Gamma}}
	{\gtypezerocell{R, S, T}}
	{\gtypeonecell{s}{R}{S}}
	{\gtypeonecell{t}{S}{T}}
	{\ctxonecell{\comprcomp{s}{t} : \horcomp{(\Gamma . s)}{(\Gamma . t)} \isorewrite \Gamma . (\horcomp{s}{t})}{\Gamma . R}{\Gamma . T}}
	{\ctxonecell{\vertcompfour{\comm{\Gamma}{s}}{(\Gamma.s \whiskerl \comm{\Gamma}{t})}{\alpha_{\Gamma.s, \Gamma.t, \proj{\Gamma}{T}}}{(\comprcomp{s}{t} \whiskerr \proj{\Gamma}{T})} \equiv \comm{\Gamma}{st} : \proj{\Gamma}{R} \rewrite (\Gamma.st) \proj{\Gamma}{T}}{\Gamma.R}{\Gamma} }
	{\ruleExtendComp}	
\ternaryRule
	{\ctxzerocell{\Gamma}}
	{\gtypezerocell{S,T}}
	{\gtypeonecell{t}{S}{T}}
	{\ctxonecell{\Gamma . 1_t \equiv 1_{\Gamma . t} : \Gamma.t \rewrite \Gamma.t}{\Gamma.S}{\Gamma.T}}
	{}
\pentRule
	{\ctxzerocell{\Gamma}}
	{\gtypezerocell{S,T}}
	{\gtypeonecell{t, t', t''}{S}{T}}
	{\gtypeonecell{\rho : t \rewrite t'}{S}{T}}
	{\gtypeonecell{\rho' : t' \rewrite t''}{S}{T}}
	{\ctxonecell{\Gamma . (\vertcomp{\rho}{\rho'}) \equiv \vertcomp{\Gamma . \rho}{\Gamma . \rho'} : \Gamma.t \rewrite \Gamma.t''}{\Gamma.S}{\Gamma.T}}
	{}
\ternaryRuleBinaryConcl
	{\ctxzerocell{\Gamma}}
	{\gtypezerocell{S,T}}
	{\gtypeonecell{t}{S}{T}}
	{\ctxonecell{\vertcomp{\left(\comprid{S} \whiskerr (\Gamma . t)\right)}{\ell_{\Gamma . t}} \equiv \vertcomp{\comprcomp{1_S}{t}}{(\Gamma . \ell_t)} : \horcomp{(\Gamma . 1_S)}{(\Gamma . t)} \rewrite \Gamma . t}{\Gamma.S}{\Gamma.T}}
        {\ctxonecell{\vertcomp{\left((\Gamma . t) \whiskerl \comprid{T}\right)}{r_{\Gamma . t}} \equiv \vertcomp{\comprcomp{t}{1_T}}{(\Gamma . r_t)} : \horcomp{(\Gamma . t)}{(\Gamma . 1_T)} \rewrite \Gamma . t}{\Gamma.S}{\Gamma.T}}
	{}
\pentRule
	{\ctxzerocell{\Gamma}}
	{\gtypezerocell{Q, R, S,T}}
	{\gtypeonecell{r}{Q}{R}}
	{\gtypeonecell{s}{R}{S}}
	{\gtypeonecell{t}{S}{T}}
	{\ctxonecell{\vertcomp{\left((\Gamma . r) \whiskerl \comprcomp{s}{t}\right)}{\vertcomp{\comprcomp{r}{(\horcomp{s}{t})}}{(\Gamma . \alpha_{r, s, t})}} \equiv {\vertcomp{\alpha_{\Gamma . r, \Gamma . s, \Gamma . t}}{\vertcomp{\left( \comprcomp{r}{s} \whiskerr (\Gamma . t) \right)}{\comprcomp{(\horcomp{r}{s})}{t}}}} : \horcomp{(\Gamma . r)}{(\horcomp{(\Gamma . s)}{(\Gamma . t)})} \rewrite \Gamma . (\horcomp{(\horcomp{r}{s})}{t})}{\Gamma.Q}{\Gamma.T}}
	{}
\pentRule
	{\ctxzerocell{\Gamma}}
	{\gtypezerocell{R, S, T}}
	{\gtypeonecell{s, s'}{R}{S}}
	{\gtypeonecell{t}{S}{T}}
	{\gtypeonecell{\rho : s \rewrite s'}{R}{S}}
	{\ctxonecell{\vertcomp{\comprcomp{s}{t}}{\Gamma . (\rho \whiskerr t)} \equiv \vertcomp{(\Gamma . \rho \whiskerr \Gamma . t)}{\comprcomp{s'}{t}} : \horcomp{(\Gamma . s)}{(\Gamma . t)} \rewrite \Gamma . (\horcomp{s'}{t})}{\Gamma.R}{\Gamma.T}}
	{}
\pentRule
	{\ctxzerocell{\Gamma}}
	{\gtypezerocell{R, S, T}}
	{\gtypeonecell{s}{R}{S}}
	{\gtypeonecell{t, t'}{S}{T}}
	{\gtypeonecell{\rho : t \rewrite t'}{S}{T}}
	{\ctxonecell{\vertcomp{\comprcomp{s}{t}}{\Gamma . (s \whiskerl \rho)} \equiv \vertcomp{(\Gamma . s \whiskerl \Gamma . \rho)}{\comprcomp{s}{t'}} : \horcomp{(\Gamma . s)}{(\Gamma . t)} \rewrite \Gamma . (\horcomp{s}{t'})}{\Gamma.R}{\Gamma.T}}
	{}
\end{rules}
\caption{Rules for comprehension}
\label{fig:rules-comprehension}
\end{figure*}

\subsection{Substitution Structure}
\label{sec:subst-struct}

Substitution is given, in the semantics, by the global and local (op)\-cleaving structure.
We reflect this into the syntax as \emph{explicit substitution},
as was also used, \eg by \cite{DBLP:conf/lics/FioreS19,DBLP:conf/popl/LicataH12} in their respective settings.

\begin{figure*}[bt] \footnotesize
\begin{rules}
\ternaryRule
{\ctxzerocell{\Gamma, \Delta}}
{\ctxonecell{s}{\Delta}{\Gamma}}
{\gtypezerocell{T}}
{\typezerocell{\Delta}{\substTy{T}{s}}}
{\ruleSubTy}

\quadRule
{\ctxzerocell{\Gamma, \Delta}}
{\ctxonecell{s}{\Delta}{\Gamma}}
{\gtypezerocell{S, T}}
{\gtypeonecell{t}{S}{T}}
{\typeonecell{\Delta}{\substTm{t}{s}}{\substTy{S}{s}}{\substTy{T}{s}}}
{\ruleSubTm}

\pentRule
{\ctxzerocell{\Gamma, \Delta}}
{\ctxonecell{s}{\Delta}{\Gamma}}
{\gtypezerocell{S, T}}
{\gtypeonecell{t, t'}{S}{T}}
{\gtypeonecell{\rho : t \rewrite t'}{S}{T}}
{\typeonecell{\Delta}{\substRed{\rho}{s} : \substTm{t}{s} \rewrite \substTm{t'}{s}}{\substTy{S}{s}}{\substTy{T}{s}}}
{\ruleSubRed}

\ternaryRule
{\ctxzerocell{\Gamma, \Delta}}
{\ctxonecell{s}{\Delta}{\Gamma}}
{\gtypezerocell{T}}
{\typeonecell{\Delta}{\subOnId{s} : 1_{\substTy{T}{s}} \isorewrite \substTm{1_T}{s}}{\substTy{T}{s}}{\substTy{T}{s}}}
{\ruleSubOnId}

\pentRule
{\ctxzerocell{\Gamma, \Delta}}
{\ctxonecell{s}{\Delta}{\Gamma}}
{\gtypezerocell{R, S, T}}
{\gtypeonecell{r}{R}{S}}
{\gtypeonecell{t}{S}{T}}
{\typeonecell{\Delta}{\subOnComp{r}{t}{s} :  \horcomp{\substTm{r}{s}}{\substTm{t}{s}}\isorewrite \substTm{(\horcomp{r}{t})}{s} }{\substTy{R}{s}}{\substTy{T}{s}}}
{\ruleSubOnComp}

\quadRule
{\ctxzerocell{\Gamma, \Delta}}
{\ctxonecell{s}{\Delta}{\Gamma}}
{\gtypezerocell{S, T}}
{\gtypeonecell{t}{S}{T}}
{\typeonecell{\Delta}{\substRed{1_t}{s} \equiv 1_{\substTm{t}{s}}: \substTm{t}{s} \rewrite \substTm{t}{s}}{\substTy{S}{s}}{\substTy{T}{s}}}
{\ruleSubRedOnId}

\pentRule
{\ctxzerocell{\Gamma, \Delta}}
{\ctxonecell{s}{\Delta}{\Gamma}}
{\gtypezerocell{S, T}}
{\gtypeonecell{t, t', t''}{S}{T}}
{\gtypeonecell{\rho : t \rewrite t'}{S}{T} \quad \quad \gtypeonecell{\rho' : t' \rewrite t''}{S}{T}}
{\typeonecell{\Delta}{\substRed{(\vertcomp{\rho}{\rho'})}{s} \equiv \vertcomp{\substRed{\rho}{s}}{\substRed{\rho'}{s}} : \substTm{t}{s} \rewrite \substTm{t''}{s}}{\substTy{S}{s}}{\substTy{T}{s}}}
{\ruleSubRedOnComp}

\quadRule
{\ctxzerocell{\Gamma, \Delta}}
{\ctxonecell{s}{\Delta}{\Gamma}}
{\gtypezerocell{S, T}}
{\gtypeonecell{t}{S}{T}}
{\typeonecell{\Delta}{{\ell_{\substTm{t}{s}}} \equiv \vertcomp{(\subOnId{s} \whiskerr \substTm{t}{s})}{\vertcomp{\subOnComp{1_S}{t}{s}}{\substRed{\ell_t}{s}}} : \horcomp{1_{\substTy{S}{s}}}{\substTm{t}{s}} \rewrite \substTm{t}{s}}{\substTy{S}{s}}{\substTy{T}{s}}}
{\ruleSubLunitor}
\quadRule
{\ctxzerocell{\Gamma, \Delta}}
{\ctxonecell{s}{\Delta}{\Gamma}}
{\gtypezerocell{S, T}}
{\gtypeonecell{t}{S}{T}}
{\typeonecell{\Delta}{{r_{\substTm{t}{s}}} \equiv \vertcomp{(\substTm{t}{s} \whiskerl \subOnId{s})}{\vertcomp{\subOnComp{t}{1_T}{s}}{\substRed{r_t}{s}}} : \horcomp{\substTm{t}{s}}{1_{\substTy{T}{s}}} \rewrite \substTm{t}{s}}{\substTy{S}{s}}{\substTy{T}{s}}}
{\ruleSubRunitor}

\quadRuleBinaryConcl
{\ctxzerocell{\Gamma, \Delta}}
{\ctxonecell{s}{\Delta}{\Gamma}}
{\gtypezerocell{Q, R, S, T}}
{\gtypeonecell{q}{Q}{R} \quad \gtypeonecell{r}{R}{S} \quad \gtypeonecell{t}{S}{T}}
{\typeonecellalt
	{\Delta}
	{
		\vertcompthree{(\substTm{q}{s} \whiskerl \subOnComp{r}{t}{s})}{\subOnComp{q}{\horcomp{r}{t}}{s}}{\substRed{\alpha_{q,r,t}}{s}} \equiv
		\vertcompthree{\alpha_{\substTm{q}{s},\substTm{r}{s},\substTm{t}{s}}}{(\subOnComp{q}{r}{s} \whiskerr \substTm{t}{s})}{\subOnComp{qr}{t}{s}}
	}
	{\substTy{Q}{s}}
}
{: \horcomp{\substTm{q}{s}}{(\horcomp{\substTm{r}{s}}{\substTm{t}{s}})} \rewrite \substTm{(\horcomp{(\horcomp{q}{r})}{t})}{s} : \substTy{T}{s}}
{\ruleSubAssoc}

\pentRule
{\ctxzerocell{\Gamma, \Delta}}
{\ctxonecell{s}{\Delta}{\Gamma}}
{\gtypezerocell{R, S, T}}
{\gtypeonecell{r}{R}{S} \quad \gtypeonecell{t, t'}{S}{T}}
{\gtypeonecell{\rho : t \rewrite t'}{S}{T}}
{\typeonecell{\Delta}{
		\vertcomp
		{\subOnComp{r}{t}{s}}
		{\substRed{(r \whiskerl \rho)}{s}}
		\equiv
		\vertcomp
		{(\substTm{r}{s} \whiskerl \substRed{\rho}{s})}
		{\subOnComp{r}{t'}{s}}
		: \horcomp{\substTm{r}{s}}{\substTm{t}{s}} \rewrite \substTm{\horcomp{r}{t'}}{s}}{\substTy{R}{s}}{\substTy{T}{s}}}
{\ruleSubLWhisker}

\pentRule
{\ctxzerocell{\Gamma, \Delta}}
{\ctxonecell{s}{\Delta}{\Gamma}}
{\gtypezerocell{R, S, T}}
{\gtypeonecell{r, r'}{R}{S} \quad \gtypeonecell{t}{S}{T}}
{\gtypeonecell{\rho : r \rewrite r'}{R}{S}}
{\typeonecell{\Delta}{\vertcomp{\subOnComp{r}{t}{s}}{\substRed{(\rho \whiskerr t)}{s}} \equiv \vertcomp{(\substRed{\rho}{s} \whiskerr \substTm{t}{s})}{\subOnComp{r'}{t}{s}} : \horcomp{\substTm{r}{s}}{\substTm{t}{s}} \rewrite \substTm{\horcomp{r'}{t}}{s}}{\substTy{R}{s}}{\substTy{T}{s}}}
{\ruleSubRWhisker}
\end{rules}
\caption{Rules for global substitution}
\label{fig:rules-global-substitution1}
\end{figure*}

\begin{figure*}[bt] \footnotesize
\begin{rules}
\binaryRule
{\ctxzerocell{\Gamma}}
{\gtypezerocell{T}}
{\isotypeonecell{\Gamma}{\subid(T)}{\substTy{T}{1_\Gamma}}{T}}
{\ruleSubId}

\ternaryRule
{\ctxzerocell{\Gamma}}
{\gtypezerocell{S, T}}
{\gtypeonecell{t}{S}{T}}
{\typeonecell{\Gamma}{\tmSubId{t} : 
		{\invsubid{}}
		{\ \substTm{t}{1_{\Gamma}}}
		{\ \subid}
		\isorewrite
		t}{S}{T}}
{\ruleTmSubId}

	\quadRule
{\ctxzerocell{\Gamma}}
{\gtypezerocell{S, T}}
{\gtypeonecell{t, t'}{S}{T}}
{\gtypeonecell{\rho : t \rewrite t'}{S}{T}}
{\gtypeonecell{
		\vertcomp{\tmSubId{t}}{\rho}
		\equiv
		\vertcomp{(\invsubid{} \whiskerl \substRed{\rho}{1_\Gamma} \whiskerr \subid)}{\tmSubId{t'}}
		:
		{\invsubid{}}
		{\ \substTm{t}{1_{\Gamma}}}
		{\ \subid}
		\rewrite
		t'
	}{S}{T}}
{\ruleSubRedId}

	\binaryRule
{\ctxzerocell{\Gamma}}
{\gtypezerocell{T}}
{\gtypeonecell{\vertcompthree
		{(\invsubid{} \whiskerl \subOnId{1_\Gamma} \whiskerr \subid)}
		{(r_{\invsubid{}} \whiskerr \subid)}
		{\subid^\rho}
		\equiv \tmSubId{1_T} : \horcompt{\invsubid{}}{\ \substTm{1_T}{1_\Gamma}}{\ \subid} \rewrite 1_T}{T}{T}}
{\ruleSubIdPrevId}

\quadRuleQuadConcl
{\ctxzerocell{\Gamma}}
{\gtypezerocell{S, R, T}}
{\gtypeonecell{t}{S}{T}}
{\gtypeonecell{r}{T}{R}}
{\gtypeonecellalt
	{\vertcomp{(\invsubid{} \whiskerl \subOnComp{t}{r}{1_\Gamma} \whiskerr \subid)}{\tmSubId{tr}} \equiv }
	{S}
	{}}%
      {\vertcompfive
        {\alpha}
	{ \left({\invsubid{}}
		{\substTm{t}{1_\Gamma}}
		\whiskerl
		\left(\vertcomp
		{\ell^{-1}_{\substTm{r}{1_\Gamma}}}
		{\left((\subid^\ell)^{-1}
			\whiskerr
			\substTm{r}{1_\Gamma}\right)}\right) \whiskerr
		{\subid}\right)}
	{\alpha}
	{\left(\invsubid{} \substTm{t}{1_\Gamma} \subid \whiskerl \tmSubId{r}\right)}
	{\left( \tmSubId{t} \whiskerr r \right)}}
{:\invsubid{} ({\substTm{t}{1_{\Gamma}}} {\ \substTm{r}{1_{\Gamma}}}) {\ \subid}\rewrite t r : R}
{}
{\ruleSubIdPrevComp}

\end{rules}
\caption{Rules for global substitution (preservation of identity)}
\label{fig:rules-global-substitution2}
\end{figure*}

\begin{figure*}[bt] \footnotesize
\begin{rules}
\quadRule
	{\ctxzerocell{\Gamma, \Delta, \Epsilon}}
	{\ctxonecell{s}{\Epsilon}{\Delta}}
	{\ctxonecell{s'}{\Delta}{\Gamma}}
	{\typezerocell{\Gamma}{T}}
	{\isotypeonecell{\Epsilon}{\subcomp{s}{s'}}{\substTy{\substTy{T}{s'}}{s}}{\substTy{T}{\horcomp{s}{s'}}}}
	{\ruleSubComp}
\pentRule
	{\ctxzerocell{\Gamma, \Delta, \Epsilon}}
	{\ctxonecell{s}{\Epsilon}{\Delta}}
	{\ctxonecell{s'}{\Delta}{\Gamma}}
	{\gtypezerocell{S, T}}
	{\typeonecell{\Gamma}{t}{S}{T}}
	{\typeonecell{\Epsilon}{\tmSubComp{t}{s}{s'} : 
		\horcompt
		{\subcomp{s}{s'}^{-1}}
		{\ \substTmtwo{t}{s'}{s}}
		{\ \subcomp{s}{s'}}
		\isorewrite
		\substTm{t}{\horcomp{s}{s'}}
			}{\substTy{S}{\horcomp{s}{s'}}}{\substTy{T}{\horcomp{s}{s'}}}}
	{\ruleTmSubComp}

	\pentRuleTernaryConcl
	{\ctxzerocell{\Gamma, \Delta, \Epsilon}}
	{\ctxonecell{s}{\Epsilon}{\Delta} \quad \quad \ctxonecell{s'}{\Delta}{\Gamma}}
	{\gtypezerocell{S, T}}
	{\gtypeonecell{t, t'}{S}{T}}
	{\gtypeonecell{\rho : t \rewrite t'}{S}{T}}
	{\typeonecellalt{\Epsilon}{
		\vertcomp{\tmSubComp{t}{s}{s'}}{\substRed{\rho}{\horcomp{s}{s'}}}
		\equiv
		\vertcomp{(\invsubcomp{s}{s'} \whiskerl {\substRedTwo{\rho}{s'}{s}} \whiskerr \subcomp{s}{s'}) }{\tmSubComp{t'}{s}{s'}}
	   }
	   {\substTy{S}{\horcomp{s}{s'}}}
	   }
	   {: \horcompt
	   {\subcomp{s}{s'}^{-1}}
	   {\ \substTmtwo{t}{s'}{s}}
	   {\ \subcomp{s}{s'}} \rewrite \substTm{t'}{\horcomp{s}{s'}} : {\substTy{T}{\horcomp{s}{s'}}}}
	   {}
	{\ruleSubRedComp}
	
\quadRuleQuadConcl
	{\ctxzerocell{\Gamma, \Delta, \Epsilon}}
	{\ctxonecell{s}{\Epsilon}{\Delta}}
	{\ctxonecell{s'}{\Delta}{\Gamma}}
	{\gtypezerocell{T}}
	{\typeonecellalt
		{\Epsilon}
		{\subcomp{s}{s'}^{-1} \whiskerl ((\substRed{\subOnId{s'}}{s} \vcomp \subOnId{s}) \whiskerr \subcomp{s}{s'} \vcomp \ell) \vcomp \subcomp{s}{s'}^{\rho} \equiv \tmSubComp{1_T}{s}{s'} \vcomp \subOnId{s s'}}
		{\substTy{T}{s s'}}}
	{}
	{: \horcompt{\subcomp{s}{s'}^{-1}}{\ \substTmtwo{1_T}{s'}{s}} {\ \subcomp{s}{s'}} \rewrite 1_{\substTy{T}{s s'}} : \substTy{T}{s s'}}
	{}
	{\ruleSubCompPrevId}

\pentRuleQuadConcl
	{\ctxzerocell{\Gamma,\Delta,\Epsilon}}
	{\ctxonecell{s}{\Epsilon}{\Delta}}
	{\ctxonecell{s'}{\Delta}{\Gamma}}
	{\gtypezerocell{S,T,R}}
	{\gtypeonecell{t}{S}{T} \hspace{2em} \gtypeonecell{r}{T}{R}}
	{\typeonecellalt
		{\Epsilon}
		{	
			\vertcomp
			{\left(
			\invsubcomp{s}{s'} \whiskerl \left(\vertcomp
			{\subOnComp{\substTm{t}{s'}}{\substTm{r}{s'}}{s}}
			{\substRed{\subOnComp{t}{r}{s'}}{s}}\right) \whiskerr \subcomp{s}{s'}
			\right)}
			{\tmSubComp{tr}{s}{s'}}
                        \equiv
                 }
		{\substTy{S}{ss'}}
	}
	{\vertcompfour
		{\alpha}
		{\left( (\invsubcomp{s}{s'}\substTmtwo{t}{s'}{s}) \whiskerl
                    \Bigl(\vertcomp{\leftunitor{\substTmtwo{r}{s'}{s}}^{-1}}{(\subcomp{s}{s'}^\ell)^{-1}} \whiskerr \substTmtwo{r}{s'}{s}\Bigr) \whiskerr
		\subcomp{s}{s'} \right)}
		{\alpha \ }
		{}
    }
    {\vertcompthree
		{\left(\tmSubComp{t}{s}{s'} \whiskerr \left(\invsubcomp{s}{s'}\substTmtwo{r}{s'}{s} \invsubcomp{s}{s'} \right) \right)}
		{\bigl(\substTm{t}{\horcomp{s}{s'}} \whiskerl \tmSubComp{r}{s'}{s}\bigr)}
		{\subOnComp{t}{r}{\horcomp{s}{s'}}}
        }
        {: \invsubcomp{s}{s'} \left(\substTm{\substTm{t}{s'}}{s} \ \substTm{\substTm{r}{s'}}{s} \right) \subcomp{s}{s'} \rewrite \substTm{t r}{s s'} : \substTy{R}{s s'}
        }
	{\ruleSubCompPrevComp}
\end{rules}
\caption{Rules for global substitution (preservation of composition)}
\label{fig:rules-global-substitution3}
\end{figure*}

\begin{figure*}[bt] \footnotesize
\begin{rules}
\quadRule
{\ctxzerocell{\Gamma, \Delta}}
{\ctxonecell{s, s'}{\Delta}{\Gamma}}
{\ctxonecell{\rho : s \rewrite s'}{\Delta}{\Gamma}}
{\gtypezerocell{T}}
{\typeonecell{\Delta}{\map{T}{\rho}}{\substTy{T}{s}}{\substTy{T}{s'}}}
{\ruleMap}	

\pentRule
{\ctxzerocell{\Gamma, \Delta}}
{\ctxonecell{s, s'}{\Delta}{\Gamma}}
{\ctxonecell{\rho : s \rewrite s'}{\Delta}{\Gamma}}
{\gtypezerocell{S,T}}
{\typeonecell{\Gamma}{t}{S}{T}}
{\typeonecell{\Delta}{ \map{t}{\rho} : \horcomp{\substTm{t}{s}}{(\map{T}{\rho})} \isorewrite \horcomp{(\map{S}{\rho})}{ \substTm{t}{s'}}}{\substTy{S}{s}}{\substTy{T}{s'}}}
{\ruleRewTm}

\pentRule
{\ctxzerocell{\Gamma, \Delta}}
{\ctxonecell{s, s'}{\Delta}{\Gamma}}
{\ctxonecell{\rho : s \rewrite s'}{\Delta}{\Gamma}}
{\gtypezerocell{S,T}}
{\typeonecell{\Gamma}{t, t'}{S}{T} \quad \typeonecell{\Gamma}{\tau : t \rewrite t'}{S}{T}}
{\typeonecell{\Delta}{ \vertcomp{\map{t}{\rho}}{(\map{S}{\rho} \whiskerl \substRed{\tau}{s'})} \equiv (\vertcomp{\substRed{\tau}{s} \whiskerr \map{T}{\rho})}{\map{t'}{\rho}}:   \horcomp{\substTm{t}{s}}{(\map{T}{\rho})} \rewrite \horcomp{(\map{S}{\rho})}{ \substTm{t'}{s'}}}{\substTy{S}{s}}{\substTy{T}{s'}}}
{\ruleRewRed}

\quadRule
{\ctxzerocell{\Gamma, \Delta}}
{\ctxonecell{s, s'}{\Delta}{\Gamma}}
{\ctxonecell{\rho : s \rewrite s'}{\Delta}{\Gamma}}
{\gtypezerocell{T}}
{\typeonecell{\Delta}{ \subOnId{s} \whiskerr \map{T}{\rho} \vcomp \map{1_T}{\rho} \vcomp (\map{T}{\rho} \whiskerl \subOnId{s'}) \vcomp \rightunitor{\map{T}{\rho}} \equiv \ell : \horcomp{1_{\substTy{T}{s}}}{(\map{T}{\rho})} \rewrite \map{T}{\rho} }{\substTy{T}{s}}{\substTy{T}{s'}}}
{\ruleRewOnId}

\pentRuleTernaryConcl
{\ctxzerocell{\Gamma, \Delta}}
{\ctxonecell{s, s'}{\Delta}{\Gamma}}
{\ctxonecell{\rho : s \rewrite s'}{\Delta}{\Gamma}}
{\gtypezerocell{S, T, R}}
{\gtypeonecell{t}{S}{T} \hspace{2em} \gtypeonecell{r}{T}{R}}
{\typeonecellalt
  {\Delta}
  {\map{(\horcomp{t}{r})}{\rho} \vcomp (\map{S}{\rho} \whiskerl \subOnComp{t}{r}{s'}^{-1} )\vcomp \alpha
    \equiv
  }
  {\substTy{S}{s}}
}
{(\subOnComp{t}{r}{s}^{-1} \whiskerr \map{T}{\rho}) \vcomp \alpha \vcomp (\substTm{t}{s} \whiskerl \map{r}{\rho}) \vcomp \alpha \vcomp (\map{t}{\rho} \whiskerr \substTm{r}{s'})}
{    : \horcomp{\substTm{(\horcomp{t}{r})}{s}}{(\map{R}{\rho})} \rewrite \horcompt{(\map{S}{\rho})}{\substTm{t}{s'}}{\substTm{r}{s'}} : \substTy{R}{s'}}
{\ruleRewOnComp}

\end{rules}
\caption{Rules for local substitution}
\label{fig:rules-local-substitution1}
\end{figure*}

\begin{figure*}[bt] \footnotesize
\begin{rules}

\ternaryRule
{\ctxzerocell{\Gamma, \Delta}}
{\ctxonecell{s}{\Delta}{\Gamma}}
{\gtypezerocell{T}}
{\typeonecell{\Delta}{\mapid{T}{s} : \map{T}{1_s} \isorewrite 1_{\substTy{T}{s}}}{\substTy{T}{s}}{\substTy{T}{s}}}
{\ruleMapId}

\quadRule
{\ctxzerocell{\Gamma, \Delta}}
{\ctxonecell{s, s', s''}{\Delta}{\Gamma}}
{\ctxonecell{\rho : s \rewrite s'}{\Delta}{\Gamma} \quad \quad \ctxonecell{\tcC : s' \rewrite s''}{\Delta}{\Gamma}}
{\gtypezerocell{T}}
{\typeonecell{\Delta}{\mapcomp{T}{\rho}{\tcC} : \map{T}{(\vertcompwrongway{\tcC}{\rho})} \isorewrite \map{T}{\rho} \cdot \map{T}{\tcC}}{\substTy{T}{s}}{\substTy{T}{s''}}}
{\ruleMapComp}

\quadRule
{\ctxzerocell{\Gamma, \Delta, \Epsilon}}
{\ctxonecell{s}{\Epsilon}{\Delta} \quad \quad \ctxonecell{s', s''}{\Delta}{\Gamma}}
{\ctxonecell{\rho : s' \rewrite s''}{\Delta}{\Gamma}}
{\gtypezerocell{T}}
{\typeonecell{\Epsilon}{\maplwhisker{T}{s}{\rho} : \map{T}{(s \whiskerl \rho)} \isorewrite \invsubcomp{s}{s'} \cdot \substTm{(\map{T}{\rho})}{s} \cdot \subcomp{s}{s''}}{\substTy{T}{\horcomp{s}{s'}}}{\substTy{T}{\horcomp{s}{s''}}}}
{\ruleMapWhiskerL}

\quadRule
{\ctxzerocell{\Gamma, \Delta, \Epsilon}}
{\ctxonecell{s, s'}{\Epsilon}{\Delta} \quad \quad \ctxonecell{s''}{\Delta}{\Gamma}}
{\ctxonecell{\rho : s \rewrite s'}{\Epsilon}{\Delta}}
{\gtypezerocell{T}}
{\typeonecell{\Epsilon}{\maprwhisker{T}{\rho}{s''} : \map{T}{(\rho \whiskerr s'')} \isorewrite \invsubcomp{s}{s''} \cdot \map{(\substTy{T}{s''})}{\rho} \cdot \subcomp{s'}{s''}}{\substTy{T}{\horcomp{s}{s''}}}{\substTy{T}{\horcomp{s'}{s''}}}}
{\ruleMapWhiskerR}

\emph{The naturality rules for \ruleMapComp, \ruleMapWhiskerL, and \ruleMapWhiskerR \xspace have been omitted.}

\end{rules}
\caption{Some rules for local substitution (preservation)}
\label{fig:rules-local-substitution2}
\end{figure*}

\begin{figure*}\footnotesize
  \begin{rules}
    \ternaryRule
    {\ctxzerocell{\Gamma, \Delta}}
    {\ctxonecell{s}{\Delta}{\Gamma}}
    {\gtypezerocell{T}}
    {\typeonecell
      {\Delta}
    { \subcompmapsubidr{T}{s} : \horcomp{\subcomp{s}{1_\Delta}} {(\map{T}{\rightunitor{s}})} \isorewrite \subid}
      {\substTyTwo{T}{s}{1_\Delta}}
      {\substTy{T}{s}}
    }
    {\ruleSubCompMapSubIdR}
	\ternaryRule
    {\ctxzerocell{\Gamma, \Delta}}
    {\ctxonecell{s}{\Delta}{\Gamma}}
    {\gtypezerocell{T}}
    {\typeonecell
      {\Delta}
    { \subcompmapsubidl{T}{s} : \horcomp{\subcomp{1_\Gamma}{s}} {(\map{T}{\leftunitor{s}})} \isorewrite \substTm{\subid}{s}}
      {\substTyTwo{T}{1_\Gamma}{s}}
      {\substTy{T}{s}}
    }
    {\ruleSubCompMapSubIdL}
    \pentRule
    {\ctxzerocell{ \Gamma, \Delta, \Epsilon, \Zeta}}
	{\ctxonecell{s}{\Zeta }{\Epsilon}}
	{\ctxonecell{t}{\Epsilon}{\Delta}}
	{\ctxonecell{u}{\Delta}{\Gamma}}
    {\gtypezerocell{T}}
    {\typeonecell
      {\Zeta}
      {\subcompcomp{T}{u}{t}{s} : \horcompt{\substRed{\subcomp{t}{u}}{s} \ }{\subcomp{s}{\horcomp{t}{u}} \ }{\map{T}{\alpha}} \isorewrite \horcomp{\subcomp{s}{t} \ }{\subcomp{\horcomp{s}{t}}{u}}}
      {\substTyThree{T}{u}{t}{s}}
      {\substTy{T}{\horcomp{\horcomp{s}{t}}{u}}}
    }
    {\ruleSubCompMapSubAssoc}

    \emph{The naturality rules for \ruleSubCompMapSubIdR, \ruleSubCompMapSubIdL, and \ruleSubCompMapSubAssoc \xspace have been omitted.}
  \end{rules}
  \caption{Some of the rules for coherence of substitution}
  \label{fig:rules-coherence-substitution}
\end{figure*}

The rules for substitution are given in Figures~\ref{fig:rules-global-substitution1}, \ref{fig:rules-global-substitution2}, \ref{fig:rules-global-substitution3}, \ref{fig:rules-local-substitution1}, \ref{fig:rules-local-substitution2}, and \ref{fig:rules-coherence-substitution}.
We distinguish them based on whether we need the global cleaving or the local opcleaving to interpret them.
There are several important observations to be made about these rules.
First, in line with our truly bicategorical approach, we do not assume the comprehension bicategory is split.
In particular, no equality between $\substTy{T}{\id}$ and $T$ is postulated.
Instead, there is an equivalence between them (see \ruleref{\ruleSubId} and \ruleref{\ruleSubComp}), and terms of these types are transported along the equivalence.

The rule \ruleref{\ruleMap} expresses that each type $T$ behaves `functorially': for each $\ctxonecell{s}{\Delta}{\Gamma}$ (\ie object in `$\hom(\Delta,\Gamma)$') we get a type $T[s]$ (\ie object in `the category of types in context $\Delta$') by \ruleref{\ruleSubTy} and for each $r: s \rewrite s'$ (i.e., morphism in `$\hom(\Delta,\Gamma)$') we get a term 
$\typeonecell{\Delta}{\map{T}{\tc}}{T[s]}{T[s']}$
(\ie morphism in `the category of types in context $\Delta$') by \ruleref{\ruleMap}. The rules \ruleref{\ruleMapId} and \ruleref{\ruleMapComp} ensure that $T[-]$ preserves identity and composition. With \ruleref{\ruleRewTm} we can then understand terms to be `natural transformations'.

\begin{remark}
For contravariant comprehension bicategories (Remark~\ref{rem:contravariant-compbicats}), the rule \ruleref{\ruleMap} would be in the opposite direction, while for isovariant comprehension bicategories, this rule would be restricted to isomorphisms in the base.
\end{remark}

\begin{remark}
Using the Grothendieck construction, we can view the rules from the perspective of fiber bicategories.
If $P : \E \rightarrow \B$ has a global cleaving and a local opcleaving, and furthermore opcartesian 2-cells are preserved under whiskering, then we obtain a trifunctor $\op{\B} \rightarrow \Bicat$, which sends every $x : \B$ to the fiber of $x$ along $P$.
The rules given in Figure~\ref{fig:rules-global-substitution1} say that any 1-cell in the base gives rise to a pseudofunctor.
For example, the rules \ruleref{\ruleSubTy}, \ruleref{\ruleSubTm}, and \ruleref{\ruleSubRed} represent the actions on objects, 1-cells, and 2-cells, respectively, while \ruleref{\ruleSubOnId} and \ruleref{\ruleSubOnComp} represent the invertible 2-cells that witness the preservation of the identity and composition of 1-cells.
The other rules in that table are the coherence laws of pseudofunctors: for example, \ruleref{\ruleSubRedOnId} is the preservation of identity 2-cells.

Figures \ref{fig:rules-global-substitution2} and \ref{fig:rules-global-substitution3} give pseudonatural equivalences expressing that the assignment of the previously mentioned pseudofunctor is actually pseudofunctorial: the identity and composition are preserved up to pseudonatural equivalence.
For example, the rules \ruleref{\ruleSubId} and \ruleref{\ruleTmSubId} in Figure~\ref{fig:rules-global-substitution2} represent the action on objects and the naturality squares.
On the other hand, \ruleref{\ruleSubRedId}, \ruleref{\ruleSubIdPrevId}, and \ruleref{\ruleSubIdPrevComp} are the usual coherencies of pseudonatural transformations.

The rules of Figure~\ref{fig:rules-local-substitution1} state that we obtain a pseudonatural transformation from a 2-cell in the base.
The action on objects and 1-cells is given by \ruleref{\ruleMap} and \ruleref{\ruleRewTm}, respectively, while \ruleref{\ruleRewRed}, \ruleref{\ruleRewOnId}, and \ruleref{\ruleRewOnComp} describe the usual coherencies of pseudonatural transformations.
The remaining rules can be found in Figure~\ref{fig:rules-local-substitution2}.
In this figure, four invertible modifications are described: \ruleref{\ruleMapId} and \ruleref{\ruleMapComp} describe the preservation of identity and composition, respectively, while \ruleref{\ruleMapWhiskerL} and \ruleref{\ruleMapWhiskerR} describe the preservation of both left and right whiskering.
Note that we left out the naturality conditions of these modifications.
In addition, two additional coherencies are required which express that all this data together forms a trifunctor.
We do not write them down, but instead, we refer the reader to Definition 3.3.1 of \cite{gurski2005algebraic}.
\end{remark}

\begin{remark}
A similar rule to \ruleref{\ruleMap} of Figure~\ref{fig:rules-local-substitution1} appears in work of \cite{johnstone1993fibrations}.
There, it is shown that a 1-cell $p$ in a 2-category is an internal fibration in a sense similar to our Definition~\ref{def:internal_sfib} if and only if a specific functor given by precomposition with $p$ has a semi-oplax right adjoint satisfying some properties, assuming that pullbacks along $p$ exist.
In Lemma 2.5 of \cite{johnstone1993fibrations}, another equivalent characterization of $p$ being a fibration is given assuming that pullbacks of $p$ exist.
This characterization uses several operations and compatibility requirements.
One of the operations is described by the following diagram:
\[
\begin{tikzcd}[column sep = large]
	{f^*e} \\
	{} & {g^*e} & e \\
	& a & b
	\arrow[""{name=0, anchor=center, inner sep=0}, "f", shift left, from=3-2, to=3-3, bend left]
	\arrow[""{name=1, anchor=center, inner sep=0}, "g"', shift right, from=3-2, to=3-3, bend right]
	\arrow["p", from=2-3, to=3-3]
	\arrow[from=1-1, to=2-3, bend left=20]
	\arrow[from=1-1, to=3-2, bend right=20]
	\arrow[from=2-2, to=3-2]
	\arrow[from=2-2, to=2-3]
	\arrow["{\widehat{p}(\alpha)}"{description}, dotted, from=1-1, to=2-2]
	\arrow["\alpha", shorten <=2pt, shorten >=2pt, Rightarrow, from=0, to=1]
\end{tikzcd}
\]
We assume that pullbacks along $p$ are given; in particular, we have $f^* e$ and $g^* e$.
The characterization of $p$ being a fibration entails, in particular, that for every 2-cell $\alpha : f \mytwocell g$ there is a 1-cell $\hat{p}(\alpha) : f^*e \rightarrow g^*e$.
This takes place in a $2$-category, though Johnstone claims that such a characterization would also hold in a bicategory. Modulo this difference, this operation is expressed by rule \ruleref{\ruleMap} of Figure~\ref{fig:rules-local-substitution1}.
\end{remark}

\section{Soundness: Interpretation in Comprehension Bicategories}
\label{sec:soundn-interpr-compr}
In this section, we give an interpretation of \BTT in any weak comprehension bicategory.
To this end, we fix a comprehension bicategory %
\textcompbicat{\D}{\chi}{\B}.
We interpret the judgments as follows.
\begin{itemize}
	\item $\Gamma \ctx$ is interpreted as an object $\sem{\Gamma}$ of $\B$.
	\item $\ctxonecell{s}{\Delta}{\Gamma}$ is interpreted as a 1-cell $\sem{s} : \sem{\Delta} \rightarrow \sem{\Gamma}$ in $\B$.
	\item $\ctxonecell{r : s \rewrite t}{\Delta}{\Gamma}$ is interpreted as a 2-cell $\sem{r} : \sem{s} \mytwocell \sem{t}$ in $\B$.
	\item $\ctxonecell{r \equiv r' : s \rewrite t}{\Delta}{\Gamma}$ is interpreted as an equality $\sem{r} = \sem{r'}$.
	\item $\Gamma \vdash T \type$ is interpreted as an object $\sem{T}$ in $\D$ over $\sem{\Gamma}$.
	\item $\typeonecell{\Gamma}{t}{S}{T}$ is interpreted as a 1-cell $\sem{t} : \sem{S} \rightarrow \sem{T}$ over the identity on $\sem{\Gamma}$.
	\item $\typeonecell{\Gamma}{r: t \rewrite t'}{S}{T}$ is interpreted as a 2-cell $\sem{r} : \sem{t} \mytwocell \sem{t'}$ over the identity 2-cell.
	\item $\typeonecell{\Gamma}{r \equiv r' : t \rewrite t'}{S}{T}$ is interpreted as an equality $\sem{r} = \sem{r'}$.
\end{itemize}

Regarding the ``bicategorical'' rules of Figures~\ref{fig:rules-base} and \ref{fig:rules-total},
each  rule is analogous to one of the operations or laws of a bicategory --- see, for instance, Definition~3.1 of \cite{ahrens_frumin_maggesi_veltri_van_der_weide_2022}. This also indicates how it is interpreted.

\subsection{Comprehension}

In this section, we interpret the rules related to comprehension of Figure~\ref{fig:rules-comprehension}.
Suppose that we have a context $\Gamma : \B$ and a type $T$ over $\Gamma$.
Its image $\chi(T)$ in $\arrows{\B}$ gives rise to an object $\Gamma . T : \B$ and a 1-cell $\pi_{\Gamma.T} : \Gamma . T \rightarrow \Gamma$ which interprets \ruleref{\ruleExtendTy}.
For a 1-cell $t$ from $S$ to $T$ over the identity, the resulting 1-cell $\chi(t)$ is part of a triangle 
\begin{equation*}
\begin{tikzcd}
  \Gamma . S \ar[rr, "\chi(t)"] \ar[dr, "\pi_{\Gamma.S}"']
  &
  &
  \Gamma.T \ar[dl, "\pi_{\Gamma.T}"]
  \\
  &
  \Gamma
\end{tikzcd}
\end{equation*}
which commutes up to invertible 2-cell.
This yields the interpretation of the rules in \ruleref{\ruleExtendTm}.
Furthermore, a reduction $r : t \rewrite t'$ is mapped by $\chi$ to a 2-cell from $\chi(t)$ to $\chi(t')$ in $\arrows{\B}$, and this is how we interpret \ruleref{\ruleExtendRed}.
The rules \ruleref{\ruleExtendId} and \ruleref{\ruleExtendComp} are interpreted by the identitor and compositor of $\chi$, respectively, while the remaining rules in Figure~\ref{fig:rules-comprehension} are satisfied because they translate to the laws that express that this data forms a pseudofunctor.

\subsection{Global Substitution}
Next, we interpret the rules for global substitution of Figure~\ref{fig:rules-global-substitution1} (in Sections~\ref{sec:interpr-ruler}, \ref{sec:interpr-rules-2}, \ref{sec:interpr-rules-3}, and \ref{sec:interpretation-sublunitor}), and the rules regarding preservation of identity under global substitution of Figure~\ref{fig:rules-global-substitution2} (in Sections~\ref{sec:interpr-rules-1}, \ref{sec:interpr-rules}, and \ref{sec:interpr-ruletms}).
The interpretation of the rules regarding preservation of composition of Figure~\ref{fig:rules-global-substitution3} is analogous to that of Figure~\ref{fig:rules-global-substitution2} and is not spelled out.

The interpretation of the rule \ruleref{\ruleSubTy} is given directly by the global cleaving.
The interpretation of the other rules require more explanation.

\subsubsection{Interpretation of \ruleSubTm}
\label{sec:interpr-ruler}

To interpret \ruleref{\ruleSubTm}, we assume that we have a substitution $s : \Delta \rightarrow \Gamma$ between contexts $\Delta$ and $\Gamma$.
We also assume that we have types $S, T$ over $\Gamma$ and a morphism $t : S \to T$ over $1_\Gamma$.
This situation is encapsulated in the following diagram, using the notation introduced in Definition~\ref{def:global-cleaving}.
\[
\begin{tikzcd}
	{\substTy{S}{s}} && S \\
	\\
	{\substTy{T}{s}} && T \\
	&& {} \\
	\Delta && \Gamma
	\arrow["s"', from=5-1, to=5-3]
	\arrow["t", from=1-3, to=3-3]
	\arrow["{\cartesianlift{s}{S}}", from=1-1, to=1-3]
	\arrow["{\cartesianlift{s}{T}}"', from=3-1, to=3-3]
\end{tikzcd}
\]
Our goal is to construct a 1-cell $\substTm{t}{s} : \substTy{S}{s} \rightarrow \substTy{T}{s}$ that lies over the identity on $\Delta$.
Since the morphism $\cartesianlift{s}{T} : \substTy{T}{s} \rightarrow T$ is cartesian, it suffices to construct a 1-cell, say, $\beta : \substTy{S}{s} \rightarrow T$ that lies over $1_\Delta \cdot s$. We then obtain the desired 1-cell as the factorization $\Factone{\cartesianlift{s}{T}}{\beta}$ of that 1-cell via $\cartesianlift{s}{T}$. (Recall that $\Factone{\_}{\_}$ was defined in Definition~\ref{def:cart-1-cell}.)
We have an invertible 2-cell $\rightunitor{s} \vcomp {\leftunitor{s}}^{-1} : s \cdot 1_\Gamma \mytwocell 1_\Delta \cdot s$, and we set $\beta \defeq \PF{(\rightunitor{s} \vcomp {\leftunitor{s}}^{-1})}{(\cartesianlift{s}{S} \cdot t)}$.
Define
\[
\substTm{t}{s} \defeq \Factone{\cartesianlift{s}{T}}{\beta}.
\]

Overall, we can summarize the construction with the following diagram.
\[
  \begin{tikzcd}[column sep = large]
    &
    \substTy{S}{s}
    \ar[r, "{\cartesianlift{s}{S}}"]
    \ar[ld, equal]
    &
    | [alias=S] | S
    \ar[rd, "t"]
    \\
    \substTy{S}{s}
    \ar[rrd, "\beta" near start, ""{name=U}, ""'{name=V}, bend left=20]     
    \ar[rd, dashed, "{\substTm{t}{s}}"']
    &
    &
    &
    T
    \ar[ld, equal]
    \\
    &
    | [alias=Ts] | \substTy{T}{s}
    \ar[r, "{\cartesianlift{s}{T}}"']
    &
    T
    \\
    \Delta
    \ar[r, "1_\Delta"']
    &
    \Delta
        \ar[r, "s"']
    &
    \Gamma
        \ar[r, "1_\Gamma"']
    &
    \Gamma
    \ar[from=S,to=U, Rightarrow, shorten=1.5em, "\cong"]
    \ar[from=Ts,to=V, Rightarrow, shorten=0.2em, "\cong"]
  \end{tikzcd}
\]

\subsubsection{Interpretation of \ruleSubRed}
\label{sec:interpr-rules-2}

To interpret \ruleref{\ruleSubRed}, we assume that we have a substitution $s : \Delta \rightarrow \Gamma$ and types $S$ and $T$ in context $\Gamma$.
We also assume that we have 1-cells $t, t' : S \to T$ over $1_\Gamma$ and a 2-cell $\rho : t \mytwocell t'$.
Consider the following diagram.
\[
\begin{tikzcd}
	{\substTy{S}{1_\Gamma}} && S \\
	\\
	{\substTy{T}{1_\Gamma}} && T \\
	\\
	\Delta && \Gamma
	\arrow["s"', from=5-1, to=5-3]
	\arrow[""{name=0, anchor=center, inner sep=0}, "t"', bend right, shift right=1, from=1-3, to=3-3]
	\arrow["{\cartesianlift{s}{S}}", from=1-1, to=1-3]
	\arrow["{\cartesianlift{s}{T}}"', from=3-1, to=3-3]
	\arrow["{\substTm{t'}{s}}", bend left, shift left=1, dashed, from=1-1, to=3-1]
	\arrow[""{name=1, anchor=center, inner sep=0}, "{t'}", bend left, shift left=1, from=1-3, to=3-3]
	\arrow["{\substTm{t}{s}}"', bend right, shift right=1, dashed, from=1-1, to=3-1]
	\arrow["\rho"', shorten <=3pt, shorten >=3pt, Rightarrow, from=0, to=1]
\end{tikzcd}
\]
Our goal is to construct a 2-cell from $\substTm{t}{s}$ to $\substTm{t}{s'}$ over the identity on $\Delta$.
The source and target, respectively, were constructed by factoring a 1-cell via a cartesian 1-cell, as follows.
\[
  \begin{tikzcd}[column sep = 3em]
    &
    \substTy{S}{s}
    \ar[r, "{\cartesianlift{s}{S}}"]
    \ar[ld, equal]
    &
    | [alias=S] | S
    \ar[rd, "t"]
    \\
    \substTy{S}{s}
    \ar[rrd, "\beta" near start, ""{name=U}, ""'{name=V}, bend left=20]     
    \ar[rd, dashed, "{\substTm{t}{s}}"']
    &
    &
    &
    T
    \ar[ld, equal]
    \\
    &
    | [alias=Ts] | \substTy{T}{s}
    \ar[r, "{\cartesianlift{s}{T}}"']
    &
    T
    \\
    \Delta
    \ar[r, "1_\Delta"']
    &
    \Delta
        \ar[r, "s"']
    &
    \Gamma
        \ar[r, "1_\Gamma"']
    &
    \Gamma
    \ar[from=S,to=U, Rightarrow, shorten=1.5em, "\cong"]
    \ar[from=Ts,to=V, Rightarrow, shorten=0.2em, "\cong"]
  \end{tikzcd}
  \quad
    \begin{tikzcd}[column sep = 3em]
    &
    \substTy{S}{s}
    \ar[r, "{\cartesianlift{s}{S}}"]
    \ar[ld, equal]
    &
    | [alias=S] | S
    \ar[rd, "t'"]
    \\
    \substTy{S}{s}
    \ar[rrd, "\beta'" near start, ""{name=U}, ""'{name=V}, bend left=20]     
    \ar[rd, dashed, "{\substTm{t'}{s}}"']
    &
    &
    &
    T
    \ar[ld, equal]
    \\
    &
    | [alias=Ts] | \substTy{T}{s}
    \ar[r, "{\cartesianlift{s}{T}}"']
    &
    T
    \\
    \Delta
    \ar[r, "1_\Delta"']
    &
    \Delta
        \ar[r, "s"']
    &
    \Gamma
        \ar[r, "1_\Gamma"']
    &
    \Gamma
    \ar[from=S,to=U, Rightarrow, shorten=1.5em, "\cong"]
    \ar[from=Ts,to=V, Rightarrow, shorten=0.2em, "\cong"]
  \end{tikzcd}
\]
To construct the desired 2-cell $\substRed{\rho}{s} : \substTm{t}{s} \mytwocell \substTm{t}{s'}$, we use Item~\ref{item:def-cartesian-1cell-2cell} of Definition~\ref{def:cart-1-cell}.
More specifically, it suffices to construct
\begin{itemize}
	\item a 2-cell $\delta$ from $1_\Delta$ to $1_\Delta$, and
	\item a 2-cell $\disp{\sigma}$ from $\beta$ to $\beta'$.
\end{itemize}
Once we have defined those, we define $\substRed{\rho}{s} \defeq \Eqtwo{\delta}{\disp{\sigma}}$.

For the first of the two, we take the identity 2-cell $1_{1_{\Delta}}$.
For the second, we take the composition of 2-cells shown in the following diagram.
\[
  \begin{tikzcd}[column sep = huge]
    \substTy{S}{s}
    \ar[rr, "\beta", ""'{name=A}, bend left=45]
    \ar[rr, "\beta'"', ""{name=B}, bend right=45]
    \ar[r, "\cartesianlift{s}{S}"]
    &
    S
    \ar[r, bend left, "t" near start, ""'{name=C}]
    \ar[r, bend right, "t'"' near start, ""{name=D}]
    &
    T
    \ar[from=A, to=1-2, Rightarrow, shorten=.3em, "\cong"']
    \ar[to=B, from=1-2, Rightarrow, shorten=.3em, "\cong"']
    \ar[from=C, to=D, Rightarrow, "\rho"]
  \end{tikzcd}
\]

The interpretation of the rules \ruleref{\ruleSubRedId} and \ruleref{\ruleSubRedComp} follows from the uniqueness of factorization 2-cells.

\subsubsection{Interpretation of \ruleSubOnId}
\label{sec:interpr-rules-3}

The rules \ruleref{\ruleSubOnId} and \ruleref{\ruleSubOnComp} are interpreted similarly, and we show the interpretation of \ruleref{\ruleSubOnId}.
Suppose that we have a substitution $s : \Delta \rightarrow \Gamma$ and a type $T$ in context $\Gamma$.
We consider two morphisms: the identity on $\substTy{T}{s}$ and $\substTm{1_T}{s}$.
These are illustrated in the following diagram.
\[
\begin{tikzcd}
	{\substTy{T}{s}} && T \\
	\\
	{\substTy{T}{s}} && T
	\arrow["{1_T}", from=1-3, to=3-3]
	\arrow["\cartesianlift{s}{S}", from=1-1, to=1-3]
	\arrow["\cartesianlift{s}{T}"', from=3-1, to=3-3]
	\arrow["{\substTm{1_T}{s}}"', bend right, shift right, from=1-1, to=3-1]
	\arrow["{1_{\substTy{T}{s}}}", bend left, shift left, from=1-1, to=3-1]
\end{tikzcd}
\]
Our goal is to construct an invertible 2-cell from $1_{\substTy{T}{s}}$ to $\substTm{1_T}{s}$.
To do so, we first recall the construction of $\substTm{1_T}{s}$.
\[
\begin{tikzcd}[column sep = large]
	&
	\substTy{T}{s}
	\ar[r, "{\cartesianlift{s}{T}}"]
	\ar[ld, equal]
	&
	| [alias=S] | T
	\ar[rd, "1_T"]
	\\
	\substTy{T}{s}
	\ar[rrd, "\beta" near start, ""{name=U}, ""'{name=V}, bend left=20]     
	\ar[rd, dashed, "{\substTm{1_T}{s}}"']
	&
	&
	&
	T
	\ar[ld, equal]
	\\
	&
	| [alias=Ts] | \substTy{T}{s}
	\ar[r, "{\cartesianlift{s}{T}}"']
	&
	T
	\\
	\Delta
	\ar[r, "1_\Delta"']
	&
	\Delta
	\ar[r, "s"']
	&
	\Gamma
	\ar[r, "1_\Gamma"']
	&
	\Gamma
	\ar[from=S,to=U, Rightarrow, shorten=1.5em, "\cong"]
	\ar[from=Ts,to=V, Rightarrow, shorten=0.2em, "\cong"]
\end{tikzcd}
\]
To construct the desired invertible 2-cell, we again use Item~\ref{item:def-cartesian-1cell-2cell} of Definition~\ref{def:cart-1-cell}.
As such, we need to construct the following:
\begin{itemize}
	\item An invertible 2-cell $\delta$ from $1_\Delta$ to $1_\Delta$; and
	\item An invertible 2-cell $\disp{\sigma}$ from $\substTm{1_T}{s} \cdot \cartesianlift{s}{T}$ to $1_{\substTy{T}{s}} \cdot \cartesianlift{s}{T}$.
\end{itemize}
With those in place, we define $\subOnId{s}^{-1} \defeq \Eqtwo{\delta}{\disp{\sigma}}$.
For $\delta$, we take the identity, while for $\disp{\sigma}$, we take the following composition.
\[
1_{\substTy{T}{s}} \cdot \cartesianlift{s}{T} \cong \cartesianlift{s}{T} \cdot 1_T \cong \beta \cong  \substTm{1_T}{s} \cdot \cartesianlift{s}{T}
\]

\subsubsection{Interpretation of \ruleSubLunitor}
\label{sec:interpretation-sublunitor}

The remaining laws stating equalities of 2-cells, such as \ruleref{\ruleSubOnComp} and \ruleref{\ruleSubIdPrevId}, are all proven in a similar way.
Our goal is to prove an equality $\disp{\alpha} = \disp{\beta}$ with 1-cells and 2-cells as in the diagram below:
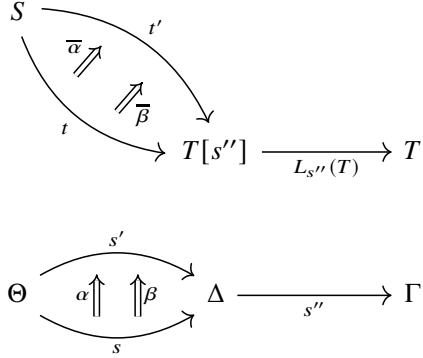
\[
\begin{tikzcd}
	S \\
	\\
	&& {\substTy{T}{s''}} && T \\
	\\
	\Epsilon && \Delta && \Gamma
	\arrow["{s''}"', from=5-3, to=5-5]
	\arrow["{\cartesianlift{s''}{T}}"', from=3-3, to=3-5]
	\arrow[""{name=0, anchor=center, inner sep=0}, "s"', bend right, shift right=1, from=5-1, to=5-3]
	\arrow[""{name=1, anchor=center, inner sep=0}, "{t'}", bend left, shift left=1, from=1-1, to=3-3]
	\arrow[""{name=2, anchor=center, inner sep=0}, "t"', bend right, shift right=1, from=1-1, to=3-3]
	\arrow[""{name=3, anchor=center, inner sep=0}, "{s'}", bend left, shift left=1, from=5-1, to=5-3]
	\arrow["{\disp{\alpha}}", shift left=4, shorten=1em, Rightarrow, from=2, to=1]
	\arrow["{\disp{\beta}}"', shift right=4, shorten=1em, Rightarrow, from=2, to=1]
	\arrow["\beta"', shift right=3, shorten=.7em, Rightarrow, from=0, to=3]
	\arrow["\alpha", shift left=3, shorten=.7em, Rightarrow, from=0, to=3]
\end{tikzcd}
\]
Here $t$ and $t'$ live over $s$ and $s'$, respectively, while $\disp{\alpha}$ and $\disp{\beta}$ live over $\alpha$ and $\beta$, respectively.
To prove the equality, we take the following steps.
\begin{enumerate}
	\item We construct an invertible 2-cell $\theta$ from $t \cdot \cartesianlift{s''}{T}$ to $t' \cdot \cartesianlift{s''}{T}$.
	\item We prove that $\alpha = \beta$.
	\item We prove that $\disp{\alpha} \whiskerr \cartesianlift{s''}{T} = \theta = \disp{\beta} \whiskerr \cartesianlift{s''}{T}$.
\end{enumerate}
From the first item, we get that both $t$ and $t'$ are cartesian factorizations of the same $S \rightarrow T$.
From the second and third item, we get that both $\disp{\alpha}$ and $\disp{\beta}$ are factorization 2-cells from $t$ to $t'$, and since such 2-cells are unique, we get $\disp{\alpha} = \disp{\beta}$.

We demonstrate this for \ruleref{\ruleSubLunitor}.
Suppose we have a substitution $s : \Delta \rightarrow \Gamma$, types $S$ and $T$ in context $\Gamma$, and a morphism $t : S \to T$ over $1_\Gamma$.
We are in the following situation.
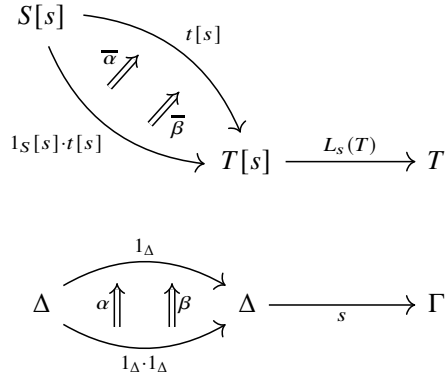
\[
\begin{tikzcd}
	{\substTy{S}{s}} \\
	\\
	&& {\substTy{T}{s}} && T \\
	\\
	\Delta && \Delta && \Gamma
	\arrow["{\cartesianlift{s}{T}}", from=3-3, to=3-5]
	\arrow["s"', from=5-3, to=5-5]
	\arrow[""{name=0, anchor=center, inner sep=0}, "{\substTm{t}{s}}", bend left, shift left=1, from=1-1, to=3-3]
	\arrow[""{name=1, anchor=center, inner sep=0}, "{\substTm{1_S}{s} \cdot \substTm{t}{s}}"', bend right, shift right=1, from=1-1, to=3-3]
	\arrow[""{name=2, anchor=center, inner sep=0}, "{1_\Delta}", bend left, shift left=1, from=5-1, to=5-3]
	\arrow[""{name=3, anchor=center, inner sep=0}, "{1_\Delta \cdot 1_\Delta}"', bend right, shift right=1, from=5-1, to=5-3]
	\arrow["{\disp{\alpha}}", shift left=4, shorten=1em, Rightarrow, from=1, to=0]
	\arrow["{\disp{\beta}}"', shift right=4, shorten=1em, Rightarrow, from=1, to=0]
	\arrow["\alpha", shift left=4, shorten=.7em, Rightarrow, from=3, to=2]
	\arrow["\beta"', shift right=4, shorten=.7em, Rightarrow, from=3, to=2]
\end{tikzcd}
\]
Here, we define $\alpha \defeq \vertcomp{(1_{1_\Delta} \whiskerr 1_\Delta)}{\ell_{1_\Delta}}$ and $\beta \defeq \vertcomp{1_{1_\Delta \cdot 1_\Delta}}{\ell_{1_\Delta}}$.
We also define $\disp{\alpha}$ and $\disp{\beta}$ as follows.
\[
\disp{\alpha} \defeq \vertcomp{(\subOnId{s}^{-1} \whiskerr \substTm{t}{s})}{\ell_{\substTm{t}{s}}}
\quad \quad
\disp{\beta} \defeq \vertcomp{\subOnComp{1_S}{t}{s}}{\substRed{\ell_t}{s}}
\]
Then by construction, both $\alpha$ and $\beta$ are equal to $\ell_{1_\Delta}$, and hence, $\alpha = \beta$.

Next we construct an invertible 2-cell from $\substTm{1_S}{s} \cdot \substTm{t}{s} \cdot \cartesianlift{s}{T}$ to $\substTm{t}{s} \cdot \cartesianlift{s}{T}$.
By construction of $\substTm{t}{s}$, we have an invertible 2-cell $\theta_1 : \substTm{t}{s} \cdot \cartesianlift{s}{T} \cong \cdot \cartesianlift{s}{S} \cdot t$.
In addition, we have an invertible 2-cell $\theta_2 : \substTm{1_S}{s} \cdot \cartesianlift{s}{S} \cong \cartesianlift{s}{S} \cdot 1_S$.
This is illustrated in the following diagram.
\[
\begin{tikzcd}
	{\substTy{S}{s}} && S \\
	\\
	{\substTy{S}{s}} && S \\
	\\
	{\substTy{T}{s}} && T
	\arrow[""{name=0, anchor=center, inner sep=0}, "t", from=3-3, to=5-3]
	\arrow[""{name=1, anchor=center, inner sep=0}, "{1_S}", from=1-3, to=3-3]
	\arrow["\cartesianlift{s}{S}", from=1-1, to=1-3]
	\arrow["\cartesianlift{s}{S}"{description}, from=3-1, to=3-3]
	\arrow["\cartesianlift{s}{T}"', from=5-1, to=5-3]
	\arrow["{\substTm{1_S}{s}}"', from=1-1, to=3-1]
	\arrow["{\substTm{t}{s}}"', from=3-1, to=5-1]
	\arrow["{\theta_1}", shorten=1.2em, Rightarrow, from=3-1, to=1]
	\arrow["{\theta_2}", shorten=1.2em, Rightarrow, from=5-1, to=0]
\end{tikzcd}
\]
Now we define an invertible 2-cell $\theta_3 : \substTm{1_S}{s} \cdot \substTm{t}{s} \cdot \cartesianlift{s}{T} \cong \cartesianlift{s}{S} \cdot t$ as follows.
\[
\theta_3 \defeq \vertcomp{\vertcomp{\alpha^{-1}}{\vertcomp{(\substTm{1_S}{s} \whiskerl \theta_2)}{\vertcomp{\alpha}{(\theta_1 \whiskerr t)}}}}{(r_\sigma \whiskerr t)}
\]
The desired invertible 2-cell $\theta : \substTm{1_S}{s} \cdot \substTm{t}{s} \cdot \cartesianlift{s}{T} \cong \substTm{t}{s} \cdot \cartesianlift{s}{T}$ is $\vertcomp{\theta_3}{\theta_2^{-1}}$.

For the last step, we only prove that $\disp{\alpha} \whiskerr \cartesianlift{s}{T} = \theta$. 
We first recall how we constructed $\subOnId{s}$. For brevity, we write $\sigma \defeq \cartesianlift{s}{S}$ and $\tau \defeq \cartesianlift{s}{T}$.

\[
\begin{tikzcd}
	{\substTy{S}{s}} && S \\
	\\
	{\substTy{S}{s}} && S
	\arrow["\sigma", from=1-1, to=1-3]
	\arrow["\sigma"', from=3-1, to=3-3]
	\arrow[""{name=0, anchor=center, inner sep=0}, "{1_S}", from=1-3, to=3-3]
	\arrow[""{name=1, anchor=center, inner sep=0}, "{1_{\substTy{S}{s}}}"', bend right=50, shift right=1, from=1-1, to=3-1]
	\arrow[""{name=2, anchor=center, inner sep=0}, "{\substTm{1_S}{s}}", bend left=50, shift left=1, from=1-1, to=3-1]
	\arrow[shorten=.5em, Rightarrow, from=1, to=2, "\subOnId{s}^{-1}"]
	\arrow["{\theta_1}"', shorten=1.2em, Rightarrow, from=3-1, to=0]
\end{tikzcd}
\]
By construction, we have that $\vertcomp{(\subOnId{s}^{-1} \whiskerr \sigma)}{\vertcomp{\ell_\sigma}{r_\sigma^{-1}}} = \theta_1$.
In particular, we have $\subOnId{s}^{-1} \whiskerr \sigma = \vertcomp{\theta_1}{\vertcomp{r_\sigma}{\ell_{\sigma}^{-1}}}$.

To show that $(\vertcomp{(\subOnId{s}^{-1} \whiskerr \substTm{t}{s})}{\ell_{\substTm{t}{s}}}) \whiskerr \tau = \vertcomp{\theta_3}{\theta_2^{-1}}$, it suffices to show that $\vertcomp{(\vertcomp{(\subOnId{s}^{-1} \whiskerr \substTm{t}{s})}{\ell_{\substTm{t}{s}}}) \whiskerr \tau}{\theta_2} = \theta_3$.
This follows from the following chain of equalities.
\begin{align*}
	& \quad \vertcomp{\alpha^{-1}}{\vertcomp{(\substTm{1_S}{s} \whiskerl \theta_2^{-1})}{\vertcomp{\alpha}{\vertcomp{(\vertcomp{(\subOnId{s}^{-1} \whiskerr \substTm{t}{s})}{\ell_{\substTm{t}{s}}}) \whiskerr \tau}{\vertcomp{\theta_2}{(r_\sigma \whiskerr t)^{-1}}}}}} \\
	&=  \vertcomp{\alpha^{-1}}{\vertcomp{(\substTm{1_S}{s} \whiskerl \theta_2^{-1})}{\vertcomp{\alpha}{\vertcomp{\vertcomp{(\subOnId{s}^{-1} \whiskerr \substTm{t}{s}) \whiskerr \tau}{(\ell_{\substTm{t}{s}}} \whiskerr \tau)}{\vertcomp{\theta_2}{(r_\sigma \whiskerr t)^{-1}}}}}} \\
	&=  \vertcomp{\alpha^{-1}}{\vertcomp{(\substTm{1_S}{s} \whiskerl \theta_2^{-1})}{\vertcomp{\subOnId{s}^{-1} \whiskerr (\substTm{t}{s} \cdot \tau)}{\vertcomp{\vertcomp{\alpha}{(\ell_{\substTm{t}{s}}} \whiskerr \tau)}{\vertcomp{\theta_2}{(r_\sigma \whiskerr t)^{-1}}}}}} \\
	&=  \vertcomp{\alpha^{-1}}{\vertcomp{\subOnId{s}^{-1} \whiskerr (\sigma \cdot t)}{\vertcomp{(1_{\substTy{S}{s}} \whiskerl \theta_2^{-1})}{\vertcomp{\vertcomp{\alpha}{(\ell_{\substTm{t}{s}}} \whiskerr \tau)}{\vertcomp{\theta_2}{(r_\sigma \whiskerr t)^{-1}}}}}} \\
	&=  \vertcomp{\subOnId{s}^{-1} \whiskerr \sigma \whiskerr t}{\vertcomp{\alpha^{-1}}{\vertcomp{(1_{\substTy{S}{s}} \whiskerl \theta_2^{-1})}{\vertcomp{\vertcomp{\alpha}{(\ell_{\substTm{t}{s}}} \whiskerr \tau)}{\vertcomp{\theta_2}{(r_\sigma \whiskerr t)^{-1}}}}}} \\
	&=  \vertcomp{(\vertcomp{\theta_1}{\vertcomp{r_\sigma}{\ell_{\sigma}^{-1}}}) \whiskerr t}{\vertcomp{\alpha^{-1}}{\vertcomp{(1_{\substTy{S}{s}} \whiskerl \theta_2^{-1})}{\vertcomp{\vertcomp{\alpha}{(\ell_{\substTm{t}{s}}} \whiskerr \tau)}{\vertcomp{\theta_2}{(r_\sigma \whiskerr t)^{-1}}}}}} \\
	&=  \vertcomp{\theta_1 \whiskerr t}{\vertcomp{({\vertcomp{r_\sigma}{\ell_{\sigma}^{-1}}}) \whiskerr t}{\vertcomp{\alpha^{-1}}{\vertcomp{(1_{\substTy{S}{s}} \whiskerl \theta_2^{-1})}{\vertcomp{\vertcomp{\alpha}{(\ell_{\substTm{t}{s}}} \whiskerr \tau)}{\vertcomp{\theta_2}{(r_\sigma \whiskerr t)^{-1}}}}}}} \\
	&=  \vertcomp{\theta_1 \whiskerr t}{\vertcomp{({\vertcomp{r_\sigma}{\ell_{\sigma}^{-1}}}) \whiskerr t}{\vertcomp{\alpha^{-1}}{\vertcomp{(1_{S[s]} \whiskerl \theta_2^{-1})}{\vertcomp{\ell_{\horcomp{\substTm{t}{s}}{\tau}}}{\vertcomp{\theta_2}{(r_\sigma \whiskerr t)^{-1}}}}}}} \\
	&=  \vertcomp{\theta_1 \whiskerr t}{\vertcomp{({\vertcomp{r_\sigma}{\ell_{\sigma}^{-1}}}) \whiskerr t}{\vertcomp{\alpha^{-1}}{\vertcomp{\ell_{\sigma \cdot t}}{\vertcomp{\theta_2^{-1}}{\vertcomp{\theta_2}{(r_\sigma \whiskerr t)^{-1}}}}}}} \\
	&=  \vertcomp{\theta_1 \whiskerr t}{\vertcomp{\vertcomp{({\vertcomp{r_\sigma}{\ell_{\sigma}^{-1}}}) \whiskerr t}{\vertcomp{\alpha^{-1}}{\ell_{\sigma \cdot t}}}}{(r_\sigma \whiskerr t)^{-1}}} \\
	&= \vertcomp{\theta_1 \whiskerr t}{\vertcomp{\vertcomp{({\vertcomp{r_\sigma}{\ell_{\sigma}^{-1}}}) \whiskerr t}{\ell_{\sigma} \whiskerr t}}{(r_\sigma \whiskerr t)^{-1}}} \\
	&=  \theta_1 \whiskerr t
\end{align*}

\subsubsection{Interpretation of \ruleSubId}
\label{sec:interpr-rules}

Next we interpret \ruleref{\ruleSubId}. Suppose we have a context $\Gamma$ and a type $T$ in $\Gamma$.
Note that we have a diagram as follows.
\[
\begin{tikzcd}
	T \\
	\\
	{\substTy{T}{1_\Gamma}} && T \\
	\\
	\Gamma && \Gamma
	\arrow["{1_\Gamma}"', from=5-1, to=5-3]
	\arrow["{1_T}", from=1-1, to=3-3]
	\arrow["{\cartesianlift{1_\Gamma}{T}}"', from=3-1, to=3-3]
\end{tikzcd}
\]
We interpret $\subid(T)$\footnote{We omit the argument $T$ in what follows.}
by $\cartesianlift{1_\Gamma}{T}$, so it suffices to show that $\cartesianlift{1_\Gamma}{T}$ is an equivalence.
To construct the inverse, we use the following diagram.
\[
\begin{tikzcd}[row sep = large, column sep = large]
  T \arrow[r, "{1_T}"]
  \arrow[rd, ""'{name=A}, "{\Factone{\cartesianlift{1_\Gamma}{T}}{1_T \cdot 1_T}}"', dashed]
  &
  T \arrow[rd, ""{coordinate, name=B} near start, "{1_T}"] 
  \\
  &
  {\substTy{T}{1_\Gamma}} \arrow[r, "{\cartesianlift{1_\Gamma}{T}}"']
  &
  T
  \\
  \Gamma \arrow[r, "{1_\Gamma}"']
  &
  \Gamma \arrow[r, "{1_\Gamma}"']
  &
  \Gamma
  \arrow["\cong"', shorten=2em, Rightarrow, from=A, to=B]
\end{tikzcd}
\]

We set $\iota \defeq \Factone{\cartesianlift{1_\Gamma}{T}}{1_T \cdot 1_T}$, and it remains to show that we have an invertible 2-cell from $\cartesianlift{1_\Gamma}{T} \cdot \iota$ to $1_T$.
To do so, we use Item~\ref{item:def-cartesian-1cell-2cell} of Definition~\ref{def:cart-1-cell}, and we consider the following diagram.
\[
\begin{tikzcd}
	{\substTy{T}{1_\Gamma}} && {\substTy{T}{1_\Gamma}} \\
	\\
	&&&& {\substTy{T}{1_\Gamma}} && T \\
	\\
	\Gamma && \Gamma && \Gamma && \Gamma
	\arrow["{1_{\substTy{T}{s}}}", from=1-3, to=3-5]
	\arrow["{\cartesianlift{1_\Gamma}{T}}"', from=3-5, to=3-7]
	\arrow["{1_\Gamma}"', from=5-3, to=5-5]
	\arrow["{1_\Gamma}"', from=5-5, to=5-7]
	\arrow["{\cartesianlift{1_\Gamma}{T} \cdot \iota}"', from=1-1, to=3-5]
	\arrow["{1_\Gamma}"', from=5-1, to=5-3]
\end{tikzcd}
\]
We need to construct an invertible 2-cell $\delta$ from $1_\Gamma \cdot 1_\Gamma$ to $1_\Gamma$ and an invertible 2-cell $\disp{\sigma}$ from $\cartesianlift{1_\Gamma}{T} \cdot \iota \cdot \cartesianlift{1_\Gamma}{T}$ to $1_{\substTy{T}{s}} \cdot \cartesianlift{1_\Gamma}{T}$.
For $\delta$, we take $\lambda$.
For $\disp{\sigma}$, we use that $\iota \cdot \cartesianlift{1_\Gamma}{T} \cong 1_T \cdot 1_T$, and thus we have the following composition of invertible 2-cells.
\begin{equation*}
\begin{split}
\cartesianlift{1_\Gamma}{T} \cdot \iota \cdot \cartesianlift{1_\Gamma}{T}
&\cong \cartesianlift{1_\Gamma}{T} \cdot 1_T \cdot 1_T\\
&\cong \cartesianlift{1_\Gamma}{T}\\
&\cong 1_{\substTy{T}{s}} \cdot \cartesianlift{1_\Gamma}{T}
\end{split}
\end{equation*}
The desired invertible 2-cell is thus defined by $\Eqtwo{\delta}{\disp{\sigma}}$.

\subsubsection{Interpretation of \ruleSubComp}
\label{sec:interpr-rules-1}

To interpret \ruleref{\ruleSubComp}, we suppose that we have substitutions $s' : \Epsilon \rightarrow \Delta$ and $s : \Delta \rightarrow \Gamma$, and a type $T$ in $\Gamma$.
Hence we have the following diagram.
\[
\begin{tikzcd}
	{\substTy{T}{s \cdot s'}} \\
	\\
	{\substTy{\substTy{T}{s'}}{s}} && {\substTy{T}{s'}} && T \\
	\\
	\Epsilon && \Delta && \Gamma
	\arrow["{s'}"', from=5-3, to=5-5]
	\arrow["s"', from=5-1, to=5-3]
	\arrow["{\cartesianlift{s \cdot s'}{T}}", from=1-1, to=3-5]
	\arrow["{\cartesianlift{s'}{T}}"', from=3-3, to=3-5]
	\arrow["{\cartesianlift{s}{\substTy{T}{s'}}}"', from=3-1, to=3-3]
\end{tikzcd}
\]
Note that both $\cartesianlift{s \cdot s'}{T}$ and $\cartesianlift{s}{\substTy{T}{s'}} \cdot \cartesianlift{s'}{T}$ are cartesian 1-cells over $s \cdot s'$, since cartesian 1-cells are closed under composition by Proposition~\ref{prop:example_carts}.
For that reason, we get an adjoint equivalence $\subcomp{s}{s'} : \substTy{\substTy{T}{s'}}{s} \rightarrow \substTy{T}{s \cdot s'}$ making the diagram below commute up to invertible 2-cell by Construction~\ref{const:equiv_carts}.
\[
\begin{tikzcd}
	{\substTy{T}{s \cdot s'}} \\
	\\
	{\substTy{\substTy{T}{s'}}{s}} && {\substTy{T}{s'}} && T \\
	\\
	\Epsilon && \Delta && \Gamma
	\arrow["{s'}"', from=5-3, to=5-5]
	\arrow["s"', from=5-1, to=5-3]
	\arrow[""{name=0, anchor=center, inner sep=0}, "{\cartesianlift{s \cdot s'}{T}}", from=1-1, to=3-5]
	\arrow["{\cartesianlift{s'}{T}}"', from=3-3, to=3-5]
	\arrow[""{name=1, anchor=center, inner sep=0}, "{\cartesianlift{s}{\substTy{T}{s'}}}"', from=3-1, to=3-3]
	\arrow["{\subcomp{s}{s'}}", dashed, from=3-1, to=1-1]
	\arrow["\cong"', shorten=1.2em, Rightarrow, from=0, to=1]
\end{tikzcd}
\]

\subsubsection{Interpretation of \ruleTmSubId}
\label{sec:interpr-ruletms}

Next we interpret \ruleref{\ruleTmSubId}, and for that, we suppose that we have a context $\Gamma$, types $S$ and $T$ over $\Gamma$, and a 1-cell $t : S \to T$ over $1_\Gamma$.
We need to construct an invertible 2-cell from $\invsubid{} \cdot \substTm{t}{1_{\Gamma}} \cdot \subid$ to $t$, and for that, it suffices to construct an invertible 2-cell from
$\substTm{t}{1_{\Gamma}} \cdot \subid$ to $\subid \cdot t$.
Recall that we defined $\subid$ to be $\cartesianlift{1_\Gamma}{T}$, and recall that we defined $\substTm{t}{1_\Gamma}$ as follows.
\[
\begin{tikzcd}[column sep = large]
	&
	\substTy{S}{s}
	\ar[r, "{\cartesianlift{1_\Gamma}{S}}"]
	\ar[ld, equal]
	&
	| [alias=S] | S
	\ar[rd, "t"]
	\\
	\substTy{S}{1_\Gamma}
	\ar[rrd, "\beta" near start, ""{name=U}, ""'{name=V}, bend left=20]     
	\ar[rd, dashed, "{\substTm{t}{1_\Gamma}}"']
	&
	&
	&
	T
	\ar[ld, equal]
	\\
	&
	| [alias=Ts] | \substTy{T}{1_\Gamma}
	\ar[r, "{\cartesianlift{1_\Gamma}{T}}"']
	&
	T
	\\
	\Gamma
	\ar[r, "1_\Gamma"']
	&
	\Gamma
	\ar[r, "1_\Gamma"']
	&
	\Gamma
	\ar[r, "1_\Gamma"']
	&
	\Gamma
	\ar[from=S,to=U, Rightarrow, shorten=1.5em, "\cong"]
	\ar[from=Ts,to=V, Rightarrow, shorten=0.2em, "\cong"]
\end{tikzcd}
\]
By composing the two invertible 2-cells depicted in this diagram, we get the desired invertible 2-cell.
The rule \ruleref{\ruleTmSubComp} is interpreted analogously.

\subsection{Local Substitution}
Next we interpret the rules in Figure~\ref{fig:rules-local-substitution1} (in Sections~\ref{sec:interpr-rulem} and \ref{sec:interpr-ruler-1}), Figure~\ref{fig:rules-local-substitution2} (in Sections~\ref{sec:interpr-rulem-1} and \ref{sec:interpr-rulem-2}), and Figure~\ref{fig:rules-coherence-substitution} (in Section~\ref{sec:interpr-rulem-3}).
For the interpretation, we use the local opcleaving.

\subsubsection{Interpretation of \ruleMap}
\label{sec:interpr-rulem}

We start with \ruleref{\ruleMap}.
Suppose we have two substitutions $s, s' : \Delta \rightarrow \Gamma$ and a 2-cell $\rho : s \Rightarrow s'$.
In addition, we assume that we have a type $T$ in context $\Gamma$.
Our goal is to construct a 1-cell $\map{T}{\rho} : \substTy{T}{s} \rightarrow \substTy{T}{s'}$.
From the local opcleaving, we obtain $\opcartesianliftmor{\rho}{\cartesianlift{s}{T}}$ over $1_\Delta \cdot s'$ as the pushforward of $\cartesianlift{s}{T}$ along $\rho \vcomp \leftunitor{s'}^{-1}$.
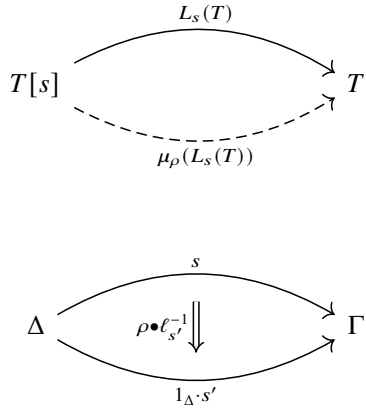
\[
\begin{tikzcd}[column sep = huge, row sep = huge]
  \substTy{T}{s} \ar[rr, bend right, dashed, "\opcartesianliftmor{\rho}{\cartesianlift{s}{T}}"'] \ar[rr, bend left, "\cartesianlift{s}{T}"]
  &&
  T
  \\
  \\
  \Delta \ar[rr, bend left, "s", ""{name=U, below}] \ar[rr, bend right, "1_\Delta \cdot s'"', ""{name=D, above}]
  &&
  \Gamma
  \arrow[Rightarrow, shorten=.5em, from=U, to=D, "\rho \vcomp \ell_{s'}^{-1}"']
\end{tikzcd}
\]
Since $\cartesianlift{s'}{T}$ is cartesian, we can factor $\opcartesianliftmor{\rho}{\cartesianlift{s}{T}}$ through it.
From that, we obtain $\map{T}{\rho}$.
\[
\begin{tikzcd}[column sep = huge, row sep = large]
	{\substTy{T}{s}} \\
	\\
	{\substTy{T}{s'}} && T
	\arrow["{\cartesianlift{s'}{T}}"', from=3-1, to=3-3]
	\arrow["{\opcartesianliftmor{\rho}{\cartesianlift{s}{T}}}", from=1-1, to=3-3]
	\arrow["{\map{T}{\rho}}"', dashed, from=1-1, to=3-1]
\end{tikzcd}
\]

\subsubsection{Interpretation of \ruleRewTm}
\label{sec:interpr-ruler-1}

Next, we give the interpretation for the rule \ruleref{\ruleRewTm}.
Suppose we have two 1-cells $s, s' : \Delta \rightarrow \Gamma$ and a 2-cell $\rho : s \rightarrow s'$.
In addition, we assume that we have types $S, T$ in context $\Gamma$ and a 1-cell $\typeonecell{\Gamma}{t}{S}{T}$.
Our goal is to construct an invertible 2-cell between $\vertcomp{\substTm{t}{s}}{(\map{T}{\rho})}$ and $\vertcomp{(\map{S}{\rho})}{ \substTm{t}{s'}}$.
For that it suffices to construct:
\begin{enumerate}
	\item An invertible 2-cell $\delta$ from $1_\Delta \cdot 1_\Delta$ to $1_\Delta \cdot 1_\Delta$; and
	\item An invertible 2-cell $\disp{\sigma}$ from $\vertcomp{\vertcomp{\substTm{t}{s}}{(\map{T}{\rho})}}{\cartesianlift{s'}{T}}$ to $\vertcomp{\vertcomp{(\map{S}{\rho})}{ \substTm{t}{s'}}}{\cartesianlift{s'}{T}}$.
\end{enumerate}
Then we define $\rewTm{t}{\rho}$ to be $\Eqtwo{\delta}{\disp{\sigma}}$.
For $\delta$ we take $1_{1_\Delta \cdot 1_\Delta}$.

To construct $\disp{\sigma}$, we first note that by construction of $\substTm{t}{s'}$, we have the following invertible 2-cell.
\[
\vertcomp{\vertcomp{(\map{S}{\rho})}{ \substTm{t}{s'}}}{\cartesianlift{s'}{T}}
\cong
\vertcomp{\vertcomp{(\map{S}{\rho})}{\cartesianlift{s'}{S}}}{t}
\]
Note that by construction of $\map{S}{\rho}$, the following triangle commutes up to invertible 2-cell.
\[
\begin{tikzcd}[column sep = huge, row sep = large]
	{\substTy{S}{s}} \\
	\\
	{\substTy{S}{s'}} && T
	\arrow["{\cartesianlift{s'}{S}}"', from=3-1, to=3-3]
	\arrow["{\opcartesianliftmor{\rho}{\cartesianlift{s}{S}}}", from=1-1, to=3-3]
	\arrow["{\map{S}{\rho}}"', from=1-1, to=3-1]
\end{tikzcd}
\]
Hence, we get
\[
\gamma_1 : \vertcomp{\vertcomp{(\map{S}{\rho})}{\cartesianlift{s'}{S}}}{t}
\cong
\vertcomp{\opcartesianliftmor{\rho}{\cartesianlift{s}{S}}}{t}.
\]
Similarly, we have an invertible 2-cell $\map{T}{\rho} \cdot \cartesianlift{s'}{S} \cong \opcartesianliftmor{\rho}{\cartesianlift{s}{T}}$, and thus we obtain
\[
\gamma_2 : \vertcomp{\vertcomp{\substTm{t}{s}}{(\map{T}{\rho})}}{\cartesianlift{s'}{T}}
\cong
\vertcomp{\substTm{t}{s}}{\opcartesianliftmor{\rho}{\cartesianlift{s}{T}}}.
\]
It thus suffices to construct an invertible 2-cell from $\vertcomp{\substTm{t}{s}}{\opcartesianliftmor{\rho}{\cartesianlift{s}{T}}}$ to $\vertcomp{\opcartesianliftmor{\rho}{\cartesianlift{s}{S}}}{t}$.
We can depict the situation with the following square.
\[
\begin{tikzcd}[column sep = huge, row sep = large]
	{\substTy{S}{s}} && S \\
	\\
	{\substTy{T}{s}} && T
	\arrow["t", from=1-3, to=3-3]
	\arrow["{\substTm{t}{s}}"', from=1-1, to=3-1]
	\arrow[""{name=0, anchor=center, inner sep=0}, "{\cartesianlift{s}{S}}", bend left=15, shift left=1, from=1-1, to=1-3]
	\arrow[""{name=1, anchor=center, inner sep=0}, "{\cartesianlift{s}{T}}"', bend right=15, shift right=1, from=3-1, to=3-3]
	\arrow[""{name=2, anchor=center, inner sep=0}, "{\opcartesianliftmor{\rho}{\cartesianlift{s}{S}}}"', bend right=15, shift right=1, from=1-1, to=1-3]
	\arrow[""{name=3, anchor=center, inner sep=0}, "{\opcartesianliftmor{\rho}{\cartesianlift{s}{T}}}", bend left=15, shift left=1, from=3-1, to=3-3]
	\arrow["\tau", shorten=.5em, Rightarrow, from=0, to=2]
	\arrow["{\tau'}"', shorten=.5em, Rightarrow, from=1, to=3]
\end{tikzcd}
\]
In this diagram, $\tau$ and $\tau'$ are the opcartesian 2-cells coming from the opcartesian lift.
Note that by construction of $\substTm{t}{s}$, the outer square of this diagram commutes.
More precisely, we have an invertible 2-cell $\theta : \substTm{t}{s} \cdot \cartesianlift{s}{T} \cong \cartesianlift{s}{S} \cdot t$.

To construct an invertible 2-cell from $\opcartesianliftmor{\rho}{\cartesianlift{s}{S}} \cdot t$ to $\substTm{t}{s} \cdot \opcartesianliftmor{\rho}{\cartesianlift{s}{T}}$, we are going to construct two opcartesian 2-cells.
First, note that we have a 2-cell
\[
\tau \whiskerr t : \cartesianlift{s}{S} \cdot t \Rightarrow \opcartesianliftmor{\rho}{\cartesianlift{s}{S}} \cdot t'
\]
that lies over $\beta_1 \defeq (\rho \vcomp l^{-1}) \whiskerr 1_\Gamma : s \cdot 1_\Gamma \Rightarrow (1_\Delta \cdot s') \cdot 1_\Gamma$.
This 2-cell is opcartesian, because opcartesian 2-cells are closed under right whiskering.

Note that we also have the 2-cell
\[
\vertcomp{\theta^{-1}}{\substTm{t}{s} \whiskerl \tau } : \cartesianlift{s}{S} \cdot t \Rightarrow \substTm{t}{s} \cdot \opcartesianliftmor{\rho}{\cartesianlift{s}{T}}
\]
which lies over $\beta_2 \defeq r \vcomp \ell^{-1} \vcomp 1_{\Gamma} \whiskerl (\rho \vcomp \ell^{-1}) : s \cdot 1_{\Gamma} \Rightarrow 1_{\Delta} \cdot (1_{\Delta} \cdot s')$.
It is opcartesian, because left whiskering also preserves opcartesian 2-cells.

We also have the following invertible 2-cell
\[
\beta_3 \defeq r \vcomp \ell^{-1}
:
(1_\Delta \cdot s') \cdot 1_\Gamma
\cong
1_{\Delta} \cdot (1_{\Delta} \cdot s'),
\]
and note that $\beta_1 \vcomp \beta_3 = \beta_2$ by naturality of the unitors.
From all of this we get the desired invertible 2-cell $\opcartesianliftmor{\rho}{\cartesianlift{s}{S}} \cdot t \cong \substTm{t}{s} \cdot \opcartesianliftmor{\rho}{\cartesianlift{s}{T}}$.

\subsubsection{Interpretation of \ruleMapId}
\label{sec:interpr-rulem-1}

To interpret \ruleref{\ruleMapId}, we assume that we have a 1-cell $s : \Delta \rightarrow \Gamma$ and a type $T$ in context $\Gamma$.
Our goal is to construct an invertible 2-cell between the following compositions.
\[
\begin{tikzcd}[row sep = large, column sep = large]
	{\substTy{T}{s}} && {\substTy{T}{s}} && T
	\arrow["{\cartesianlift{s}{T}}", from=1-3, to=1-5]
	\arrow["{1_{\substTy{T}{s}}}", shift right=3, from=1-1, to=1-3, bend left=10, shift left=4]
	\arrow["{\map{T}{1_s}}"', shift left=3, from=1-1, to=1-3, bend right=10, shift right=4]
\end{tikzcd}
\]
To do so, we need to construct:
\begin{enumerate}
	\item An invertible 2-cell $\delta$ from $1_\Delta$ to $1_\Delta$; and
	\item An invertible 2-cell $\disp{\sigma}$ from $\map{T}{1_s} \cdot \cartesianlift{s}{T}$ to $\cartesianlift{s}{T}$.
\end{enumerate}
The desired 2-cell is then defined to be $\Eqtwo{\delta}{\disp{\sigma}}$.
For $\delta$ we take $1_{1_\Delta}$.

To construct $\disp{\sigma}$, we first note that by construction of $\map{T}{\rho}$, the following triangle commutes up to an invertible 2-cell $\tau : \map{T}{\rho} \cdot \cartesianlift{s'}{T} \cong \opcartesianliftmor{\rho}{\cartesianlift{s}{T}}$.
\[
\begin{tikzcd}[column sep = huge, row sep = large]
	{\substTy{T}{s}} \\
	\\
	{\substTy{T}{s'}} && T
	\arrow["{\map{T}{\rho}}"', from=1-1, to=3-1]
	\arrow["{\cartesianlift{s'}{T}}"', from=3-1, to=3-3]
	\arrow[""{name=0}, "{\opcartesianliftmor{\rho}{\cartesianlift{s}{T}}}", from=1-1, to=3-3]
	\arrow["\tau"', shorten=1em, Rightarrow, from=3-1, to=0]
\end{tikzcd}
\]
Next we recall that $\opcartesianliftmor{\rho}{\cartesianlift{s}{T}}$ was constructed as the following pushforward.
\[
\begin{tikzcd}[row sep = large, column sep = large]
	{\substTy{T}{s}} && T
	\arrow[""{name=0}, bend right, "{\opcartesianliftmor{1_s}{\cartesianlift{s}{T}}}"',  dashed, from=1-1, to=1-3]
	\arrow[""{name=1}, "{\cartesianlift{s}{T}}", bend left, from=1-1, to=1-3]
	\arrow[shorten=.5em, Rightarrow, dashed, from=1, to=0, "\theta"]
	\\
	\\
	\Delta \ar[rr, bend left, "s", ""{name=U, below}] \ar[rr, bend right, "1_\Delta \cdot s"', ""{name=D, above}]
	& &
	\Gamma
	\arrow[Rightarrow, shorten=.3em, from=U, to=D, "\leftunitor{s}^{-1}"]
\end{tikzcd}
\]
Since $\leftunitor{s}^{-1}$ is invertible, the opcartesian lift $\theta$ is invertible as well.
Hence, we define $\disp{\sigma} \defeq \vertcomp{\tau}{\theta^{-1}}$.
We interpret \ruleref{\ruleMapComp} similarly.

\subsubsection{Interpretation of \ruleMapWhiskerL}
\label{sec:interpr-rulem-2}

To interpret \ruleref{\ruleMapWhiskerL} and \ruleref{\ruleMapWhiskerR}, we use the same idea, and we need to use that we assumed that opcartesian 2-cells are closed under whiskering.
We give the precise details for \ruleref{\ruleMapWhiskerL}.
Suppose we have a 1-cell $s : \Epsilon \rightarrow \Delta$, a 2-cell $\rho : s' \Rightarrow s''$ where $s', s'' : \Delta \rightarrow \Gamma$, and a type $T$ in context $\Gamma$.
Our goal is to construct an invertible 2-cell from $\map{T}{(s \whiskerl \rho)}$ to $\invsubcomp{s}{s'} \cdot \substTm{(\map{T}{\rho})}{s} \cdot \subcomp{s}{s''}$.
Since $\map{T}{\rho}$ was constructed as a cartesian factorization, it suffices to construct:
\begin{enumerate}
	\item An invertible 2-cell $\delta$ from $1_\Delta$ to $1_\Delta \cdot 1_\Delta \cdot 1_\Delta$; and
	\item An invertible 2-cell $\disp{\sigma}$ from $\map{T}{(s \whiskerl \rho)} \cdot \cartesianlift{s \cdot s''}{T}$ to $\invsubcomp{s}{s'} \cdot \substTm{(\map{T}{\rho})}{s} \cdot \subcomp{s}{s''} \cdot \cartesianlift{s \cdot s''}{T}$.
\end{enumerate}
With those in place, the desired invertible 2-cell is defined to be $\Eqtwo{\delta}{\disp{\sigma}}$.
We define $\delta$ to be $\ell_{1_\Delta}^{-1} \cdot \ell_{1_\Delta \cdot 1_\Delta}^{-1}$.

For the construction of $\disp{\sigma}$, let us start by recalling the construction of $\map{T}{(s \whiskerl \rho)}$.
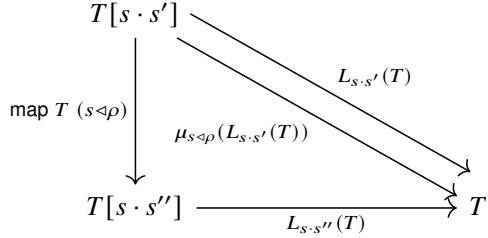
\[
\begin{tikzcd}[column sep = huge, row sep = large]
	{\substTy{T}{s \cdot s'}} \\
	\\
	{\substTy{T}{s \cdot s''}} && T
	\arrow["{\cartesianlift{s \cdot s''}{T}}"', from=3-1, to=3-3]
	\arrow["{\map{T}{(s \whiskerl \rho)}}"', from=1-1, to=3-1]
	\arrow["{\cartesianlift{s \cdot s'}{T}}", shift left=4, from=1-1, to=3-3]
	\arrow["{\opcartesianliftmor{s \whiskerl \rho}{\cartesianlift{s \cdot s'}{T}}}"', from=1-1, to=3-3]
\end{tikzcd}
\]
Next we inspect the definition of $\subcomp{s}{s''}$.
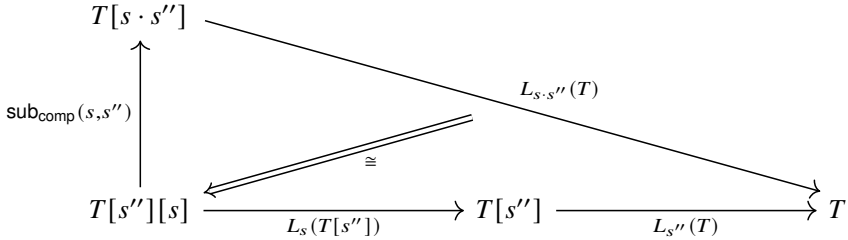
\[
\begin{tikzcd}[column sep = huge, row sep = large]
	{\substTy{T}{s \cdot s''}} \\
	\\
	{\substTy{\substTy{T}{s''}}{s}} && {\substTy{T}{s''}} && T
	\arrow[""{name=0, anchor=center, inner sep=0}, "{\cartesianlift{s \cdot s''}{T}}", shift left=2, from=1-1, to=3-5]
	\arrow["{\cartesianlift{s''}{T}}"', from=3-3, to=3-5]
	\arrow["{\cartesianlift{s}{\substTy{T}{s''}}}"', from=3-1, to=3-3]
	\arrow["{\subcomp{s}{s''}}", from=3-1, to=1-1]
	\arrow["\cong", shorten <=15pt, Rightarrow, from=0, to=3-1]
\end{tikzcd}
\]
From this, we already get the following invertible 2-cell.
\begin{align*}
\gamma_1 : & \invsubcomp{s}{s'} \cdot \substTm{(\map{T}{\rho})}{s} \cdot \subcomp{s}{s''} \cdot \cartesianlift{s \cdot s''}{T}
  \\
           & \cong
  \\
           & \invsubcomp{s}{s'} \cdot \substTm{(\map{T}{\rho})}{s} \cdot \cartesianlift{s}{\substTy{T}{s''}} \cdot \cartesianlift{s''}{T}
\end{align*}
We also inspect the construction of $\substTm{(\map{T}{\rho})}{s}$, from which we get the following invertible 2-cell.
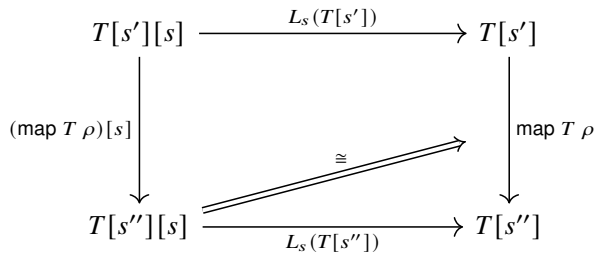
\[
\begin{tikzcd}[column sep = huge, row sep = large]
	{\substTy{\substTy{T}{s'}}{s}} && {\substTy{T}{s'}} \\
	\\
	{\substTy{\substTy{T}{s''}}{s}} && {\substTy{T}{s''}}
	\arrow[""{name=0, anchor=center, inner sep=0}, "{\map{T}{\rho}}", from=1-3, to=3-3]
	\arrow["{\cartesianlift{s}{\substTy{T}{s''}}}"', from=3-1, to=3-3]
	\arrow["{\cartesianlift{s}{\substTy{T}{s'}}}", from=1-1, to=1-3]
	\arrow["{\substTm{(\map{T}{\rho})}{s}}"', from=1-1, to=3-1]
	\arrow["\cong", shorten >=16pt, Rightarrow, from=3-1, to=0]
\end{tikzcd}
\]
As such, we get another invertible 2-cell as follows.
\begin{align*}
\gamma_2 : & \invsubcomp{s}{s'} \cdot \substTm{(\map{T}{\rho})}{s} \cdot \cartesianlift{s}{\substTy{T}{s''}} \cdot \cartesianlift{s''}{T}
  \\
           & \cong
  \\
           & \invsubcomp{s}{s'} \cdot \cartesianlift{s}{\substTy{T}{s'}} \cdot \map{T}{\rho} \cdot \cartesianlift{s''}{T}.
\end{align*}
Let us recall the construction of $\map{T}{\rho}$ as well.
\[
\begin{tikzcd}[column sep = huge, row sep = large]
	{\substTy{T}{s'}} \\
	\\
	{\substTy{T}{s''}} && T
	\arrow["{\map{T}{\rho}}"', from=1-1, to=3-1]
	\arrow["{\cartesianlift{s''}{T}}"', from=3-1, to=3-3]
	\arrow[""{name=0} near end, "{\opcartesianliftmor{\rho}{\cartesianlift{s'}{T}}}", from=1-1, to=3-3]
	\arrow["\cong"', shorten=1em, Rightarrow, from=3-1, to=0]
\end{tikzcd}
\]
From this, we obtain yet another invertible 2-cell.
\begin{align*}
  \gamma_3 : & \invsubcomp{s}{s'} \cdot \substTm{(\map{T}{\rho})}{s} \cdot \cartesianlift{s}{\substTy{T}{s''}} \cdot \cartesianlift{s''}{T}
  \\
             & \cong
  \\
             & \invsubcomp{s}{s'} \cdot \cartesianlift{s}{\substTy{T}{s'}} \cdot \opcartesianliftmor{\rho}{\cartesianlift{s''}{T}}
\end{align*}
Hence, we achieve our goal if we construct an invertible 2-cell of the following form.
\[
\gamma_4 :\invsubcomp{s}{s'} \cdot \cartesianlift{s}{\substTy{T}{s'}} \cdot \opcartesianliftmor{\rho}{\cartesianlift{s'}{T}}
\cong
\opcartesianliftmor{s \whiskerl \rho}{\cartesianlift{s \cdot s'}{T}}
\]
We can depict this situation as follows.
\[
\begin{tikzcd}[column sep = huge, row sep = large]
	{\substTy{T}{s \cdot s'}} &&&& T \\
	\\
	{\substTy{\substTy{T}{s'}}{s}} && {\substTy{T}{s'}} && T \\
	\\
	\Epsilon && \Delta && \Gamma
	\arrow["s"', from=5-1, to=5-3]
	\arrow[""{name=0, anchor=center, inner sep=0}, "{s'}", bend left=15, shift left=1, from=5-3, to=5-5]
	\arrow[""{name=1, anchor=center, inner sep=0}, "{s''}"', bend right=15, shift right=1, from=5-3, to=5-5]
	\arrow[""{name=2, anchor=center, inner sep=0}, "{\cartesianlift{s'}{T}}", bend left=15, shift left=1, from=3-3, to=3-5]
	\arrow[""{name=3, anchor=center, inner sep=0}, "{\opcartesianliftmor{\rho}{\cartesianlift{s'}{T}}}"', bend right=15, shift right=1, from=3-3, to=3-5]
	\arrow["{\invsubcomp{s}{s'}}"', from=1-1, to=3-1]
	\arrow["{\cartesianlift{s}{\substTy{T}{s'}}}"', from=3-1, to=3-3]
	\arrow[""{name=4, anchor=center, inner sep=0}, "{\cartesianlift{s \cdot s'}{T}}", bend left=15, shift left=1, from=1-1, to=1-5]
	\arrow[""{name=5, anchor=center, inner sep=0}, "{\opcartesianliftmor{s \whiskerl \rho}{\cartesianlift{s \cdot s'}{T}}}"', bend right=15, shift right=1, from=1-1, to=1-5]
	\arrow["\rho", shorten=.5em, Rightarrow, from=0, to=1]
	\arrow["\tau", shorten=.5em, Rightarrow, from=2, to=3]
	\arrow["{\tau'}", shorten=.7em, Rightarrow, from=4, to=5]
        \ar[from=1-5, to=3-5, equal]
\end{tikzcd}
\]
Here $\tau$ and $\tau'$ are the 2-cells coming from the opcartesian lifts.

Since whiskering preserves opcartesian 2-cells, the following 2-cell is opcartesian.
\begin{align*}
(\invsubcomp{s}{s'} \cdot \cartesianlift{s}{\substTy{T}{s'}}) \whiskerl \tau
\enspace : \enspace
  & \invsubcomp{s}{s'} \cdot \cartesianlift{s}{\substTy{T}{s'}} \cdot \cartesianlift{s'}{T}
    \\
  & \Rightarrow
    \\
  & \invsubcomp{s}{s'} \cdot \cartesianlift{s}{\substTy{T}{s'}} \cdot \opcartesianliftmor{\rho}{\cartesianlift{s'}{T}}
\end{align*}
Now we unfold the definition of $\subcomp{s}{s'}$, and from that we get the following invertible 2-cell.
\[
\begin{tikzcd}[column sep = huge, row sep = large]
	{\substTy{T}{s \cdot s'}} \\
	\\
	{\substTy{\substTy{T}{s'}}{s}} && {\substTy{T}{s'}} && T
	\arrow[""{name=0, anchor=center, inner sep=0}, "{\cartesianlift{s \cdot s'}{T}}", bend left=15, shift left=0, from=1-1, to=3-5]
	\arrow["{\cartesianlift{s'}{T}}"', from=3-3, to=3-5]
	\arrow["{\cartesianlift{s}{\substTy{T}{s'}}}"', ""{name=T}, from=3-1, to=3-3]
	\arrow["{\invsubcomp{s}{s'}}"', from=1-1, to=3-1]
	\arrow["\cong", shorten=1.2em, Rightarrow, from=0, to=3-3]
\end{tikzcd}
\]
Since invertible 2-cells are opcartesian and opcartesian 2-cells are closed under composition, we get an opcartesian 2-cell
\[
\theta : \cartesianlift{s \cdot s'}{T}
\Rightarrow 
\invsubcomp{s}{s'} \cdot \cartesianlift{s}{\substTy{T}{s'}} \cdot \opcartesianliftmor{\rho}{\cartesianlift{s'}{T}}.
\]
Since we already had an opcartesian 2-cell $\tau' : \cartesianlift{s \cdot s'}{T} \Rightarrow \opcartesianliftmor{s \whiskerl \rho}{\cartesianlift{s \cdot s'}{T}}$, we get an invertible 2-cell between the codomains of $\theta$ and $\tau'$.
Hence, we obtain an invertible 2-cell
\[
\gamma_5 : \invsubcomp{s}{s'} \cdot \cartesianlift{s}{\substTy{T}{s'}} \cdot \opcartesianliftmor{\rho}{\cartesianlift{s'}{T}}
\cong 
\opcartesianliftmor{s \whiskerl \rho}{\cartesianlift{s \cdot s'}{T}}.
\]
By chaining all the invertible 2-cells, we get the desired 2-cell $\disp{\sigma}$.

\subsubsection{Interpretation of \ruleSubCompMapSubIdR}
\label{sec:interpr-rulem-3}
Finally, we show how to interpret \ruleref{\ruleSubCompMapSubIdR}.
Suppose that we have objects $\Gamma$ and $\Delta$, an object $T$ over $\Gamma$, and a 1-cell $s : \Delta \rightarrow \Gamma$.
Our goal is to construct an invertible 2-cell from $\horcomp{\subcomp{s}{1_\Delta}} {(\map{T}{\rightunitor{s}})}$ to $\subid$.
To do so, it suffices to construct an invertible 2-cell from $\subcomp{s}{1_\Delta} \cdot (\map{T}{\rightunitor{s}}) \cdot \cartesianlift{s}{T}$ to $\subid \cdot \cartesianlift{s}{T}$.
Our situation is depicted in the following diagram.
\[
\begin{tikzcd}
	& {\substTy{T}{s \cdot 1_\Delta}} \\
	{\substTyTwo{T}{s}{1_\Delta}} && {\substTy{T}{s}} & T
	\arrow["{\cartesianlift{s}{T}}"', from=2-3, to=2-4]
	\arrow["\subid"', from=2-1, to=2-3]
	\arrow["{\subcomp{s}{1_\Delta}}", from=2-1, to=1-2]
	\arrow["{\map{T}{\rightunitor{s}}}", from=1-2, to=2-3]
\end{tikzcd}
\]
Recall that $\subid : \substTyTwo{T}{s}{1_\Delta} \rightarrow \substTy{T}{s}$ is defined to be $\cartesianlift{1_\Delta}{\substTy{T}{s}}$.
By construction, we have an invertible 2-cell
\[
\varphi_1 :
\map{T}{\rightunitor{s}} \cdot \cartesianlift{s}{T}
\mytwocell
\opcartesianliftmor{\rightunitor{s}}{\cartesianlift{s \cdot 1_{\Delta}}{T}} .
\]
In addition, since $\map{T}{\rightunitor{s}}$ was constructed as an opcartesian lift, we have the following invertible 2-cell.
\[
\varphi_2 :
\cartesianlift{s \cdot 1_{\Delta}}{T}
\mytwocell
\opcartesianliftmor{\rightunitor{s}}{\cartesianlift{s \cdot 1_{\Delta}}{T}}
\]
Note that $\varphi_2$ is invertible: this is because it is the opcartesian lift of an invertible 2-cell.
By construction of $\subcomp{s}{1_\Delta}$, we have the following invertible 2-cell.
\[
\varphi_3 :
\subcomp{s}{1_\Delta} \cdot \cartesianlift{s \cdot 1_{\Delta}}{T}
\mytwocell
\cartesianlift{1_{\Delta}}{\substTy{T}{1_\Delta}} \cdot \cartesianlift{s}{T} .
\]
The composition $\varphi_3 \vcomp \varphi_2 \vcomp \varphi_1^{-1}$ yields the desired 2-cell.

We omit the description of the verification of several equalities.
All in all, we can state the following theorem.

\begin{theorem}[Soundness]
\label{thm:soundness}
We can interpret \BTT in any weak comprehension bicategory.
\end{theorem}

\section{Variations on Syntax}
\label{sec:variations-syntax}

\BTT is a very complicated type theory, and might not be feasible to implement or use in practice.
Its main purpose is to serve as a framework for studying specialized syntax and the corresponding semantics.
Based on a user's goal, they might adopt some of the following simplifications.

\begin{enumerate}[label=\bfseries V\arabic*:, align=left,
  ref={V\arabic*},topsep=0pt]
\item \textbf{Strictness.} The rules in Figures~\ref{fig:rules-base} and \ref{fig:rules-total}
  are aimed at bicategories.
  When working with strict 2-categories instead, the unitors and associators would become equalities, and as a result, rules for inverse laws, naturality, and the pentagon and triangle equations are not needed.
  For this variant, an equality judgment on contexts, types, substitutions, and terms would be added to \BTT.
  \label{var:strict}
\item \textbf{Splitness.} We could assume the comprehension bicategory to be split in addition to being strict; thus, in particular the rules \ruleref{\ruleSubId} and \ruleref{\ruleSubComp} would collapse into ordinary equalities.
  \label{var:split}

\item \textbf{Undirected TT.} We could add a rule postulating inverses of reductions. Semantically, this would amount to working in groupoid-enriched categories.
  \label{var:undir}

\item \textbf{Proof-irrelevant reductions.} Our syntax, and the semantics, allow us to distinguish parallel reductions (2-cells). We could instead ``truncate'' them, by moving to \proofirrelevant reductions, making the judgmental equality on them superfluous.
This would yield a directed analog to the judgments of \MLTT;
semantically, it corresponds to working in poset-enriched categories instead of general bicategories.
\label{var:proof-irr}

\end{enumerate}

We have developed comprehension bicategories with the explicit goal of encompassing previously defined interpretations of higher-dimensional and directed type theory.
In the following remarks we summarize the relationship with two previous works.

\begin{remark}[Comparison to \cite{DBLP:journals/mscs/Garner09}]
  \label{rem:comparison-garner}
  To summarize the differences between Garner's comprehension 2-categories \cite[]{DBLP:journals/mscs/Garner09} and our comprehension bicategories:
  the former are full, strict, split, undirected (\ie locally groupoidal), and incorporate type constructors.
\end{remark}

\begin{remark}[Licata and Harper's work]
\cite{LicataH11} interpret a type in context $\Gamma$ as a strict 1-functor from the category interpreting $\Gamma$ into a 1-category $\CAT$ of categories.
Formally, they thus consider the slice bicategory $\Cat/\CAT$, where $\Cat$ is the bicategory of categories, in which $\CAT$ is assumed to be a 0-cell.
The ``domain'' pseudofunctor $\dom : \Cat/\CAT \to \Cat$ carries the structure of a global cleaving and local opcleaving;
this is more generally the case for any ``domain'' pseudofunctor $\dom : \B/a \to \B$ (\coqfile{Bicategories.DisplayedBicats.ExamplesOfCleavings}{DomainCleaving}).
To construct the comprehension pseudofunctor, one needs to use specifics of the category of strict categories, namely the Grothendieck construction of functors into $\CAT$.
\end{remark}

\section{Terms in Directed Type Theory}
\label{sec:terms}

The notion of term in \BTT, and the interpretation of these terms, is different from the notion of term studied in the syntax and semantics by \cite{LicataH11} or \cite{DBLP:journals/entcs/North19}.
Specifically, our terms correspond to 1-cells in the (total) category of types, and are interpreted as such.
Terms in the sense of Licata and Harper, and North, instead correspond to particular 1-cells in the category of contexts, specifically, to sections of the canonical projections $\pi_{\Gamma,A} : \Gamma.A \to \Gamma$.

One might attempt to reconcile these two, on the syntactic side, by
\begin{enumerate}
\item adding a unit type to \BTT; and
\item asking for context morphisms to be built from terms, that is, from type morphisms. (Semantically, this corresponds to asking for $\chi$ to be full.)
\end{enumerate}

\noindent
However, note that most of our comprehension bicategories are \emph{not} full.
Furthermore, this syntactic modification would still leave a difference between the interpretations of terms given by Licata and Harper, and North, on the one hand, and our interpretation on the other hand.
Specifically,
in Example~\ref{ex:opfibincat-compcat}, terms are interpreted as functors that preserve opcartesian cells; by the Grothendieck construction, terms can equivalently be interpreted as \emph{pseudonatural} transformations. In contrast, terms are interpreted as \emph{lax} natural transformations by others \cite[]{DBLP:journals/entcs/North19,LicataH11}.
Hence, there is a mismatch between how terms are interpreted in comprehension bicategories compared to other interpretations.

We can analyze this mismatch more closely, and suggest a potential way of reconciling it, by looking again at the intended interpretation of two-dimensional and directed type theory in categories via the opcleaving model.
Taking this mismatch seriously leads us in the direction of directed type theory, where there can be different flavors of terms. For example, in work by \cite{nuyts_dtt} terms can be used isovariantly, covariantly, and contravariantly.
In the opcleaving model (Example~\ref{ex:opfibincat-compcat}), these different notions of terms would correspond to different kinds of natural transformations: pseudonatural transformations, lax natural transformations, and oplax natural transformations.

To reconcile this difference, one would need to modify the syntax and the definition of comprehension bicategory.
In the syntax, one would need to add a term judgment for every kind of term present in directed type theory.
Accordingly, one would also add the corresponding operations like substitution.
The notion of comprehension bicategory would also require slight modifications for this purpose; one necessary requirement would be that $\chi$ preserves both cartesian 1-cells and opcartesian 2-cells.
The reason for this change, is that this would be necessary to interpret substitution of a certain class of terms.
More precisely, in the opfibration model, lax transformations can be identified with sections of the projections $\int \D \rightarrow \CC$. 
To interpret substitution for these, we need that certain squares in the arrow bicategory are pullbacks; to guarantee that, one requires that $\chi$ preserves cartesian 1-cells.

It is an open problem to develop categorical semantics for type theories that feature terms with different variances.

\section{Conclusion}

We have introduced the notion of comprehension bicategory, inspired by the notion of comprehension category.
From this semantic notion, we have extracted a two-dimensional core syntax for dependent types, terms, and reductions, and an interpretation of that syntax in comprehension bicategories. In future work, we hope to accompany this soundness result with a completeness result.

Our work is very general. We hope that it gives future investigations into two-dimensional and directed type theories a firm foundation. For instance, as outlined in Section~\ref{sec:judg-and-typal}, in future work we will extend our structural rules with variances and a suitable hom-type former à la \cite{DBLP:journals/entcs/North19}.

\section{Acknowledgements}
  We gratefully acknowledge the work by the Coq development team in providing the Coq proof assistant and surrounding infrastructure, as well as their support in keeping UniMath compatible with Coq.
  We are very grateful to Dan Licata and Bob Harper for their help in understanding their interpretation and how it relates to our framework.
We thank the anonymous referees of the short version, and of the present long version, as well as editor Richard Garner, for their insightful comments and helpful advice.
This work was partially funded by EPSRC under agreement number EP/T000252/1.
This material is based upon work supported by the Air Force Office of Scientific Research under award number FA9550-21-1-0334.
This research was supported by the NWO project ``The Power of Equality'' OCENW.M20.380, which is financed by the Dutch Research Council (NWO).

\section*{Competing Interests}

Ahrens is employed at Delft University of Technology,
North is employed at Utrecht University, and
Van der Weide is employed at Radboud University.

\bibliographystyle{msclike}
\bibliography{literature-mscs}

\end{document}